\documentclass[12pt]{article}
\usepackage{amssymb,amsmath}
\usepackage{graphicx,psfrag}

\newcommand{\bea}{\begin{eqnarray}}
\newcommand{\eea}{\end{eqnarray}}
\newcommand{\simgt}{\hbox{ \raise3pt\hbox to 0pt{$>$}\raise-3pt\hbox{$\sim$} }}
\newcommand{\simlt}{\hbox{ \raise3pt\hbox to 0pt{$<$}\raise-3pt\hbox{$\sim$} }}

\newcommand{\clfn}{\setcounter{footnote}{0}}

\begin{document}

\begin{titlepage}
\title{\bf 
\Large
Family Gauge Symmetry as an Origin of\\ 
Koide's Mass Formula and
Charged Lepton Spectrum
\vspace{7mm}}
\author{
Y.~Sumino
\\ \\ \\ Department of Physics, Tohoku University\\
Sendai, 980--8578 Japan
}
\date{}
\maketitle
\thispagestyle{empty}
\vspace{-3.5truein}
\begin{flushright}
{\bf TU--833}\\
{\bf Dec. 2008}
\end{flushright}
\vspace{4truein}
\begin{abstract}
\noindent
{\small
Koide's mass formula is an empirical
relation among the charged lepton masses which
holds with a striking precision.
We present a model of charged lepton sector 
within an effective field theory
with
$U(3)\times SU(2)$ family gauge symmetry, which predicts
Koide's formula within the present experimental accuracy.
Radiative corrections as well as other corrections to
Koide's mass formula have been taken into account.
We adopt a known mechanism, through which
the charged lepton spectrum is determined by
the vacuum expectation value of a 9--component
scalar field $\Phi$.
On the basis of this mechanism,
we implement the following mechanisms into our model:
(1) The radiative correction induced by family gauge interaction
cancels the QED
radiative correction to Koide's mass formula, assuming a
scenario in which the $U(3)$ family gauge symmetry
and $SU(2)_L$ weak gauge symmetry
are unified at $10^2$--$10^3$~TeV scale;
(2)
A simple potential of $\Phi$ invariant under $U(3)\times SU(2)$
leads to a realistic
charged lepton spectrum, consistent
with the experimental values, assuming that
Koide's formula is protected;
(3) 
Koide's formula is stabilized
by embedding 
$U(3)\times SU(2)$ symmetry in a larger symmetry group.
Formally fine tuning of parameters in the model is circumvented
(apart from two exceptions)
by appropriately connecting the charged lepton spectrum
to the boundary (initial) conditions of the model at the
cut--off scale.
%Physics above the cut--off scale is
%speculated
%that may lead to these boundary conditions.
We also disucss some phenomenological implications.
}
\end{abstract}
\vfil

\end{titlepage}

%%%%%%%%%%%%%%%%%%%%%%%%%%%%%%%%%%%%%%%%%%%%%%%%%%%%%%%%%
\section{Introduction}
\label{s1}
\clfn

Among various properties of elementary particles,
the spectra of the quarks and leptons exhibit unique patterns, and
their origin still remains as a profound mystery.
Within the Standard Model (SM) of elementary particles,
the origin of
the masses and mixings of the quarks and leptons is attributed to
their interactions with the (as yet hypothetical) Higgs boson.
Namely, these are 
the Yukawa interaction in the case of 
the charged leptons and
quarks, and possibly the interaction represented by
dimension--5 operators in the case of
the left--handed neutrinos.
Even if these interactions will be
confirmed experimentally
in the future, since the
couping constants of these interactions are free parameters
of the theory, the underlying mechanism how the texture of these couplings
is determined would remain unrevealed.

There have been many attempts to  approach
the mystery of the fermion masses by identifying empirical relations
among the observed fermion masses and
exploring underlying physics that would lead to such relations.
In particular, Koide's mass formula is an empirical relation
among the charged lepton masses given by \cite{Koide:1982wm}
\bea
\frac{\sqrt{m_e}+\sqrt{m_\mu}+\sqrt{m_\tau}}{\sqrt{m_e+m_\mu+m_\tau}}
=\sqrt{\frac{3}{2}}\, ,
\label{KoideMF}
\eea
which holds with a striking precision.
In fact, substituting the present experimental values of the charged lepton
masses \cite{Amsler:2008zz}, 
the formula is valid within the present experimental accuracies.
The relative experimental
error of the left-hand side (LHS) of eq.~(\ref{KoideMF}) is dominated
by 
%$-2\sqrt{m_\mu/m_\tau}\,\delta m_\tau/m_\tau$
$\frac{1}{\sqrt{6}}\,(m_\mu/m_\tau)^{1/2}(\Delta m_\tau/m_\tau)$
($\Delta m_\tau$ is the experimental error of $m_\tau$) and
is of order $10^{-5}$.
A simple mnemonic of the relation (\ref{KoideMF}) is that the angle between
the two vectors $(\sqrt{m_e},\sqrt{m_\mu},\sqrt{m_\tau})$ 
and $(1,1,1)$ equals $45^\circ$ \cite{Foot:1994yn}.

Given the remarkable accuracy with which Koide's mass formula holds,
many speculations have been raised as to existence of some
physical origin behind
this mass formula 
\cite{Koide:1983qe,Foot:1994yn,Koide:2005nv,Li:2006et,Xing:2006vk,Ma:2006ht,Rosen:2007rt}.
Despite the attempts to find its origin, 
so far no realistic model or mechanism has been found
which predicts Koide's mass formula within the required accuracy.
The most serious problem one
faces in finding a realistic model or mechanism 
is caused by the QED radiative correction \cite{Xing:2006vk}.
Even if one postulates some mechanism at a high
energy scale 
that leads to this mass relation, 
the charged lepton
masses receive the 1--loop
QED radiative corrections given by
\bea
m^{\rm pole}_i = \left[
1+\frac{\alpha}{\pi}\left\{
\frac{3}{4}\log\left(
\frac{\mu^2}{\bar{m}_i(\mu)^2}
\right) +1
\right\}
\right]\,
\bar{m}_i(\mu) \, .
\label{QED1Lcorr}
\eea
$\bar{m}(\mu)$ and $m^{\rm pole}$ denote the 
running mass defined in the modified--minimal--subtraction scheme
($\overline{\rm MS}$ scheme) and the pole mass, respectively;
$\mu$ represents the
renormalization scale.
It is the pole mass  that is measured in experiments.
Suppose  $\bar{m}_i(\mu)$ 
(or the corresponding Yukawa couplings $\bar{y}_i(\mu)$)
satisfy the relation
(\ref{KoideMF}) at a high energy scale $\mu\gg M_W$.
Then $m_i^{\rm pole}$
do not satisfy the same relation \cite{Li:2006et,Xing:2006vk}: 
Eq.~(\ref{KoideMF})
is corrected by approximately 0.1\%, which is 120 times 
larger than the present experimental error.
Note that this correction
originates only from the term $-3\alpha/(4\pi) \times\bar{m}_i \, \log(\bar{m}_i^2)$
of eq.~(\ref{QED1Lcorr}), since the other terms, which are of the form
${\rm const.}\times\bar{m}_i$, do not affect the relation
(\ref{KoideMF}).
This is because, the latter corrections only change the length of the
vector $(\sqrt{m_e},\sqrt{m_\mu},\sqrt{m_\tau})$ but not the direction.
We also note 
that 
$\log(\bar{m}_i^2)$ results from the fact that $\bar{m}_i$
plays a role of an infrared (IR) cut--off in the
loop integral.

The 1--loop weak correction is of the form
${\rm const.}\times\bar{m}_i$ in the leading order of
$\bar{m}_i^2/M_W^2$ expansion;
the leading non--trivial correction is
${\cal O}(G_F \bar{m}_i^3/\pi)$ whose effect is smaller than the 
current experimental accuracy.
Other radiative corrections within the SM
(due to Higgs and would-be Nambu--Goldstone bosons) are also  negligible.

Thus, 
if there is indeed a physical origin to Koide's mass formula at
a high energy scale,
we need to account for a correction to the relation
(\ref{KoideMF})
that cancels
the QED correction.
Since such a correction is absent up to the scale of
${\cal O}(M_W)$ to our present knowledge, it must originate from a higher scale.
Then, there is a difficulty in explaining why
the size of such a correction 
 should coincide accurately with the
size of the QED correction which arises from much lower scales.
There are also other less serious, but important questions that are often asked:
(1) Why do not quark masses satisfy the same or a similar relation?
(2) In Koide's formula the three lepton masses appear symmetrically.
Then why
is there a hierarchy among these masses, $m_e\ll m_\mu \ll m_\tau$?
If there is indeed a physical origin to Koide's mass formula,
there must be reasonable answers to all of these questions.

Among various existing models which attempt to explain origins
of Koide's mass formula, we find 
a class of models particularly
attractive \cite{Koide:1989jq,Koide:1995pb}.
These are the models which predict the mass matrix of 
the charged leptons to be
proportional to the square of 
the vacuum expectation value (VEV)
of a 9--component
scalar field (we denote it as $\Phi$) written in 
a 3--by--3 matrix form:
\bea
{\cal M}_\ell \propto \langle \Phi \rangle \langle \Phi \rangle
\, .
\label{MasPhisq}
\eea
Thus, $(\sqrt{m_e},\sqrt{m_\mu},\sqrt{m_\tau})$ is proportional
to the diagonal elements of $\langle \Phi \rangle$ in the
basis where it is diagonal.
The VEV $\langle \Phi \rangle$
is determined by minimizing the potential of scalar fields
in each model.
Hence, the origin of Koide's formula is
attributed to the specific form of the potential which realizes
this relation in the vacuum configuration.
Up to now, no model is complete with respect to symmetry:
Every model requires either
absence or strong suppression of
some of the terms in the potential (which are allowed by
the symmetry of that model),
without 
justification.

In this paper, we study possible connections between 
family (horizontal) gauge symmetries and
Koide's formula and the charged lepton spectrum.
These will be discussed within the context
of an effective field theory
(EFT) which is valid below some cut--off scale.
In particular we address the following points:
\begin{enumerate}
\renewcommand{\labelenumi}{(\roman{enumi})}
\item
We propose a possible
mechanism for cancellation of the QED
radiative correction to Koide's mass formula.
\item
We propose a mechanism that produces the
charged lepton spectrum, which is hierarchical and 
approximates the experimental values, under the
assumption that 
Koide's formula is protected by some other mechanism.
\item
We present a model of charged lepton sector
based on $U(3)\times SU(2)$ family gauge
symmetry,
incorporating the mechanisms (i)(ii).
A new mechanism that 
stabilizes Koide's formula is incorporated
in this model.
\end{enumerate}
(Among these, we have reported
the main point of (i) separately in \cite{Sumino:2008hu}.)

In our study
we adopt the mechanism eq.~(\ref{MasPhisq}) for generating the charged
lepton masses at tree level of EFT, for the following
reasons.
First, the mechanism allows for transparent and
concise perturbative analyses of models, which
is crucial in keeping radiative corrections under control.
This may be contrasted with models with other mass
generation mechanisms, such as
dynamical symmetry breaking
or composite lepton models, which typically involve strong
interactions.
Secondly, since $\Phi$ is renormalized multiplicatively,
the structure of radiative corrections becomes simple,
as opposed to cases in which VEVs of more than one scalar
fields contribute to the
charged lepton spectrum.
In short, this type of mass generation mechanism is pertinent to serious
analyses of radiative corrections to Koide's formula, which is a
distinguished aspect of this study.

We alert in advance
that we do not solve the hierarchy problem or fine
tuning problem of the electroweak scale.
We cannot explain how to stabilize the electroweak symmetry--breaking
scale against other higher scales included in our model.
Solution to this problem is beyond the scope of this paper.

The paper is organized as follows.
In Sec.~2, we explain philosophy of our analysis using
EFT and argue for its validity and usefulness.
We also give a brief overview of the ideas presented in this paper.
In Sec.~\ref{s2}, we explain the mechanism for cancelling the QED corrections to 
Koide's formula.
In Sec.~\ref{s3}, we present a potential for generating a realistic
charged lepton spectrum, assuming that Koide's formula is protected.
In Sec.~\ref{s4}, we analyze a minimal potential whose vacuum
corresponds to a desired lepton spectrum.
In Sec.~\ref{s5}, we extend the potential by including another field.
In Sec.~\ref{s6}, we introduce a higher--dimensional operator which generates 
the lepton masses and compute corrections to Koide's formula.
In Sec.~\ref{s7}, we discuss the energy scales and unsolved questions
in our model.
In Sec.~\ref{s8}, we discuss phenomenological implications of our model.
In Sec.~\ref{s9} summary and discussion are given.
Technical details are collected in Appendices.

\section{EFT Approach and Brief Overview of the Model}
\clfn

Throughout this paper, we consider an EFT which
is valid up to some cut--off scale denoted by $\Lambda\,\,(\gg M_W)$.
In this EFT, we assume that the charged lepton
masses are induced by a higher--dimensional operator
\bea
{\cal O}=\frac{\kappa(\mu)}{\Lambda^2}\, 
\bar{\psi}_{Li}\, \Phi_{ik}\, \Phi_{kj}\, \varphi \, e_{Rj} \, 
\label{HigerDimOp1}
\eea
(or by other similar operators, as will be described later).
Here, $\psi_{Li}=(\nu_{Li},e_{Li})^T$ denotes the left--handed lepton 
$SU(2)_L$ doublet
of the $i$--th generation;
$e_{Rj}$ denotes the right--handed charged lepton
of the $j$--th generation;
$\varphi$ denotes the Higgs doublet field.
They are respectively assigned to the standard representations of
the SM gauge group.
By contrast, a 9--component scalar field $\Phi$
is absent in the SM and a singlet under the SM gauge group.
We suppressed all the indices except for the generation (family)
indices $i,j,k=1,2,3$.
(Summation over repeated indices is understood throughout
the paper unless otherwise stated.)
The dimensionless Wilson coefficient of this operator is
denoted as $\kappa(\mu)$.
Once $\Phi$ acquires a VEV, the operator $\cal O$ will
effectively be rendered to
the Yukawa interactions of the SM; 
after the Higgs field 
also acquires a VEV, $\langle\varphi\rangle=(0,v_{\rm ew}/\sqrt{2})^T$
with $v_{\rm ew}\approx 250$~GeV, the operator will induce the
charged--lepton mass matrix of the form
eq.~(\ref{MasPhisq}) at tree level:
\bea
{\cal M}_\ell^{\rm tree} = \frac{\kappa\,v_{\rm ew}}{\sqrt{2}\Lambda^2} 
\langle \Phi \rangle \langle \Phi \rangle
\, .
\label{MellatTree}
\eea
% Later we introduce a symmetry which prohibits
% the dimension-4 Yukawa interactions
% $y_{ij}\,\bar{\psi}_{Li}\varphi e_{Rj}$.
For a moment, let us assume that the dimension--4 Yukawa interactions
$y_{ij}\,\bar{\psi}_{Li}\varphi e_{Rj}$ are prohibited by
some mechanism.
This will be imposed explicitly
by a symmetry in our model to be discussed through Secs.~3--9.

We now explain philosophy of our analysis using EFT.
Conventionally a more standard approach for explaining Koide's mass formula has been
to construct models within
renormalizable theories.
%In comparison, it 
%is certainly a retreat to make an analysis within EFT.
Nevertheless, the long history since the discovery of Koide's formula shows 
that it is quite difficult to construct
a viable renormalizable model for explaining Koide's relation.
It is likely that we are missing some essential hints to achieve this goal, 
if the relation is not a sheer coincidence.
In this paper we will show that, within EFT, explanation of Koide's
formula is possible by largely avoiding fine tuning
of parameters.
Consistency conditions (with respect to symmetries of the theory)
can be satisfied relatively easily in EFT, or in other words, they can be 
replaced by reasonable boundary conditions of EFT at the cut--off scale
$\Lambda$ without conflicting symmetry requirements of the theory.
(See Sec.~\ref{s4}.)
Even under this less restrictive theoretical constraints,
we may learn some important hints concerning the relation between
the lepton spectrum and family symmetries.
These are the role of
specific family gauge symmetry in canceling the QED correction,
the role of family symmetry in stabilizing Koide's mass relation, or
the role of family symmetry in realizing a realistic 
charged lepton spectrum consistently with experimental values.
These properties do not come about separately but 
are closely tied with each other.
These features do not seem to depend on details of more fundamental
theory above the cut--off scale $\Lambda$ but rather on some general
aspects of family symmetries and their breaking patterns.
Thus, we consider that our approach based on EFT would be useful
even in the case
in which physics above the scale $\Lambda$ is obscure and may
involve some totally
unexpected ingredients -----
as it was the case with chiral perturbation theory
before the discovery of QCD.

Before discussing radiative corrections within EFT, one would be worried
about effects of higher--dimensional operators suppressed in higher powers
of $1/\Lambda$.
Indeed, using the values of tau mass and the electroweak symmetry breaking
scale $v_{\rm ew}$, one readily finds that 
$v_3/\Lambda \simgt 0.1$ ($v_i$ are the diagonal elements 
of $\langle \Phi \rangle$ in the basis where it is diagonal).
Hence, naive dimensional analysis indicates that there would be
corrections to Koide's formula of order 10\% even at tree level.
We now argue that this is not necessarily the case within the scenario
under consideration.
We may
divide the corrections into two parts.
These are (i) $1/\Lambda^n$ corrections to the operator
$\cal O$ of eq.~(\ref{HigerDimOp1}) (the operator which
reduces to the SM
Yukawa interactions after $\Phi$ is replaced by
its VEV), and
(ii) $1/\Lambda^n$ corrections to the  VEV of $\Phi$.

\begin{figure}[t]\centering
\includegraphics[width=6cm]{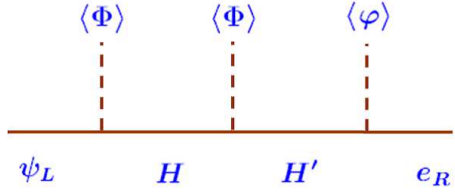}
\caption{\small 
Diagram which induces the higher--dimensional operator
${\cal O}=\frac{\kappa(\mu)}{\Lambda^2}\, 
\bar{\psi}_{Li}\, \Phi_{ik}\, \Phi_{kj}\, \varphi \, e_{Rj} $.
\label{see-saw}}
\end{figure}

Concerning the corrections (i), we may consider the following
example.\footnote{
A similar mechanism is used in \cite{Koide:1989jq,Koide:1995xk}
to induce $\cal O$; corrections by higher--order terms in $1/\Lambda^n$
have not been discussed, however.
}
Suppose that the operator $\cal O$ is induced
from the interactions
\bea
{\cal L}=y_1 \,\bar{\psi}_{Li}\Phi_{ij} H_{Rj} + M\, \bar{H}_{Ri} H_{Li} + 
y_2\, \bar{H}_{Li} \Phi_{ij} H'_{Rj} + M' \bar{H}'_{Ri} H'_{Li} +
y_3\, \bar{H}'_{Li} \varphi e_{Ri}
+ ({\rm h.c.})
\eea
through the diagram shown in Fig.~\ref{see-saw},
after fermions $H_{L,R}$ and $H'_{L,R}$ have been integrated out.
Fermions $H_{L,R}$ and $H'_{L,R}$ are assigned to appropriate
representations of the SM gauge group such that the above interactions
become gauge singlet.
For instance, in the case that 
$v_3/M' \simgt 3$, $y_1,y_2,y_3\approx 1$ and
$v_{\rm ew}/M'<3\times 10^{-3}$, 
one finds,
 by computing the mass eigenvalues,\footnote{
Since the values of $m_\tau$ and $v_{\rm ew}$ are known,
once we choose the values
of $v_3/M'(\simgt 3)$ and $y_1,y_2,y_3(\approx 1)$, the value of
$v_3/M(\simlt 0.03)$ will be fixed.
%We assume that $v_{\rm ew}/M'$ is a small parameter
%($<3\times 10^{-3}$).
Then the mass eigenvalues corresponding to 
the SM charged leptons
can be computed in series expansion
in the small parameters
$v_{\rm ew}/M'$, $v_i/M$ and $v_i^2/(MM')=\sqrt{2}m_i/v_{\rm ew}$.
}
that the largest correction to the lepton spectrum 
eq.~(\ref{MellatTree}) arises from the operator
$\displaystyle
-\frac{y_1^3y_2^3y_3}{2M^3M'^3}\,
\bar{\psi}_L \Phi^6 \varphi e_R
$; its contribution to the
tau mass is
$\delta m_\tau/m_\tau = (m_\tau/v_{\rm ew})^2
\approx 5\times 10^{-5}$.
This translates to a correction to Koide's relation of $3\times 10^{-6}$,
due to the suppression factor
$\frac{1}{\sqrt{6}}\,(m_\mu/m_\tau)^{1/2}(\delta m_\tau/m_\tau)$.
Thus, this is an example of underlying mechanism that generates the operator
$\cal O$ without generating higher--dimensional operators conflicting the current
experimental bound.
If we introduce even more (non--SM) fermions to generate
the leading--order operator $\cal O$, one can always find a pattern of
spectrum of these fermions, for which higher--dimensional operators
are sufficiently suppressed, since the number of 
adjustable parameters increases.
(Another example of underlying mechanism
may be the one proposed in \cite{Ma:2006ht}, based on the
idea of \cite{Koide:2005ep}.)
In general, sizes of higher--dimensional operators 
depend heavily on underlying dynamics above the cut--off scale.

\begin{figure}[t]\centering
\psfrag{Phi}{\hspace{0mm}$\Phi$}
\psfrag{psiL}{\hspace{0mm}$\psi_L$}
\psfrag{eR}{\hspace{0mm}$e_R$}
\includegraphics[width=13cm]{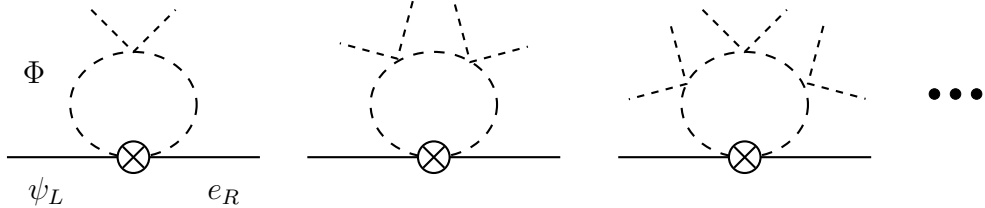}
\caption{\small 
EFT 1--loop diagrams which generate higher--dimensional operators
contributing to the charged lepton spectrum.
Dashed line represents $\Phi$;
$\otimes$ represents
the higher--dimensional
operator which generate the charged
lepton masses at tree--level
[corresponding to ${\cal O}$ of eq.~(\ref{HigerDimOp1})].
\label{1LoopAnalyInEFT}}
\end{figure}

Let us restrict ourselves within EFT.
If we introduce only the operator $\cal O$, by definition this is the only
contribution to the charged lepton spectrum at tree level.
Whether loop diagrams induce higher--dimensional operators
which violate Koide's relation is an important question,
and a detailed analysis is necessary.
This is the subject of the present study, where
the result depends on the mechanisms how Koide's formula is satisfied and
how the charged lepton spectrum is determined, even within EFT.
The conclusion is as follows.
Within the model to be discussed in Secs.~\ref{s2}--\ref{s7},
the class of 1--loop diagrams shown
in Fig.~\ref{1LoopAnalyInEFT} do not generate operators that violate Koide's relation
sizably; see Sec.~\ref{s6}.
(There is another type of 1--loop diagrams that possibly
cancels the QED correction; see Sec.~\ref{s2}.)
In fact, we do not find any loop--induced higher--dimensional
operators, which violate Koide's relation in conflict with
the current experimental bound.

Concerning the corrections (ii),
in our analysis
we introduce specific family gauge symmetries and their breaking patterns
such that 
the corrections (ii) are suppressed.
% In this regard, in the subsequent sections
% we will try to make clear the motivation which led us
% to consider the specific mechanisms.

Since the above example of underlying mechanism that suppresses
higher--dimensional operators
is simple, and since suppression of loop--induced 
$1/\Lambda^n$ corrections
within EFT 
provides a non--trivial cross check of theoretical consistency,
we believe that our approach based on EFT has a certain
justification and would be useful as a basis for considering
more fundamental models.
\medbreak

In the
rest of this section, we present a brief overview of the
basic ideas of the analysis to be given through Secs.~\ref{s2}--\ref{s7}, 
in order to facilitate reading.
Our analysis starts from
investigating a possibility that the radiative correction
generated by a family gauge symmetry cancels the QED correction
to Koide's formula (Sec.~\ref{s2}).
We find that
$U(3)\simeq SU(3)\times U(1)$ family gauge symmetry has a 
unique property
in this regard.
In fact, if $\psi_L$ and $e_R$ are assigned to 
mutually conjugate representations of this symmetry group,
the $U(3)$ radiative correction has the same form as the QED
correction with opposite sign.
In particular, if the gauge coupling of $U(3)$ family symmetry
$\alpha_F=g_F^2/(4\pi)$
satisfies the relation
$\alpha(m_\tau)\approx \frac{1}{4}\alpha_F(g_Fv_3)$,
both corrections cancel.
We speculate that this relation would be realized
within a scenario in which 
$U(3)$ family gauge symmetry is unified with $SU(2)_L$ 
gauge symmetry
at $10^2$--$10^3$~TeV scale, although we need to fine
tune the unification scale within an accuracy of
factor 3.

The non--trivial form of the
radiative correction by the $U(3)$ gauge interaction is
dictated by the $U(3)$ symmetry and its breaking pattern
induced by the VEV $\langle \Phi \rangle$.
In particular,  multiplicative renormalizability of 
$\langle \Phi \rangle$ ensures that the correction to
Koide's formula is
independent of the renormalization scale $\mu$ of
the effective potential of $\Phi$.
Namely, the charged lepton pole masses are determined,
up to a common multiplicative constant,
directly
by the form of the
effective potential renormalized at an arbitrary
high scale $\mu$ ($\leq \Lambda$),
and we may ignore the QED 
and $U(3)$ radiative corrections altogether.
For our purpose, it is most convenient 
to take this scale to be $\mu=\Lambda$.
In this part of our analysis,
we assume that $\langle \Phi \rangle$ can be brought to 
a diagonal form
by symmetry transformation, and also that
Koide's relation for
the diagonal elements,
\bea
\frac{v_1(\mu)+v_2(\mu)+v_3(\mu)}
{\sqrt{v_1(\mu)^2+v_2(\mu)^2+v_3(\mu)^2}}
=\sqrt{\frac{3}{2}}\, ,
\label{relvi1}
\eea
is satisfied.

In the second step, we search for an effective potential
for which the eigenvalues of $\langle \Phi \rangle$
satisfy the relation (\ref{relvi1}) and reproduce
the experimental values of the mass ratios
$v_1:v_2:v_3 = \sqrt{m_e}:\sqrt{m_\mu}:\sqrt{m_\tau}$
 (Secs.~\ref{s3} and \ref{s4}).
If we choose the renormalization scale to be $\mu=\Lambda$,
radiative corrections to the effective potential
essentially vanish within EFT, or in other words, 
the form of the effective potential at this scale
is determined by physics above the scale $\Lambda$ as
boundary (initial) conditions of EFT.
Hence, our goal is to find 
an effective potential which
satisfies the boundary conditions without conflicting
symmetry requirements of the theory.
Although it may seem an easy task, it still involves
fairly non-trivial analyses.

We impose $U(3)\times SU(2)$ family symmetry as a 
symmetry of EFT.
The motivation of this choice
is that it is the symmetry possessed by
the simplest higher--dimensional operator analyzed in
the first step.
It turns out, however, that this symmetry is not large enough
to constrain the form of the effective potential
sufficiently.
We therefore further
assume a symmetry enhancement.
Namely, we assume that
above the cut--off scale $\Lambda$ there is an $SU(9)\times U(1)$ 
gauge symmetry, and this
symmetry is spontaneously broken to
$U(3)\times SU(2)$ below the cut--off scale.
The symmetry $SU(9)\times U(1)$ is motivated by
a geometrical interpretation of Koide's relation
eq.~(\ref{relvi1}).
Within this scenario, we still need to
introduce an additional scalar field
$X$
in order to realize a desirable vacuum configuration.
%Under $SU(9)$, $\Phi$ is in the fundamental representation,
%while $X$ is in the 
%second--rank symmetric representation and 
%constrained to be unitary.
Thus, we analyze the vacuum of the
general potential of $\Phi$ and $X$.
(Details of the analysis are rather technical.)
The conclusion is that in a finite region of the parameter space of the
potential, Koide's relation is satisfied by the eigenvalues of 
$\langle \Phi \rangle$.
Furthermore, the eigenvalues can be made consistent with 
the experimental values of the charged lepton masses without 
fine tuning of parameters.
These are realized in the case that
certain hierarchical relations among the parameters of the
potential are satisfied, and these
relations do not conflict the requirement
of the assumed symmetry and symmetry enhancement.
We speculate on possible physics scenario above the
cut--off scale that may lead to (part of) these hierarchical relations.

So far, these desirable features are satisfied by the eigenvalues
of $\langle \Phi \rangle$.
There remains, however, 
a problem that $\langle \Phi \rangle$ cannot be
brought to a diagonal form by the $U(3)\times SU(2)$
symmetry transformation, and this contradicts the
assumption made in the first step.
To remedy this difficulty, we introduce yet another field
$\Sigma_Y$ such that, with an appropriate
potential with $\Phi$,
it can generate an appropriate
higher--dimensional operator necessary to produce
the charged lepton masses.
(Secs.~\ref{s5} and \ref{s6}.)
Although the potential and the
higher--dimensional operator involving $\Sigma_Y$
do not conflict the requirement
of the assumed symmetry and symmetry enhancement,
these would be the most unsatisfactory
part of our model.
This is because
it is difficult to speculate 
any plausible
scenario above the cut--off scale,
which would lead to these potential and 
higher--dimensional operator.

With all these setups, it is possible to compute the radiative
corrections which are induced by the diagrams
shown in Fig.~\ref{1LoopAnalyInEFT}.
Due to the specific form of the effective potential of
$\Phi$ and $X$, corrections to Koide's formula turn out
to be quite suppressed, as long as the aforementioned
hierarchical conditions between parameters of the
potential are satisfied.
As already mentioned, this serves as a non--trivial
consistency check of the model as an EFT.

There are a few unsolved questions and incompleteness of the
present model and these are discussed in Secs.~\ref{s7}
and \ref{s9}.

\section{Radiative Correction by Family Gauge 
Interaction}
\label{s2}
\clfn

\begin{figure}[t]\centering
\psfrag{psi}{\hspace{0mm}$\psi_{L}$}
\psfrag{e}{\hspace{0mm}$e_{R}$}
\psfrag{phi1}{\hspace{0mm}$\langle\Phi\rangle$}
\psfrag{phi2}{\hspace{0mm}$\langle\Phi\rangle$}
\psfrag{+subleading}{\hspace{-3mm}$+~{\cal O}(\frac{1}{N_F})$}
\includegraphics[width=16cm]{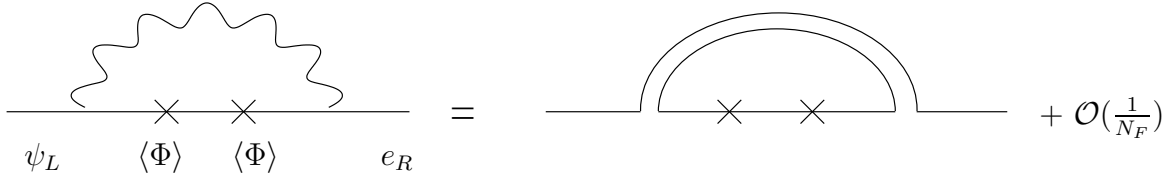}
\caption{\small 
Diagram for the 1--loop correction by the family gauge bosons
to the operator ${\cal O}$ when both
$\psi_L$ and $e_R$ are in the $\bf 3$ of $SU(3)$.
The diagram on the
right--hand--side shows the flow of family charge in the
leading contribution
of the $1/N_F$ expansion ($N_F=3$);
closed loop corresponds to ${\rm tr}(\langle \Phi \rangle \langle \Phi \rangle)$.
\label{1LoopGraph1}}
\end{figure}

In this section
we introduce family gauge symmetries and consider radiative corrections
to the mass matrix eq.~(\ref{MellatTree})
by the family gauge interaction.
First we consider the case, in which the 
family gauge group is $SU(3)$
and both $\psi_L$ and $e_R$ are assigned to the $\bf 3$ (fundamental
representation) 
of this symmetry group.
%In fact, a number of models with SU(3)
%family gauge symmetry have been proposed (whether they
%predict Koide's mass formula or not), 
%in which the left--handed and right--handed
%SM fermions are both in the 
%$\bf 3$.
We readily see, however, that with this choice of representation, Koide's
formula will receive a severe radiative correction unless the 
family gauge
interaction is strongly suppressed.
In fact, the 1--loop diagram shown in Fig.~\ref{1LoopGraph1} induces an effective 
operator
\bea
{\cal O} '\sim \frac{\alpha_F}{\pi}\times{\kappa}\,
\bar{\psi}_{Li} \, \varphi \, e_{Ri} 
\times \frac{\langle\Phi\rangle_{jk}
\langle\Phi\rangle_{kj}
}{\Lambda^2} 
%\times {\rm tr}\left(\langle\Phi\rangle
%\langle\Phi\rangle\right)
\, ,
\eea
hence corrections universal to all the charged--lepton masses,
$(\delta m_e, \delta m_\mu, \delta m_\tau) \propto (1,1,1)$,
are induced.
This is due to the fact that the dimension--4
operator $\bar{\psi}_{Li} \, \varphi \, e_{Ri} $
is not prohibited by symmetry.
Here, $\alpha_F=g_F^2/(4\pi)$ denotes the gauge
coupling constant
of the family gauge interaction.
As  noted above, corrections which are proportional to individual
masses do not affect Koide's formula;
oppositely, the universal correction violates Koide's formula rather strongly.
In order that the correction to Koide's formula cancel the QED correction,
a naive estimate shows that 
$\alpha_F/\pi$ should be order $10^{-5}$, provided that
the cut--off $\Lambda$ is not too large and that the above operator
${\cal O}'$ is absent at tree level.
If ${\cal O}'$ exists at tree level, there should be a
fine tuning between the tree--level and 1--loop
contributions.
The situation is similar if the family symmetry is 
$O(3)$ and 
both $\psi_L$ and $e_R$ are in the $\bf 3$,
which is also a typical assignment
in existing models.
In these cases \cite{Antusch:2007re}
we were unable to find any sensible reasoning 
for the cancellation between the QED correction and
the correction induced by family gauge interaction, other than 
to regard the cancellation as just
a pure coincidence.
Hence, we will not investigate these choices of
representation further.

%Appearance of corrections universal to all three
%masses can be circumvented, as well as 
%a remarkable resemblance of the
%radiative correction to the QED correction
%follows,
%if $\psi_L$ is assigned to $\bf 3$ and
%$e_R$ to $\bar{\bf 3}$ (or {\it vice versa}) of $U(3)$
%family gauge group.
In the case that $\psi_L$ is assigned to $\bf 3$ and
$e_R$ to $\bar{\bf 3}$ (or {\it vice versa}) of $U(3)$
family gauge group,
(i) the dimension--4 operator
$\bar{\psi}_{Li} \, \varphi \, e_{Ri} $ is
prohibited by symmetry, and hence
corrections universal to all the three
masses do not appear; and
(ii) marked resemblance of the
radiative correction to the QED correction
follows.
We show these points explicitly in a specific setup.

We denote the generators for the fundamental representation
of $U(3)$ by $T^\alpha$
($0\leq\alpha\leq 8$), which satisfy
\bea
{\rm tr}\left(T^\alpha T^\beta\right)=\frac{1}{2}\, \delta^{\alpha\beta} 
~~~;~~~
T^\alpha = {T^\alpha}^\dagger
\, .
\label{U3generators}
\eea
$T^0$ is the generator of $U(1)$,  hence  it is proportional to the 
identity matrix, while $T^a$ ($1\leq a \leq 8$) are the generators of
$SU(3)$.
Here and hereafter, $\alpha,\beta,\gamma,\dots$ represent $U(3)$
indices $0,\dots,8$, while $a,b,c,\dots$ represent $SU(3)$ indices
$1,\dots,8$.
The explicit forms of $T^\alpha$ are given in Appendix~\ref{appA}.

We assign 
$\psi_L$ to the representation
$({\bf 3},1)$, where $\bf 3$ stands
for the $SU(3)$ representation and 1 for the $U(1)$ charge,
while $e_R$ is assigned to $(\bar{\bf 3},-1)$.
Under $U(3)$, the 9--component field
$\Phi$ transforms as three $({\bf 3},1)$'s.
Explicitly the transformations of these fields 
are given by
\bea
\psi_L \to U \, \psi_L \, ,
~~~
e_R \to U^* \, e_R \, ,
~~~
\Phi \to U \, \Phi 
~~;~~~
U = \exp \left(i\theta^\alpha T^\alpha\right)\, ,
~~
U\,U^\dagger={\bf 1}\, .
\label{U3transf}
\eea

We assume that the 
charged--lepton mass matrix is induced by a higher--dimensional
operator ${\cal O}^{(\ell)}$ similar to ${\cal O}$ in eq.~(\ref{HigerDimOp1}).
%The (tree--level) 
%mass matrix induced by such an operator
%can be brought to a diagonal form by a bi--unitary transformation
%in the family space.
We further assume that $\langle \Phi \rangle$ can be brought to a
diagonal form %at the same time 
in an appropriate basis.
%Thus, in this basis the higher--dimensional operator, after
%$\Phi$ and $\varphi$ acquire VEVs, can be written
%as
Thus, in this basis ${\cal O}^{(\ell)}$, after
$\Phi$ and $\varphi$ acquire VEVs, turns to the lepton mass terms 
as
\bea
&&
{\cal O}^{(\ell)}\to \bar{\psi}_{L}\, {\cal M}_\ell^{\rm tree} \,
e_R\, ,
%\frac{\kappa^{(\ell)}(\mu)}{\Lambda^2}\,
%\bar{\psi}_{L}\, \Phi_d^2 \, \langle\varphi\rangle \, e_{R} \, ,
%\label{HigherDimOp3}
%\\
%&&
~~~~~
~~~~~
{\cal M}_\ell^{\rm tree} =
\left(\begin{array}{ccc}
m_e^{\rm tree}&0&0\\
0&m_\mu^{\rm tree}&0\\
0&0&m_\tau^{\rm tree}
\end{array}\right)
=\frac{\kappa^{(\ell)}(\mu)\,v_{\rm ew}}{\sqrt{2}\Lambda^2} \, \Phi_d(\mu)^2\, ,
\label{DiagonalMassMat}
\eea
where
\bea
\Phi_d (\mu)=
\left(\begin{array}{ccc}
v_1(\mu)&0&0\\
0&v_2(\mu)&0\\
0&0&v_3(\mu)
\end{array}\right) \, ,
~~~~~ v_i(\mu) >0
\, .
\label{Phid}
\eea
When all $v_i$ are different,
$U(3)$ symmetry is completely broken by $\langle \Phi \rangle=\Phi_d$,
and the spectrum of the $U(3)$ gauge bosons is determined 
by $\Phi_d$.

Note that the operator ${\cal O}$ in eq.~(\ref{HigerDimOp1})
is {\it not} invariant under the $U(3)$ transformations
eq.~(\ref{U3transf}).
As an example of ${\cal O}^{(\ell)}$, one may consider 
\bea
{\cal O}^{(\ell)}_1=
\frac{\kappa^{(\ell)}(\mu)}{\Lambda^2}\,
\bar{\psi}_{L}\, \Phi\, \Phi^T\, \varphi \, e_{R} \, .
\label{exampleO1}
\eea
It is invariant under a larger symmetry  
$U(3)\times SU(2)$, under which
$\Phi$ transforms as $\Phi\to U\Phi O^T$ ($O\,O^T = {\bf 1}$).
In this case, 
we need to assume, for instance, that the $SU(2)$ symmetry is gauged 
and spontaneously
broken at a high energy 
scale before the breakdown of the $U(3)$ symmetry, in order to
eliminate massless Nambu--Goldstone bosons and to
suppress mixing of the $U(3)$ and $SU(2)$ gauge bosons.
A more elaborate example %\footnote{
%One may consider 
%$\frac{\kappa^{(\ell)}(\mu)}{\Lambda^2}\,
%\bar{\psi}_{L}\, \Phi\, \Phi^T\, \varphi \, e_{R} $
%as a simple example.
%It is invariant under a larger symmetry  
%$U(3)\times SU(2)$ in which
%$\Phi$ transforms as $\Phi\to U\Phi O^T$ ($O\,O^T = {\bf 1}$).
%We omit the discussion how the above assumptions
%can be satisfied
%in this case.
%} 
of the higher--dimensional operator,
which is consistent with the symmetry and
satisfies eqs.~(\ref{DiagonalMassMat}) and (\ref{Phid}), will be given in Sec.~\ref{s6}.
In any case, 
the properties of ${\cal O}^{(\ell)}$ given by
eqs.~(\ref{DiagonalMassMat}) and (\ref{Phid})
are sufficient for computing the radiative correction
by the $U(3)$ gauge bosons to the mass matrix,
without an explicit form of ${\cal O}^{(\ell)}$.
\begin{figure}[t]
\begin{center}
\begin{tabular}{c}
\psfrag{psi}{\hspace{0mm}$\psi_{L}$}
\psfrag{e}{\hspace{0mm}$e_{R}$}
\psfrag{phi1}{\hspace{0mm}$\Phi_d$}
\psfrag{phi2}{\hspace{0mm}$\Phi_d$}
\includegraphics[width=13cm]{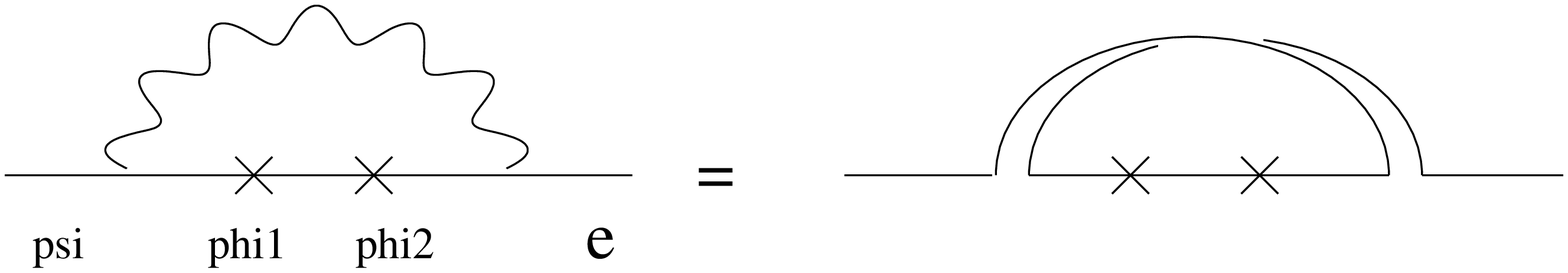}
\vspace{-5mm}
\\
(a)\vspace{5mm}\\
%{\raise30pt\hbox{(a)}}\\
\psfrag{psi}{\hspace{0mm}$\psi_{L}$}
\psfrag{psi2}{\hspace{0mm}$\psi_{L}$}
\includegraphics[width=13cm]{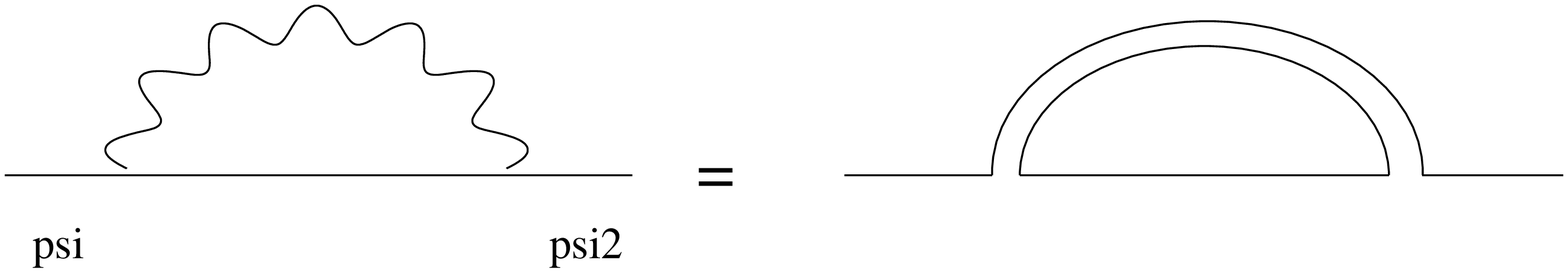}
\vspace{-7mm}
\\
(b)\vspace{5mm}\\
\psfrag{psi}{\hspace{0mm}$e_{R}$}
\psfrag{psi2}{\hspace{0mm}$e_{R}$}
\includegraphics[width=13cm]{1LoopGraph2.eps}
\vspace{-7mm}
\\
(c)\vspace{5mm}\\
 \end{tabular}
\caption{\small 
1--loop diagrams contributing to $\delta m_i^{\rm pole}$
when 
$\psi_L$ and $e_R$ are in the $({\bf 3},1)$ and $(\bar{\bf 3},-1)$,
respectively,
of $SU(3)\times U(1)$.
The right--hand--sides show flows of family charge.
(a) Correction of the form $\bar{\psi}_L \, \delta {\cal M} \, e_R$:
charge flow is connected in one line, showing multiplicative
renormalization,
(b) correction of the form $\bar{\psi}_L \!\not\! p \,Z_\psi \, \psi_L$, and
(c) correction of the form $\bar{e}_R \!\not\! p \,Z_e\, e_R$.
\label{1LoopGraph2abc}}
\end{center}
\end{figure}

We take the $U(1)$ and $SU(3)$ gauge coupling
constants to be the same:
\bea
\alpha_{U(1)}=\alpha_{SU(3)}=\alpha_F\, .
\label{UnivCouplings}
\eea
We compute the radiative correction in Landau gauge,
which is known to be convenient for computations in theories
with spontaneous symmetry breaking.
From the diagrams shown in 
Figs.~\ref{1LoopGraph2abc}(a)(b)(c), we find
\bea
&&
\delta m^{\rm pole}_i =
-\frac{3\,\alpha_F}{8\,\pi}\left[
\log\left(
\frac{\mu^2}{v_i(\mu)^2}
\right) + c
\right] \,
{m}_i(\mu) \, ,
\label{alphaFcorr}
\\
&&
{m}_i(\mu) = 
\frac{\kappa^{(\ell)}(\mu)\,v_{\rm ew}}{\sqrt{2}\Lambda^2} \, v_i(\mu)^2 \, .
\label{mmu}
\eea
Here, $c$ is a constant independent of $i$.
The Wilson coefficient
$\kappa^{(\ell)}(\mu)$ is defined in $\overline{\rm MS}$ scheme.
$v_i(\mu)$'s are defined as follows:
The VEV of $\Phi$ at renormalization
scale $\mu$, $\Phi_d(\mu)=\langle \Phi(\mu)\rangle$
given by eq.~(\ref{Phid}),
is determined by minimizing the 1--loop effective potential
in Landau gauge
(although we do not discuss the explicit form of the effective
potential in this section);
$\Phi$ is renormalized in $\overline{\rm MS}$ scheme.
We ignored terms suppressed by
${m}_i^2/v_j^2(\ll 1)$ in the above expression.
Note that the
pole mass is renormalization--group invariant and
gauge independent.
Therefore, the above expression is
rendered gauge--independent if we
express $v_i(\mu)$ in terms of gauge--independent parameters, 
%such as $\overline{\rm MS}$ couplings in the effective potential.
such as coupling constants defined in on--shell scheme.

The coefficient of $\log \mu^2$ is determined by the sum of the
anomalous dimension of the Wilson coefficient $\kappa^{(\ell)}(\mu)$ and
twice of the wave--function renormalization of $\Phi$.
(The former is gauge independent, while the latter is not.)
The term $\log v_i^2$ originates from the role
of the gauge boson masses as an IR cut--off of the loop integral, hence
it reflects the spectrum of the gauge bosons.
The sign in front of $\log \mu^2$ is opposite to that of
the QED correction eq.~(\ref{QED1Lcorr}),
which results from the fact that $\psi_L$ and $e_R$ have the same
QED charges but mutually
conjugate (opposite) $U(3)$ charges.

%As well known, technically the computation is simple in
%Landau gauge.
In Landau 
gauge, 
the diagrams in Figs.~\ref{1LoopGraph2abc}(b)(c)
are finite and 
flavor  independent, i.e.\ proportional to $\delta_{ij}$ in 
terms of the family indices; 
hence they contribute only to the constant $c$.
Apart from this constant,
the difference between the QED correction and the correction
(\ref{alphaFcorr}) resides in the factors
\bea
%\alpha\, \left[
%\log \left(\frac{\mu}{{{\cal M}_\ell^{\rm tree}}}\right)^2 
%\right]_{ij}
\alpha\, \bar{m}_i\,\log \left(\frac{\mu^2}{\bar{m}_i^2}\right) \delta_{ij}
~~~~~~~{\rm vs.}~~~~~~
-\alpha_F \, \left[
T^\alpha {\cal M}_\ell^{\rm tree}\, {T^\beta}^* \left\{
\log\left(\frac{\mu^2}{M_F^2}\right)
\right\}_{\alpha\beta}
\right]_{ij} 
\eea
in the QED self--energy diagram and the diagram
in Fig.~\ref{1LoopGraph2abc}(a), respectively.
(No sum over $i$ is taken in the former factor.)
One may easily identify the factor 2 difference in the coefficients
of $\log\mu^2$ using the Fierz identity
\bea
(T^\alpha)_{ij}\,({T^\alpha}^*)_{kl}=
(T)^\alpha_{ij}\,({T^\alpha})_{lk}=
\frac{1}{2}\,\delta_{ik}\,\delta_{lj}
\, .
\label{FierzId}
\eea
From this identity, 
it follows that the operator 
${\cal O}^{(\ell)}$ is multiplicatively
renormalized; see family charge flow in
Fig.~\ref{1LoopGraph2abc}(a).
$(M_F^2)_{\alpha\beta}$ denotes the mass matrix of the family gauge bosons.
After diagonalization, one obtains the spectrum of 
family gauge bosons as
\bea
&&
\frac{1}{2}\,(M^2_F)_{\alpha\beta} \, f_\mu^\alpha \, f_\mu^\beta
\equiv g_F^2\, {\rm tr}(\Phi_0^\dagger T^\alpha T^\beta \Phi_0)
\, f_\mu^\alpha \, f_\mu^\beta
\nonumber\\&&
~~~~~~~~~~~~~~~~~~~~
=\frac{g_F^2}{2}\Biggl[
v_1^2 \, ({\cal F}_\mu^1)^2 + v_2^2 \, ({\cal F}_\mu^1)^2 +
\frac{1}{2}(v_1^2+v_2^2)\{  ({\cal F}_\mu^3)^2 +  ({\cal F}_\mu^4)^2 \}
+ v_3^2 \, ({\cal F}_\mu^5)^2 
\nonumber\\&&
~~~~~~~~~~~~~~~~
~~~~~~~~~~~
+
\frac{1}{2}(v_1^2+v_3^2)\{  ({\cal F}_\mu^6)^2 +  ({\cal F}_\mu^7)^2 \}+
\frac{1}{2}(v_2^2+v_3^2)\{  ({\cal F}_\mu^8)^2 +  ({\cal F}_\mu^9)^2 \}
\Biggr] \, .
\eea
The mass eigenstates ${\cal F}_\mu^i$ are labelled in the order of their masses, which
are given by
\bea
&&
\begin{array}{l}\displaystyle
{\cal F}_\mu^1 = \frac{f_\mu^0}{\sqrt{3}}+\frac{f_\mu^3}{\sqrt{2}}+\frac{f_\mu^8}{\sqrt{6}},
~~~
{\cal F}_\mu^2 = \frac{f_\mu^0}{\sqrt{3}}-\frac{f_\mu^3}{\sqrt{2}}+\frac{f_\mu^8}{\sqrt{6}},
~~~
{\cal F}_\mu^5 = \frac{f_\mu^0-\sqrt{2}\,f_\mu^8}{\sqrt{3}},
\\
\rule[0mm]{0mm}{7mm}
{\cal F}_\mu^{3,4} =f_\mu^{1,2},
~~~~~~~
{\cal F}_\mu^{6,7} =f_\mu^{4,5},
~~~~~~~
{\cal F}_\mu^{8,9} =f_\mu^{6,7}\, .
\end{array}
\eea
Hence,
\bea
 f_\mu^\alpha\,T^\alpha =
\left(
\begin{array}{ccc}
\frac{1}{\sqrt{2}}{\cal F}_\mu^1&
-\frac{i}{2}({\cal F}_\mu^3+i{\cal F}_\mu^4)&
-\frac{i}{2}({\cal F}_\mu^6+i{\cal F}_\mu^7)\\
\rule[0mm]{0mm}{5mm}
\frac{i}{2}({\cal F}_\mu^3-i{\cal F}_\mu^4)&
\frac{1}{\sqrt{2}}{\cal F}_\mu^2&
-\frac{i}{2}({\cal F}_\mu^8+i{\cal F}_\mu^9)\\
\rule[0mm]{0mm}{5mm}
\frac{i}{2}({\cal F}_\mu^6-i{\cal F}_\mu^7)&
\frac{i}{2}({\cal F}_\mu^8-i{\cal F}_\mu^9)&
\frac{1}{\sqrt{2}}{\cal F}_\mu^5
\end{array}
\right)\,.
\label{fmuTmu}
\eea

The form of the radiative correction given by eqs.~(\ref{alphaFcorr})
and (\ref{mmu})
is constrained by symmetries and
their breaking patterns.
As the diagonal elements of the VEV,
$v_3>v_2>v_1>0$, are successively turned on,
gauge symmetry is broken according to the 
pattern:
\bea
U(3) \to U(2)\to U(1)\to \mbox{nothing} \, .
\label{SymBreakPat}
\eea
At each stage, the gauge bosons corresponding to 
the broken generators
acquire masses and decouple.
Furthermore, the vacuum $\Phi_d$ and the 
family gauge interaction
respect a global $U(1)_{V1}\times U(1)_{V2}\times U(1)_{V3}$
symmetry generated by
\bea
\psi_L \to U_d\,\psi_L  ,
~
e_R \to U_d^*\, e_R ,
~
\Phi_d \to U_d\, \Phi_d \,U_d^* 
\eea
with
\bea
U_d=
\left(\begin{array}{ccc}
e^{i\phi_1}&0&0\\
0&e^{i\phi_2}&0\\
0&0&e^{i\phi_3}
\end{array}\right) \, 
~~~~~~;~~~~~~~
\phi_i \in {\bf R}
\, .
\label{U1VcubeTr}
\eea
The operator ${\cal O}^{(\ell)}$ after symmetry breakdown,
 eq.~(\ref{DiagonalMassMat}),
is not invariant under this transformation but the
variation can be absorbed into a redefinition of $v_i$'s.
As a result, the lepton mass matrix has a following
transformation property:
\bea
{\cal M}_\ell \biggr|_{v_i\to v_i \exp(i\phi_i)}
= U_d\, {\cal M}_\ell \, U_d^* \, .
\label{TransfU1Vcube}
\eea
This is satisfied including the 1--loop radiative correction.
The symmetry breaking pattern eq.~(\ref{SymBreakPat})
and the above transformation property 
constrain the form of the radiative correction
to $\delta m_i^{\rm pole} \propto v_i^2 [\log (|v_i|^2) + {\rm const.}]$,
where the constant is independent of $i$.
Note that $|v_i|^2$ in the argument of logarithm originate from
the gauge boson masses, which are invariant 
under $v_i \to v_i \exp(i\phi_i)$.

The universality of the $SU(3)$ and $U(1)$ gauge couplings 
eq.~(\ref{UnivCouplings}) is necessary to
guarantee the above symmetry breaking pattern eq.~(\ref{SymBreakPat}).
One may worry about validity of the assumption for the universality,
since the two couplings are renormalized differently in general.
The universality can be ensured approximately if these two 
symmetry groups are embedded
into a simple group down to a scale close to the relevant scale.
There are more than one ways to achieve this.
A simplest way would be to embed $SU(3)\times U(1)$
into $SU(4)$.
It is easy to verify that the $\bf 4$ of $SU(4)$ decomposes into
$({\bf 3}, -\frac{1}{2})\oplus ({\bf 1},\frac{3}{2})$ under $SU(3)\times U(1)$.
Hence, the $\bf 6$ (second-rank antisymmetric representation)
and $\bar{\bf 6}$ 
of $SU(4)$, respectively, include $(\bar{\bf 3},-1)$ and $({\bf 3}, 1)$.

Within the effective theory under consideration, the QED correction
to the pole mass is given 
just as in eq.~(\ref{QED1Lcorr}) with $\bar{m}_i(\mu)$
replaced by $m_i(\mu)$.
Recall that corrections of the form ${\rm const.}\times
{m}_i$ do not affect Koide's formula.
Then, noting $\log v_i^2 = \frac{1}{2}\log {m}_i^2 + \mbox{const.}$, 
one observes that if a relation between the QED and family gauge coupling constants
\bea
\alpha = \frac{1}{4}\, \alpha_F
\label{relalphas}
\eea
is satisfied,
the 1--loop radiative correction induced by family gauge interaction
cancels the 1--loop QED correction to Koide's mass formula.
%(We may identify
%$\bar{m}_i(\mu)$ with $m_i(\mu)$ up to the order of our interest.)
%{\it Why? The latter is gauge-dependent whereas the former is not! The
%gauge-dependent term is ${\cal O}(\alpha\alpha_F)$.
%Furthermore, the expression of \delta m^pole is valid only in Landau gauge.}) 

In fact, with the relation (\ref{relalphas}),
cancellation holds for all the leading logarithms generated
by renormalization group:
the coefficient of $\log\mu^2$ of the QED correction is determined
by the 1--loop anomalous dimension of the running mass, while the
coefficient of $\log\mu^2$ of eq.~(\ref{alphaFcorr}) is determined by
the anomalous dimension of the Wilson coefficient $\kappa^{(\ell)}$ and twice
of the
wave--function renormalization of $\Phi$;
they are resummed in the same way
by 1--loop renormalization group equations.
The renormalization group evolution and the symmetry breaking pattern
eq.~(\ref{SymBreakPat})
in the scale range
across the family gauge boson masses
dictate how
$\log m_i^2$'s induced by family gauge interaction are resummed.
The renormalization group evolution and the 
same symmetry breaking pattern in the QED sector dictate the
$\log m_i^2$ resummation of the QED correction,
in the scale range across
the lepton masses.
If $m_i\log m_i^2$ cancel at 
1--loop, $\log m_i^2$ dependences in all the leading logarithms 
$m_i[\alpha \log(\mu^2/m_i^2)]^n$
also cancel.
On the other hand, effects of the running of $\alpha$
and $\alpha_F$ do not cancel.
It is 
related to the question which we stated in the Introduction:
What are the relevant scales for the
coupling constants in the relation (\ref{relalphas})?
The scale of $\alpha$ is determined by the
lepton masses, while the scale of $\alpha_F$ is determined by 
the family gauge boson masses, which should be much
higher than the electroweak 
scale.

Suppose the relation (\ref{relalphas}) is satisfied.
Then
\bea
m_i^{\rm pole} \propto v_i(\mu)^2
\label{RelPoleMassVEVs}
\eea
holds
including the leading logarithms generated by the running of
$\kappa^{(\ell)}$ and $v_i$'s.
This is valid for any value of $\mu$.
This means,
if $v_i(\mu)$'s satisfy 
\bea
\frac{v_1(\mu)+v_2(\mu)+v_3(\mu)}
{v_0(\mu)}
=\sqrt{\frac{3}{2}} 
~~~~~;~~~~~
v_0(\mu) =\sqrt{v_1(\mu)^2+v_2(\mu)^2+v_3(\mu)^2}
\label{relvi}
\eea
at some scale $\mu$, Koide's formula is satisfied
at any scale $\mu$.
This is a consequence of the fact that $\Phi$
is multiplicatively renormalized.
Generally, the form of the effective potential varies with scale $\mu$.
If the relation (\ref{relvi}) is realized at some scale as a
consequence of a specific nature of the effective potential
(in Landau gauge), the same
relation holds
automatically at any scale.
Although these statements are formally true, 
physically one should consider scales only above the family
gauge boson masses, since decoupling of
the gauge bosons is not encoded 
in $\overline{\rm MS}$
scheme.
For our purpose, it is most appropriate to use eq.~(\ref{RelPoleMassVEVs}) to
relate the charged lepton pole masses with the VEV at the cut--off
scale, i.e.\ $\mu=\Lambda$, which sets a boundary (initial) 
condition of the effective theory.

The advantages of choosing Landau gauge in our computation are
two folds:
(1) The computation of the 1--loop
effective potential for the determination of
$\langle \Phi \rangle$ becomes particularly simple
(as well known in computations of the
effective potential in various models); in particular
there is no ${\cal O}(\alpha_F)$ correction to the effective
potential.
(2) The lepton wave--function renormalization is finite;
as a consequence, 
the diagrams in Figs.~\ref{1LoopGraph2abc}(b)(c) 
are independent of $\langle\Phi(\mu)\rangle$
and independent of flavor.
Due to the former property, there is no ${\cal O}(\alpha_F)$ correction to
the relation eq.~(\ref{relvi}) if it is satisfied at tree level.
Due to the latter property, 
$\delta m_i^{\rm pole}$ is determined essentially by
the diagram in Fig.~\ref{1LoopGraph2abc}(a) and a simple relation
to $\langle\Phi(\mu)\rangle$ follows.
%$m_i(\mu)$ defined 
%from $\langle\Phi(\mu)\rangle$ 
%in Landau gauge [eq.~(\ref{mmu})] is conceptually close 
%to the $\overline{\rm MS}$ mass $\bar{m}_i(\mu)$ defined
%in the low energy effective theory (QED).\footnote{
%We neglect flavor mixing effects due to weak interaction,
%since the masses of the left-handed neutrinos are very small.
%}

Let us comment on gauge dependence of our prediction.
If we take another gauge and express
the radiative correction $\delta m_i^{\rm pole}$
in terms of $\langle\Phi(\mu)\rangle$, 
the coefficient of $\log (\mu^2/\langle\Phi\rangle^2)$ changes, and
other non--trivial flavor dependent corrections 
are induced.
Suppose the relation eq.~(\ref{relvi}) is satisfied at tree
level.\footnote{
To simplify the argument we consider only those gauges in which
tree--level vacuum configuration is gauge independent, such as the class
of gauges considered in \cite{DelCima:1999gg}.
}
The VEV $\langle\Phi\rangle$ in another gauge
receives an ${\cal O}(\alpha_F)$ correction,
which induces a correction to eq.~(\ref{relvi}) at  ${\cal O}(\alpha_F)$.
These additional corrections to $\delta m_i^{\rm pole}$ at
${\cal O}(\alpha_F)$ should cancel altogether 
if they are reexpressed in terms of the tree--level $v_i$'s which
satisfy eq.~(\ref{relvi}),
since 
the ${\cal O}(\alpha_F)$ correction to the relation (\ref{relvi}) 
vanishes in Landau gauge.
General analyses of gauge dependence of 
the effective potential may be found in \cite{Dolan:1974gu}.

Now we speculate on a possible scenario how the relation
(\ref{relalphas}) may be satisfied.
Since the relevant scales involved in $\alpha$ and $\alpha_F$
are very different, we are unable to 
avoid assuming some
accidental factor (or parameter tuning) to achieve this condition.
Instead we seek for an indirect evidence which indicates such an
accident has occurred in Nature.
The relation (\ref{relalphas}) shows that the value of
$\alpha_F$ is close to that of the
weak gauge coupling constant $\alpha_W$, since $\sin^2\theta_W(M_W)$ is close
to $1/4$.
In fact, within the SM, $\frac{1}{4}\,\alpha_W(\mu)$ approximates $\alpha(m_\tau)$
at scale $\mu \sim 10^2$--$10^3$~TeV.
Hence, if the electroweak $SU(2)_L$ gauge
group and the $U(3)$ family gauge
group are unified 
around this scale, naively we expect that
\bea
\alpha \approx \frac{1}{4}\, \alpha_F
\label{UnifCond}
\eea
is satisfied.
Since $\alpha_W$ runs relatively slowly in the SM, 
even if the unification scale is varied within a factor of 3,
Koide's mass formula is satisfied within the present experimental
accuracy.
This shows the level of parameter tuning required in this scenario.

We may generalize our setup and see how the radiative correction
alters.
If $\psi_L$ and $e_R$ are assigned to
$({\bf 3},Q_{\psi_L})$ and $(\bar{\bf 3},Q_{e_R})$, respectively,  
the correction
eq.~(\ref{alphaFcorr}) generalizes to
\bea
&&
\delta m^{\rm pole}_i =
-\frac{\alpha_F}{8\pi}
\left[
(Q_{\psi_L}\!\!-\!Q_{e_R}\!+\!1)
\log\left(
\frac{\mu^2}{v_i(\mu)^2}
\right) + c'
\right] \,
{m}_i(\mu) \, ,
\label{Gen1}
\eea
where $c'$ is a flavor--independent constant.
Thus, the form $m_i\log m_i^2$ is maintained.
This is not the case if we vary the $U(1)$ charge of
$\Phi$, which violates the breaking pattern of gauge symmetry
eq.~(\ref{SymBreakPat})
strongly.\footnote{
Even in the case in which only the $U(1)$ charges of $\psi_L$ and $e_R$ are
varied, this symmetry breaking pattern is violated but only softly through
the gauge interaction of $\psi_L$ and $e_R$.
By contrast, varying the $U(1)$ charge of
$\Phi$ affects the spectrum of the gauge bosons.
}
The form $m_i\log m_i^2$ is maintained in 
yet another generalization, in which 
$U(3)\times U(3)$ 
symmetry is gauged.
We introduce another field $\Sigma:({\bf 1}, 0, {\bf 6}, 2)$
under $SU(3)\times U(1)\times SU(3)\times U(1)$.
The symmetry transformations are given by
$\psi_L \to U\psi_L$, $e_R \to U^*e_R$,
$\Phi \to U\Phi V^\dagger$, $\Sigma\to V\Sigma V^T$
with $U=\exp(i\theta^\alpha T^\alpha)$,
$V=\exp(i\tilde{\theta}^\alpha T^\alpha)$.
We assume that
$\langle\Sigma\rangle = v_\Sigma {\bf 1}$ with $v_\Sigma\ll v_1,v_2,v_3$,
and that the lepton masses are generated by a higher--dimensional operator
\bea
{\cal O}^{(\ell)}_2 = 
\frac{\kappa^{(\ell)}(\mu)}{\Lambda^3}\,
\bar{\psi}_{L}\, \Phi\, \Sigma \, \Phi^T\, \varphi \, e_{R} \, .
\label{exampleO2}
\eea
For the assignment
$\psi_L:({\bf 3},1,{\bf 1},0)$ and $e_R:(\bar{\bf 3},-1,{\bf 1},0)$,
the radiative correction reads
\bea
&&
\delta m^{\rm pole}_i =
-\frac{3}{8\pi}
\frac{\alpha_F^4}{\alpha_F^2+{\alpha_F^{\,\prime}}^2}
\left[
\log\left(
\frac{\mu^2}{v_i(\mu)^2}
\right) + c''
\right] \,
{m}_i(\mu) \, ,
\label{Gen2}
\eea
where $\alpha_F$ and $\alpha_F'$ denote, respectively, the
gauge couplings of the first $U(3)$ and second $U(3)$ 
symmetries.\footnote{
If $\alpha_F=\alpha_F'$, 
$SU(3)\times SU(3)\times U(1)_A$ can be
embedded into $SU(6)$.
In this case, $\psi_L$ and $e_R$ can be assigned to
the ${\bf 6}$ and $\bar{\bf 6}$ of $SU(6)$, respectively.
The remaining $U(1)_V$, corresponding to the lepton number, 
is unbroken, so it may be taken as a global symmetry.
}
Thus, the coefficient of $m_i\log m_i^2$ varies
in different setups.
Accordingly the condition for the cancellation of the QED
correction changes from eq.~(\ref{relalphas}).
We need to seek for 
other possible scenarios which lead to such conditions,
or maybe to let the cancellation be a sheer coincidence.
The level of fine tuning required for the coupling(s) is about
1\% to meet the present expermental accuracy of Koide's formula.

In the rest of this paper, we do not consider these generalizations.
We adhere to eq.~(\ref{relalphas}), assuming the scenario in which
$SU(2)_L$ and $U(3)$ gauge symmetries are unified at
around $10^2$--$10^3$~TeV.
In this paper we do not construct a model which incorporates this
unification scenario.
We simply assume that this unification scenario 
is realized in the underlying full theory, in which
the unification scale is at or around
the cut--off scale $\Lambda$ of our effective theory;
we further assume that the hierarchy
between $v_3$ and $\Lambda(>v_3)$ is mild;
see discussions in Secs.~2 and \ref{s7}.

\section{Potential Minimum and Charged Lepton Spectrum}
\label{s3}
\clfn

The analysis in the previous section indicates relevance of the
$U(3)$ family gauge symmetry in relation to the charged lepton
spectrum and Koide's mass formula.
In this section we study the potential of $\Phi$
invariant under this family symmetry and its classical
vacuum.
In particular, we propose a mechanism for generating a realistic
charged lepton spectrum, assuming that Koide's mass relation is
protected.
For later convenience,
we express components of  $\Phi$ using $T^\alpha$, defined in 
eqs.~(\ref{U3generators}) and (\ref{Texplicit}),
as the basis:
\bea
\Phi =  \Phi^\alpha \, T^\alpha \, .
\eea
In general $\Phi^\alpha$ takes a complex value.

The largest symmetry 
that can be imposed on
the higher--dimensional operator ${\cal O}^{(\ell)}$
is $U(3)\times U(3)$. 
[An example is given in eq.~(\ref{exampleO2}).]
We may consider the
potential of $\Phi$ consistent with this symmetry, 
allowing only operators with dimension 4 or less.
A general analysis shows that, for any choice of the parameters (couplings) of 
this potential, the classical vacuum $\langle \Phi \rangle$,
 after its diagonalization,
does not satisfy the relation (\ref{relvi})
\cite{Li:1973mq}.
Namely, there is no classical vacuum that leads to Koide's mass formula.
If we impose a smaller symmetry on the potential of $\Phi$, it
is possible to tune the parameters in the potential and realize the
relation (\ref{relvi}) as well as a realistic charged lepton spectrum.
We were, however, unable to find a sensible reasoning for tuning the
parameters with an accuracy
necessary to realize Koide's mass formula.
\begin{figure}[t]\centering
\psfrag{45}{\hspace{0mm}$45^\circ$}
\psfrag{Phi0}{\hspace{0mm}$\Phi^0T^0$}
\psfrag{Phia}{\hspace{0mm}$\Phi^aT^a$}
\psfrag{Phimu}{\hspace{0mm}$\Phi^\alpha T^\alpha$}
\includegraphics[width=6cm]{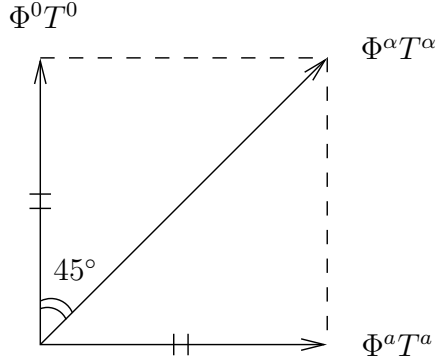}
\caption{\small 
Geometrical interpretation of \protect{eq.~(\ref{KFrelation})}.
\protect{Eq.~(\ref{U3generators})} defines the inner product in a 9--dimensional
real vector space spanned by the basis $\{ T^\alpha \}$.
Since $\Phi^0 T^0$, $\Phi^a T^a$ and $\Phi=\Phi^\alpha T^\alpha$
form an isosceles right triangle,
the angle between $T^0$ and $\Phi$ is $45^\circ$.
This is
Koide's formula in the basis where $\Phi$ is diagonal
\cite{Foot:1994yn}.
\label{GeometricInt}}
\end{figure}

We may reverse the argument partially and
search for a realistic vacuum within a
restricted set of configurations.
Namely, in view of the high accuracy with which Koide's mass formula
is realized in Nature,
it may make sense to
assume that this mass relation is protected by some
mechanism.
(An example of such a mechanism will be given in the next section.)
We assume that the vacuum configuration satisfies
\bea
(\Phi^0)^2= \Phi^a\, \Phi^a
~~~~~;~~~~~
\Phi^\alpha \in {\bf R}\, 
\label{KFrelation}
\eea
in an appropriate basis allowed by the
symmetry.
In this case, the relation (\ref{relvi}) is satisfied
by the eigenvalues of $\Phi$ \cite{Koide:1989jq}; see Fig.~\ref{GeometricInt}.
Then we minimize the potential of $\Phi$ within the configurations
which satisfy this condition.
Eq.~(\ref{KFrelation}) imposes one condition among the
three masses of leptons.
Apart from the overall normalization of the
spectrum, there remains only one
free parameter, which should be fixed by minimizing the
potential.
Since the condition (\ref{KFrelation}) or the relation (\ref{relvi})
treats the three mass eigenvalues symmetrically, {\it a priori}
it seems difficult to generate a hierarchical spectrum.
If we impose the $U(3)\times U(3)$ symmetry to the potential
 of $\Phi$, there is
no vacuum corresponding to a realistic lepton spectrum.
We find that in the case of the $U(3)\times SU(2)$ symmetry, 
a realistic spectrum follows from the vacuum of
a simple potential.

In the rest of this section we study a classical vacuum of the potential of
$\Phi$ which is invariant under the $U(3)\times SU(2)$ transformation
\bea
\Phi \to U\,\Phi\,O^T
~~~~;
~~~~
U\,U^\dagger = O\,O^T={\bf 1} 
\, .
\label{U3SU2Transf}
\eea
Up to dimension 4, there are only 4 independent invariant
operators.
We parametrize the potential as
\bea
V(\Phi) = V_{\Phi 1}(\Phi) +V_{\Phi 2} (\Phi)+V_{\Phi 3}(\Phi)
\eea
where 
\bea
&&
V_{\Phi 1} (\Phi)= \lambda 
\left[ {\rm tr}(\Phi^\dagger\,\Phi) - v^2 \right]^2 \, ,
\label{VPhi1}
\\ &&
V_{\Phi 2}(\Phi) = \varepsilon_{\Phi 2} \,
{\rm tr}(\Phi^\dagger\,\Phi\,\Phi^\dagger\,\Phi) \, , 
\label{VPhi2}
\\ &&
V_{\Phi 3}(\Phi) = \varepsilon_{\Phi 3} \,
{\rm tr}(\Phi\,\Phi^T\,\Phi^*\,\Phi^\dagger) \, .
\label{VPhi3}
\eea
The 4 independent parameters $\lambda$, $v^2$, $\varepsilon_{\Phi 2}$
and $\varepsilon_{\Phi 3}$ are real.
These potentials are classified according to the symmetries:
Since 
${\rm tr}(\Phi^\dagger\,\Phi) = 
\frac{1}{2} \, {\Phi^\alpha}^*\Phi^\alpha$,
$V_{\Phi 1} $ is invariant under $SO(18)$;
$V_{\Phi 2} $ is invariant under $SU(3)\times SU(3)\times U(1)$;
$V_{\Phi 3} $ is invariant under $U(3)\times SU(2)
\simeq SU(3)\times SU(2)\times U(1)$.

We assume the condition
(\ref{KFrelation}).
When 
$\varepsilon_{\Phi 2}=0$ and $\lambda,v^2,\varepsilon_{\Phi 3}>0$,
the configuration which
minimizes $V(\Phi) $ under this condition
corresponds to a charged lepton spectrum
very close to the
experimentally observed one.
Let us describe the details of this configuration.
Using the transformation (\ref{U3SU2Transf}),
any $\Phi$
can be brought to a form parametrized by 6 real parameters.
Without loss of generality, we can choose
$( \Phi^0, \Phi^2, \Phi^3, \Phi^4, \Phi^6, \Phi^8 )$ as the
real parameters, while
$\Phi^1,\Phi^5,\Phi^7$ are set to zero.\footnote{
Since any $\Phi$ can be diagonalized by a bi--unitary transformation,
 $\Phi_d = U \Phi V^\dagger$, $\Phi$ can be brought to an hermite
matrix by a $U(3)$ transformation as
$\Phi'=V^\dagger U\Phi = V^\dagger \Phi_d V$, i.e.\
${\Phi^\alpha}' \in {\bf R}$.
Noting that
$({\Phi^2}',{\Phi^5}',{\Phi^7}')$ transforms as the ${\bf 3}$ of the
diagonal subgroup $SU(2)_V\subset U(3)\times SU(2)$,
we may set ${\Phi^5}', {\Phi^7}'=0$ using this
transformation.
Using a residual degree of
freedom, which rotates
$({\Phi^1}',{\Phi^3}')$ as a real doublet of
$O(2)$, we can set ${\Phi^1}'=0$.
}
Then it is straightforward (but cumbersome)
to minimize the potential
$V_{\Phi 1} +V_{\Phi 3} $ under the condition (\ref{KFrelation}).
One finds the global minimum at the configuration
\bea
\Phi_0:\left\{
\begin{array}{l}
(\Phi_0^0,\Phi_0^2,\Phi_0^8)=v_0\, (1, \sqrt{1-x_0^2},x_0)\\
\mbox{other}~\Phi_0^\alpha=0
\rule[0mm]{0mm}{7mm}
\end{array}
\right.
\label{Phi0}
\eea
where\footnote{
$x_0$ is a real solution to the equation
$8\,x^3+4\,x-\sqrt{2}=0$.
}
\bea
&&
x_0 = \frac{(\sqrt{129}+9)^{1/3}-(\sqrt{129}-9)^{1/3}}
{2^{7/6}\cdot 3^{2/3}}
=0.2997... \, ,
\label{x0}
\\&&
v_0 = v \, \left[ 1+
\frac{1-3\sqrt{2}x_0+4x_0^2}{24}\,
\frac{\varepsilon_{\Phi 3}}{\lambda}
\right]^{-\frac{1}{2}}  
\approx \, 
\frac{v}{\sqrt{
 1+
0.003656\,
({\varepsilon_{\Phi 3}}/{\lambda})
}}
\, .
\label{v0A}
\eea
There are no other degenerate vacua except those which
are connected to $\Phi_0$ by the $U(3)\times SU(2)$
transformation (\ref{U3SU2Transf}).
Note that there is a residual $U(1)_{T^2}$ symmetry 
corresponding to the transformation
\bea
\Phi \to \exp(i\theta T^2) \, \Phi \, \exp(-i\theta T^2) \, ,
\label{residualU1}
\eea
which keeps the above vacuum invariant.
This is a subgroup of  $U(3)\times SU(2)$.

The three eigenvalues of $\Phi_0$ are given by
\bea
&&
(v_1,v_2,v_3)=
%\left(
%\frac{1}{\sqrt{6}}+\frac{x_0}{2\sqrt{3}}-\frac{\sqrt{1-x_0^2}}{2}, \,
%\frac{1}{\sqrt{6}}-\frac{x_0}{\sqrt{3}}, \,
%\frac{1}{\sqrt{6}}+\frac{x_0}{2\sqrt{3}}+\frac{\sqrt{1-x_0^2}}{2} \,
%\right)
%\\&&
\frac{v_0}{6}\left(\mbox{$
\sqrt{6}+\sqrt{3}\,x_0-3\sqrt{1\!-\!x_0^2}, \,
\sqrt{6}-2\sqrt{3}\,x_0, \,
\sqrt{6}+\sqrt{3}\,x_0+3\sqrt{1\!-\!x_0^2} \,
$}
\right)
%\\&&
%\left(\mbox{$
%\frac{1}{6}(\sqrt{6}+\sqrt{3}x_0-3\sqrt{1-x_0^2}), \,
%\frac{1}{\sqrt{6}}-\frac{x_0}{\sqrt{3}}, \,
%\frac{1}{6}(\sqrt{6}+\sqrt{3}x_0+3\sqrt{1-x_0^2}) \,
%$}
%\right)
\nonumber
\\&&
\rule[0mm]{0mm}{5mm}
~~~~~~~~~~~~~
\approx
v_0 \, (0.01775,0.2352,0.9718) \, .
\label{eigenvaluesvi}
\eea
The corresponding experimental values read
\bea
(\sqrt{m_e},\sqrt{m_\mu},\sqrt{m_\tau})
\approx \sqrt{m_\Sigma}\, (0.01647,0.2369,0.9714)\, ,
\label{exprootmi}
\eea
where $m_\Sigma=m_e+m_\mu+m_\tau$.
We pay particular attention to the value of 
$v_3/v_0$,
which approximates the 
corresponding experimental value with an accuracy
of $4\times 10^{-4}$.
Since the constraint (\ref{KFrelation}) treats the
three eigenvalues symmetrically, 
some kind of fine tuning should be inherent in this vacuum configuration
corresponding to a hierarchical spectrum.
Indeed, this is reflected to the fact that, if the value of $v_3/v_0$
is varied slightly from the above value under
the condition (\ref{KFrelation}), variations of $v_1/v_0$ and $v_2/v_0$ are
fairly enhanced.
(Note that the values of $v_1/v_0$ and $v_2/v_0$ are fixed by
$v_3/v_0$.)
As a result, a tiny perturbation to the potential can
bring all $v_i/v_0$ to be consistent with the experimental values.
For instance, turning on $V_{\Phi 2}$ with
$\varepsilon_{\Phi 2}/\varepsilon_{\Phi 3}\approx -6\times 10^{-3}$
will achieve this.
This feature is indifferent to details of perturbations:
they can be any mixture of $V_{\Phi 2}$, higher--dimensional operators,
and radiatively induced potentials (log potentials).

The following comparison may
 illustrate markedness of the above configuration.
When this configuration is the zeroth--order vacuum,
correct orders of magnitude
of $m_e/m_\Sigma$, $m_\mu/m_\Sigma$, $m_\tau/m_\Sigma$ are reproduced
if perturbations are sufficiently small.
By contrast,
when the zeroth--order value of
$v_3/v_0$ is in less accurate agreement with
the experimental value,
a fine tuning of perturbative contributions is necessary
even to reproduce 
the mass ratios $m_i/m_\Sigma$ with correct orders of magnitude.\footnote{
For instance, if $v_3/v_0\approx 0.9856$ ($1.5\%$ difference from the 
experimental value), $m_e$ and $m_\mu$ are predicted to be the same,
$m_e/m_\Sigma=m_\mu/m_\Sigma$.
}

We find it quite intriguing that the vacuum
of such a simple potential, which respects the $U(3)$ family
symmetry, selects this particular value 
of $v_3/v_0$ very close to the realistic value.
Noting the relation (\ref{RelPoleMassVEVs})
between the lepton pole masses and the VEV 
of $\Phi$ at high energy scales,
the above feature may suggest that the potential
takes a form
$V(\Phi) \approx V_{\Phi 1}(\Phi) + V_{\Phi 3}(\Phi)$
at the cut--off scale $\mu=\Lambda$.

At this stage, it is unclear what mechanism protects
the condition (\ref{KFrelation}).
Furthermore, it is unclear why $V_{\Phi 2}$
should be so much suppressed compared to
$V_{\Phi 3}$, 
$|\varepsilon_{\Phi 2}/\varepsilon_{\Phi 3}|\simlt 10^{-2}$.
Naively, one would expect that
radiative corrections induce $V_{\Phi 2}$
at least with a similar order of magnitude as $V_{\Phi 3}$.
In the next section, we will present a possible mechanism
or scenario to solve these problems 
(not completely but at least in such a way
to circumvent fine tuning of parameters).

\section{A Minimal Potential}
\label{s4}
\clfn

In this section, we present a potential of scalar fields,
possibly minimal in its content, which realizes
$\langle \Phi \rangle \approx \Phi_0$, defined in
eq.~(\ref{Phi0}), at its classical vacuum.
This is discussed within an effective theory which has $U(3)\times SU(2)$
family gauge symmetry, valid below the  cut--off scale $\Lambda$.
The assignment to $SU(3)\times SU(2)\times U(1)$ of the fields,
which are already introduced in the previous sections, reads
\bea
&& ~~~~~~
\psi_L:({\bf 3},{\bf 1},1), ~~~~~~
e_R:(\bar{\bf 3},{\bf 1},-1), ~~~~~~
\Phi:({\bf 3},{\bf 3},1) , ~~~~~~
\varphi : ({\bf 1},{\bf 1},0) ,
\eea
with the transformation properties
\bea
&&
\psi_L \to U \, \psi_L  , ~~~~~~
e_R \to U^* \, e_R , ~~~~~~
\Phi \to U \, \Phi \, O^T , ~~~~~~
\varphi \to \varphi \, ,
\\ &&
\rule[0mm]{0mm}{6mm}
U= \exp (i\theta^\alpha T^\alpha)\, ,
~~~
O=\exp (2i\tilde{\theta}^x T^x) 
~~(x=2,5,7)
~~~~;~~~~
U\, U^\dagger = O\, O^T = {\bf 1} \, .
\eea
Furthermore, we assume that
above the cut--off scale $\Lambda$ there is an $SU(9)\times U(1)$ 
gauge symmetry and that this
symmetry is spontaneously broken to
$U(3)\times SU(2)$ below the cut--off scale.

Let us describe the 
assignment of the fields to the group
$SU(9)\times U(1)$.
$\Phi$ is assigned to $({\bf 9},1)$; its
transformation is given by
$\Phi^\alpha \to \tilde{U}^{\alpha\beta}\Phi^\beta$
with a 9--by--9 unitary matrix\footnote{
For instance,
$2\,{\rm tr}(\Phi^\dagger\Phi)={\Phi^\alpha}^*\Phi^\alpha$ 
is invariant under $SU(3)\times SU(2)\times U(1)$
as well as $SU(9)\times U(1)$.
}
 $\tilde{U}^{\alpha\beta}$.
$\psi_L$ is included in $(\overline{\bf 36},1)$ (the ${\bf 36}$ is the
second--rank antisymmetric representation), which
decomposes into $(\bar{\bf 6},{\bf 3},1)\oplus ({\bf 3},{\bf 1},1)\oplus
({\bf 3},{\bf 5},1)$ after the symmetry breakdown;
similarly $e_R$ is included in $({\bf 36},-1)$.
$\varphi$ is a singlet under $SU(9)\times U(1)$.

In order to realize a desirable vacuum configuration,
we introduce another field $X$, which is in the representation
$({\bf 45},Q_X)$ (the ${\bf 45}$ is the
second--rank symmetric representation) and is unitary.
It can be represented by a 9--by--9 unitary symmetric matrix:
\bea
X^{\alpha\beta}=X^{\beta\alpha},
~~~~~
X^{\alpha\gamma}\,{X^{\beta\gamma}}^* = \delta^{\alpha\beta} 
~~~~~;~~~~~
X^{\alpha\beta}\to
\tilde{U}^{\alpha\rho}\,X^{\rho\sigma}\,\tilde{U}^{\beta\sigma} \, .
\label{transfX}
\eea
$X$ decomposes into
$X_S^1({\bf 6},{\bf 1},Q_X)\oplus X_S^5({\bf 6},{\bf 5},Q_X)\oplus
X_A(\bar{\bf 3},{\bf 3},Q_X)$
after the symmetry breakdown.
See Appendix~\ref{appB} for the decomposition 
of $X$ under $U(3)\times SU(2)$.

We may summarize
the essence of how to realize a 
vacuum, which satisfies eq.~(\ref{KFrelation}), 
as follows.
If the VEV of $X$ takes a value
\bea
%\langle X^{\alpha\beta} \rangle = 
X_0^{\alpha\beta} =
[{\rm diag.}(-1,+1,\cdots ,+1) ]_{\alpha\beta} =
-2\, \delta^{\alpha 0}\, \delta^{\beta 0}+\delta^{\alpha\beta} \, ,
\label{X0}
\eea
an $SU(9)\times U(1)$--invariant condition 
\bea
\Phi^\alpha \, {X^{\alpha\beta}}^* \, \Phi^\beta = 0 \, 
\label{simplecond}
\eea
reduces to the first condition in eq.~(\ref{KFrelation})
at $X=X_0$.
The second condition in eq.~(\ref{KFrelation}) can be realized
by maximizing ${|\Phi^0|}^2$
upon fixing the value of ${\Phi^\alpha}^*\Phi^\alpha$ and
imposing the first condition of eq.~(\ref{KFrelation});
see Appendix~\ref{appC1}.
These conditions can be met at the classical vacuum of the potential
of the scalar fields under consideration, with an appropriate choice of parameters
in the potential.
We may avoid fine tuning 
of the parameters, except for
the one related to stabilization of the electroweak scale.

In what follows we do not discuss any details of the theory above the
scale $\Lambda$.
Rather we use general properties of 
$SU(9)\times U(1)$ gauge symmetry to infer
boundary conditions to be imposed at the scale $\Lambda$.
We also investigate boundary 
conditions at this scale
 required from the low--energy side phenomenologically, 
consistently with symmetry requirements.

We study the potential and its vacuum
of the scalar fields, $\Phi$, $X$ and $\varphi$.
First we analyze the potential of a specific form
(or with a specific choice of parameters of the potential), which incorporates an
essential part of our model.
Later we extend the potential to more general forms.
The potential we analyze reads
\bea
V(\Phi,X)=V_{\Phi 1} + V_{\Phi 3} + V_{X1} + V_{K1} + V_{\Phi X 1}\, ,
\label{PotEssence}
\eea
where $V_{\Phi 1}$ and $V_{\Phi 3}$ are defined in eqs.~(\ref{VPhi1}) and
(\ref{VPhi3}), respectively, and the other potentials are defined by
\bea
&&
V_{X1} = \varepsilon_{X1} \, v^4 \, {\rm
tr}(T^\alpha\,T^\rho\,T^\beta\,T^\sigma)\,
X^{\alpha\beta}\,{X^{\rho\sigma}}^* \, ,
\label{VX1}
\\ && 
\rule[0mm]{0mm}{5mm}
V_{K1} = \varepsilon_{K1} \, \bigl|
\Phi^\alpha \, {X^{\alpha\beta}}^* \, \Phi^\beta 
\bigr|^2 \, ,
\label{VK1}
\\ && 
\rule[0mm]{0mm}{5mm}
V_{\Phi X 1}=  - \varepsilon_{\Phi X 1} \, v^2 \, 
{\rm tr}(T^\alpha \, T^\beta \, \Phi^\dagger \, T^\rho \, T^\sigma \, \Phi ) \,
{X^{\alpha \sigma}}^* \, {X^{\beta\rho}} \, .
\label{VPhiX1}
\eea
All the parameters of the potential,
$\lambda$, $\varepsilon_{\Phi 3} $,
$\varepsilon_{X1}$, $\varepsilon_{K} $,
$\varepsilon_{\Phi X 1}$, $v$, are taken to be positive.
Note that since the field
$X$ is unitary, it is dimensionless.
The physical scale of its VEV is determined by the kinetic term of
$X$, which is normalized as $f_X^2\,|(D_\mu X)^{\alpha\beta}|^2$.
Thus, the physical scale of the VEV of $X$ is ${\cal O}(f_X)$.
We choose $f_X$ to be much smaller than $v$ (the scale of $\langle \Phi
\rangle$), such that the spectrum of the family gauge bosons is determined
predominantly by $\langle \Phi \rangle$.
(See discussion in Sec.~\ref{s7}.)

One may verify the following properties of the potential:
\begin{itemize}
\item
The global minimum of $V_{X1}$ is at $X=X_0$, 
defined by eq.~(\ref{X0}).
Degenerate configurations are only those which are connected to $X_0$ by
the $SU(3)\times SU(3)\times U(1)$ transformation
(the symmetry transformation of $V_{X1}$).
\item
$V_{K1}$ is minimized if eq.~(\ref{simplecond}) holds.
This equation reduces to the condition $(\Phi^0)^2=\Phi^a\Phi^a$ in the
     case that 
$X=X_0$.
\item
If $X=X_0$, $V_{\Phi X 1}\sim - \, \varepsilon_{\Phi X 1}     {|\Phi^0|}^2$ 
up to a term that can be absorbed in $V_{\Phi 1}$:
\bea
V_{\Phi X 1}\Bigr|_{X=X_0}=
- \varepsilon_{\Phi X 1} \, v^2 \,
\left[ \, 
\frac{5}{8}\, {|\Phi^0|}^2 + \frac{1}{18}\,{\Phi^\alpha}^* \Phi^\alpha
\right] \, .
\eea
\item
If the constraints $(\Phi^0)^2=\Phi^a\Phi^a$ and
${\Phi^\alpha}^*\Phi^\alpha=2v_0^2(>0)$ are imposed,
$V_{\Phi X1}|_{X=X_0}$ is minimized when $\Phi^\alpha$'s
have a common phase $\theta \in {\bf R}$, namely
$e^{-i\theta}\Phi^\alpha \in {\bf R}$ for all $\alpha$; see Appendix~\ref{appC1}.
\item
$\Phi=\Phi_0$, defined by eq.~(\ref{Phi0}), is the classical vacuum of $V_{\Phi 3}$
under the constraints 
${\Phi^\alpha}^*\Phi^\alpha=2v_0^2$,
$(\Phi^0)^2=\Phi^a\Phi^a$ and $\Phi^\alpha\in {\bf R}$.
\item
If the constraints $(\Phi^0)^2=\Phi^a\Phi^a$ and
${\Phi^\alpha}^*\Phi^\alpha=2v_0^2$ are imposed,
the first derivative of $V_{\Phi 3}$ vanishes,\footnote{
     It can be shown, for instance, from the invariance of both $V_{\Phi 3}$ and $\Phi_0$
     under the $U(1)_{T^2}$ transformation
eq.~(\ref{residualU1}) and two
$Z_2$ transformations given by
$\Phi^\alpha \to (P_i^{\alpha\beta}\Phi^\beta)^*$
     with $P_1={\rm diag.}(+1,-1,+1,+1,-1,+1,+1,-1,+1)$ and
     $P_2={\rm diag.}(+1,-1,+1,+1,+1,-1,-1,+1,+1)$.
     } 
$\partial V_{\Phi 3}/\partial \Phi^\alpha=
\partial V_{\Phi 3}/\partial {\Phi^\alpha}^*=
0$, at $\Phi=\Phi_0$.
This is not trivial:
In general there may be a non--zero
derivative in an imaginary direction,
since $\Phi_0$ is determined assuming $\Phi^\alpha\in {\bf R}$.
\item
All terms in $V(\Phi, X)$ except $V_{\Phi 3}$ is invariant under
     $SU(3)\times SU(3)\times U(1)
(\supset U(3)\times SU(2))
$, while
the variation of $V_{\Phi 3}$ 
is positive semi--definite at $\Phi=\Phi_0$. 
Namely,
if $\Phi_0'=U_1\Phi_0 U_2^\dagger$ $(U_1U_1^\dagger=U_2U_2^\dagger={\bf 1})$, 
$V_{\Phi 3}(\Phi_0')\geq V_{\Phi 3}(\Phi_0)$;
see Appendix~\ref{appC2}.
\end{itemize}
Due to these properties, the classical vacuum
of $V(\Phi,X)$ in the limit 
$\varepsilon_{\Phi 3},\varepsilon_{\Phi X1}\ll
\varepsilon_{K1}, \varepsilon_{X1} $
is given by $\Phi=e^{i\theta}\Phi_0$ and $X=X_0$
up to a $U(3)\times SU(2)$ transformation, provided
that $\varepsilon_{\Phi X1}/\varepsilon_{\Phi 3}$ exceeds a critical value
to assure the
reality condition on $\Phi^\alpha$
(up to a common phase):
\bea
\frac{\varepsilon_{\Phi X1}}{\varepsilon_{\Phi 3}} > 0.02164\dots \, .
\label{RealityCond}
\eea
We note that the definition of $v_0$ should be modified, including the
effect of $V_{\Phi X 1}$, from eq.~(\ref{v0A}) to
\bea
&&
v_0 = v \, 
\left[
1+\frac{53}{144}\,\frac{\varepsilon_{\Phi X 1}}{\lambda}
\right]^{\frac{1}{2}}
\left[ 1+
\frac{1-3\sqrt{2}x_0+4x_0^2}{24}\,
\frac{\varepsilon_{\Phi 3}}{\lambda}
\right]^{-\frac{1}{2}}  
\nonumber\\&&
~~~
\approx \, v\,
\sqrt{
\frac{
1+0.3681({\varepsilon_{\Phi X 1}}/{\lambda})
}
{ 1+
0.003656\,
({\varepsilon_{\Phi 3}}/{\lambda})
}}
\, .
\label{v0modified}
\eea
The operators $V_{\Phi 3}$ and $V_{\Phi X1}$,
whose couplings need to be suppressed, are non--invariant
under $SU(9)\times U(1)$.
This is a key property of our model which allows
us to circumvent fine tuning, as we will discuss shortly.
As far as the charged lepton masses are
concerned,
the phase $\theta$ can be removed by redefining
the phases of $\psi_L$ and $e_R$.
Hence, we set $\theta=0$ in the following analysis for
simplicity.\footnote{
The degeneracy of the vacua parametrized by $\theta$ originates
from an accidental $U(1)_\Phi$ global symmetry of the potential $V(\Phi, X)$,
under which 
the overall phase of $\Phi$ is rotated independently of $X$.
The degeneracy will be lifted if we include in the potential
operators which break this accidental symmetry.
}

The ${\cal O}(\varepsilon_{\Phi 3})$ and
${\cal O}(\varepsilon_{\Phi X1})$ corrections to the
vacuum configuration can be computed.
In an appropriate basis, these are given by
\bea
&&
{\delta\Phi^0} =
{v_0}\,
\left[
\left(\frac{1}{16\,\varepsilon_{K1}} + \frac{3}{13\,\varepsilon_{X1}}\right)
\bar{\varepsilon}_{\Phi}
+\frac{6-5\sqrt{2}\,x_0}{78\,\varepsilon_{X1}}\,\varepsilon_{\Phi X1}
\right]\, ,
%~~~~~~~
\label{Corr2VacConfig1}
\\&&
{\delta\Phi^2}=-{\textstyle  \sqrt{1-x_0^2}}\,\,{\delta\Phi^0}
,
~~~~~~~
{\delta\Phi^8}=-x_0\,{\delta\Phi^0}
,
\label{Corr2VacConfig1b}
\\&&
\delta X^{02}=\delta X^{20}=
2{\textstyle  \sqrt{1-x_0^2}}\, 
\left(
\frac{3}{13\,\varepsilon_{X1}}\,
\bar{\varepsilon}_{\Phi}
+\frac{5-2\sqrt{2}\,x_0}{78\,\varepsilon_{X1}}\,\varepsilon_{\Phi X1}
\right) \, ,
%~~~~~~~ ~~~~~~~
\label{Corr2VacConfig2}
\\&&
\delta X^{08}=\delta X^{80}=
2\,x_0\, 
\left(
\frac{3}{13\,\varepsilon_{X1}}\,
\bar{\varepsilon}_{\Phi}
+\frac{1+2\sqrt{2}\,x_0-8\,x_0^2}{78\,\varepsilon_{X1}}\,\varepsilon_{\Phi X1}
\right) \, ,
\label{Corr2VacConfig2b}
\\&&
\mbox{all other}~~\delta \Phi^\alpha,\,\delta X^{\alpha\beta}=0
\, ,
\label{Corr2VacConfig3}
\eea
where
\bea
\bar{\varepsilon}_{\Phi}=
\frac{5}{8}\,\varepsilon_{\Phi X1} - 
\frac{\sqrt{6}\,v_1v_3\,(v_1+v_3)}{v_0^3} \, \varepsilon_{\Phi 3}
\approx
\frac{5}{8}\,(\varepsilon_{\Phi X1} - 0.06690\, \varepsilon_{\Phi 3}
)
\, ,
\eea
and $v_i$'s are given by eq.~(\ref{eigenvaluesvi}).
Hence, violation of Koide's mass formula is expected to be 
${\cal O}(\varepsilon_\Phi/\varepsilon_{K1})$ or ${\cal O}(\varepsilon_\Phi/\varepsilon_{X1})$,
where $\varepsilon_\Phi$ represents $\varepsilon_{\Phi X1} $ or $\varepsilon_{\Phi 3} $.
The explicit expression of the charged lepton spectrum
including the above corrections
depends on the precise form of the higher--dimensional operator
${\cal O}^{(\ell)}$ which generates the lepton masses.
%; c.f.\ Sec.~\ref{s2}.
Naively one expects that $\bar{\varepsilon}_\Phi/\varepsilon_{K1}, \, 
\bar{\varepsilon}_\Phi/\varepsilon_{X1}, \,
{\varepsilon}_{\Phi X}/\varepsilon_{X1}
\simlt 10^{-5}$ should be satisfied, in order
to meet the experimental
accuracy of Koide's formula.
[Compare with the estimates below eq.~(\ref{DeltaNum}).]
%have $\langle \Phi \rangle \approx \Phi_0$.

Next we 
consider the potential of $\Phi$ and $X$ in general and
examine conditions necessary for realizing
$\langle \Phi \rangle \approx \Phi_0$ and
$\langle X \rangle \approx X_0$.
Noting that $X$ is unitary,
the potential  invariant under $SU(9)\times U(1)$ can be written as
\bea
V^{SU(9)\times U(1)} _{\,\Phi X}=
\sum_{n,m\geq 0} C_{nm}\,
({\Phi^\alpha}^*\Phi^\alpha)^n\, 
\bigl| \Phi^\beta \, {X^{\beta\gamma}}^* \, \Phi^\gamma \bigr|^{2m}
\, .
\label{SU9invPot}
\eea
When the coefficients $C_{nm}$ are appropriately chosen
(without fine tuning), $V^{SU(9)\times U(1)} _{\,\Phi X}$ can  have
a minimum at ${\Phi^\alpha}^*\Phi^\alpha>0$
and $\Phi^\beta \, {X^{\beta\gamma}}^* \, \Phi^\gamma=0$.
These are satisfied by $\Phi=\Phi_0$ and $X=X_0$.

All the other operators are non--invariant under
$SU(9)\times U(1)$.
We separate them into three categories:
those which depend only on $X$ ($V_X$),
those which depend only on $\Phi$ ($V_\Phi$),
and those which depend on both $\Phi$ and $X$ ($V_{\Phi X}$).
Requirements to
each of them are as follows:
\begin{itemize}
\item
We can show that the first derivative of $V_X$ vanishes
at $X=X_0$ if $CP$ invariance is preserved; see Appendix~\ref{appC4}.
This means that, assuming $CP$ invariance, 
$X=X_0$ can be the global minimum of $V_X$
in a certain domain of the parameter space 
(spanned by the parameters in $V_X$),
without fine tuning of parameters.\footnote{ 
One example is the case in which $V_{X1}$ gives a dominant contribution
in $V_X$, 
although other operators need not be suppressed
by orders of magnitude.
This is because, contributions of other operators cannot create
a non--zero derivative at $X=X_0$, and the position of the 
global minimum
is altered only if their contributions are large enough to
create a global minimum at another configuration.
%It is also likely that operators other than $V_{X1}$
%have the global minima at $X=X_0$.
}
\item
Up to dimension 4,
$V_{\Phi}$ consists only of $V_{\Phi 2}$ and $V_{\Phi 3}$; see Sec.~\ref{s3}.
Since effects of higher--dimensional operators are expected
to be suppressed, $V_\Phi$ will be minimized at
$\Phi \approx \Phi_0$ 
if $\varepsilon_{\Phi 2}\ll \varepsilon_{\Phi 3}
\ll \varepsilon_K, \varepsilon_X$ and
$\Phi^\alpha \, {X_0^{\alpha\beta}} \, \Phi^\beta=0$.
Here, $\varepsilon_K/v$ and $\varepsilon_X/v$ 
represent typical magnitudes of
the second derivatives of $V^{SU(9)\times U(1)} _{\,\Phi X}$
and $V_X$, respectively, at their minima.
\item
The contribution of $V_{\Phi X}$
needs to be suppressed as compared to
those of $V^{SU(9)\times U(1)} _{\,\Phi X}$ and $V_X$.
If we can treat $V_{\Phi X}$ as a perturbation,
we may substitute $X=X_0$ and
${\Phi^\alpha }^*\Phi^\alpha=v_0^2$ in the lowest--order
approximation.
Then $V_{\Phi X}$ becomes dependent only on $\Phi^0$ and
$\Phi^x$ $({x=2,5,7} )$.
The role of the operators dependent only on $\Phi^0$
is similar to $V_{\Phi X1}$;
their total contribution should not be too small compared to
that of $V_{\Phi 3}$ and should enforce the reality 
condition on $\Phi^\alpha$; c.f.\ eq.~(\ref{RealityCond}).
The role of the operators dependent on $\Phi^x$
is similar to $V_{\Phi 2}$;
in order to suppress corrections to the lepton spectrum,
contributions of these operators need to be suppressed
compared to that of $V_{\Phi 3}$.
\end{itemize}

Thus, under appropriate conditions, 
$\langle \Phi \rangle \approx \Phi_0$ can be
realized with a more general potential than
the specific potential eq.~(\ref{PotEssence}).
Coefficients of certain operators need to be suppressed
compared to the others.
One may estimate typical orders of magnitudes of 
hierarchies required in
the constraints, from the 
analysis of the specific
potential $V(\Phi,X)$, which serves as a reference
case.
Moreover, in principle it is straightforward
to compute corrections to the vacuum configuration
similar to eqs.~(\ref{Corr2VacConfig1})--(\ref{Corr2VacConfig3})
for a more general potential.

Let us comment on $CP$ invariance.
We may assume that
either $CP$ invariance is broken explicitly (but weakly)
or it is broken spontaneously.
In the former case,
since there is no observed $CP$ asymmetry in the lepton sector,
we may assume effects of the explict breaking are very small
and will not affect our argument given above significantly.
In the latter case,
since $CP$ asymmetry resides only in the
Yukawa interaction in the SM, we may attribute the 
Kobayashi-Maskawa $CP$ phase to 
the VEV of the scalar field, which is presumably existent to give masses
to the quarks, while keeping all the interactions in the
$U(3)\times SU(2)$ effective theory $CP$--invariant.
As we do not discuss quark sector at all in this paper,
this argument is rather ambiguous.
In passing, we note that all the operators in 
the potential $V(\Phi,X)$ [eq.~(\ref{PotEssence})]
are $CP$--invariant; see Appendix~\ref{appC3} for 
the $CP$ transformations.

Furthermore, the Higgs field $\varphi$ needs to be incorporated
in the potential.
Since $\varphi$ is a singlet under $U(3)\times SU(2)$, 
it can be included effectively by replacing
the coefficients of the operators in the above discussion by 
functions (polynomials) of $\varphi^\dagger\varphi$, 
e.g.\ $C_{nm} \to C_{nm}(\varphi^\dagger\varphi)$.
Hence, the conditions on the coefficients are the same
as above when evaluated at 
$\varphi^\dagger\varphi =v_{\rm ew}^2(\ll v_0^2)$.
On the other hand, the VEV of $\varphi$ is determined
from the same potential after substituting
$\Phi\approx\Phi_0$ and $X\approx X_0$, whose expansion about the minimum
should take a form
$~{\rm const.} + \lambda_\varphi \, (\varphi^\dagger\varphi -v_{\rm ew}^2)^2
+ \cdots$.

It is appropriate to regard the conditions discussed above
as those to be imposed 
on the Wilson coefficients
in the effective potential (in Landau gauge) {\it renormalized 
at the cut--off scale $\mu=\Lambda$}.
Recall that, as we discussed in Sec.~\ref{s2}, we may relate the 
charged lepton spectrum directly to the vacuum configuration
of the effective potential at $\mu=\Lambda$.
The advantage of choosing $\mu=\Lambda$ is 
that certain fine tuning can be avoided in this way.
Let us describe how it works.

According to the argument above, $\varepsilon_{\Phi 3}$ is required
to be much smaller than $\varepsilon_K$ or $\varepsilon_{X}$
in order to suppress corrections to Koide's formula.
Furthermore,  in Sec.~\ref{s3} we have seen that $\varepsilon_{\Phi 2}$
should be much smaller than $\varepsilon_{\Phi 3}$ to
generate a realistic charged lepton spectrum.
Hence, $\varepsilon_{\Phi 2}/\varepsilon_K$ and
$\varepsilon_{\Phi 2}/\varepsilon_{X}$ should be quite small,
of order $10^{-5}$ or less.
On the other hand,  
the 1--loop correction by family gauge interaction to the effective potential 
induces $V_{\Phi 2}$.
This indicates that a natural size of $\varepsilon_{\Phi 2}$ is order 
$\alpha_F^2 \sim 10^{-3}$ or larger within the $U(3)\times SU(2)$
effective theory, assuming the relation (\ref{relalphas}).
Thus, in order to realize $\langle\Phi\rangle\approx\Phi_0$,
a fine tuning of $\varepsilon_{\Phi 2}$ seems to be
requisite (provided magnitudes of $\varepsilon_K$ and $\varepsilon_{X}$
are moderate).
This argument, however, does not apply at the cut--off scale $\mu=\Lambda$:
Since $SU(9)\times U(1)$ symmetry forbids $V_{\Phi 2}$ and
$V_{\Phi 3}$ in the theory above
the scale $\Lambda$, 
both $\varepsilon_{\Phi 2}$ and $\varepsilon_{\Phi 3}$ are expected to
be suppressed at $\mu=\Lambda$ 
in the $U(3)\times SU(2)$ theory.
They are determined by the matching conditions
at $\mu=\Lambda$.
Radiative corrections within the $U(3)\times SU(2)$ effective theory
essentially do not exist at this scale.

Another advantage of choosing $\mu=\Lambda$ in the
effective potential
is that $SU(9)\times U(1)$ symmetry breaking effects
on $\langle X \rangle = X_0$ are also expected to be suppressed.
In other words, the wave function renormalizations are common to
$X_A$, $X_S^1$ and $X_S^5$, to a good approximation.
It helps to keep the first condition of eq.~(\ref{KFrelation}) precise,
which follows from eqs.~(\ref{X0}) and (\ref{simplecond}).

The sizes of Wilson coefficients of 
operators non--invariant under $SU(9)\times U(1)$ 
at $\mu=\Lambda$ depend on the dynamics 
how the breakdown of $SU(9)\times U(1)$ gauge symmetry
occurs  in the theory above
the scale $\Lambda$.
%Their sizes may
%range by orders of magnitude. 
%{\it Example?}
%In particular, they
%are not constrained by the usual argument of fine tuning\footnote{
%Here we mean that
%a strong cancellation
%between tree--level contributions and contributions from radiative
%corrections is regarded as unnatural.
%}
%within the $U(3)\times SU(2)$ theory.
For example, one can imagine cases in which
these operators are proportional to (powers of)
a VEV of some scalar field  which
breaks $SU(9)\times U(1)$ symmetry.
Then,
an operator whose dimension is $n$ would have a coefficient 
of order $g\,\Lambda^k/M^{n+k-4}$,
where $g$ is a combination of coupling constants,
$\Lambda$ is a typical scale of the scalar VEV, 
$k$ is the power of the VEV, and
$M$ represents an $SU(9)\times U(1)$--invariant mass scale
much larger than $\Lambda$.
There are no evident conflicts between this naive estimate and
the conditions on the Wilson coefficients
which we derived above,
presuming that $g$ can be small but cannot be much
larger than unity.
For instance, applying the estimate
to the parameters of $V(\Phi,X)$ and
$V_{\Phi 2}$, we find
\bea
&&
\mbox{$SU(9)\times U(1)$ invariant}:
~~~~~~~~~
\lambda, \varepsilon_{K1} \simlt {\cal O}(1) ,
~~~~~~~~~~
\lambda v^2 \simlt M ;
\\&&
\mbox{$SU(9)\times U(1)$ non--invariant}:
~~~
\varepsilon_{X1} \simlt \frac{f_X^2\Lambda^2}{v^4},
~~~
\varepsilon_{\Phi 2}, \varepsilon_{\Phi 3}
\simlt \frac{\Lambda^k}{M^k},
~~~
\varepsilon_{\Phi X1}\simlt \frac{f_X^2}{v^2} \frac{\Lambda^k}{M^k} ,
~~~~
\eea
which are compatible with the desired hierarchy of the parameters
$\varepsilon_{\Phi 2}\ll \varepsilon_{\Phi 3},\varepsilon_{\Phi X1}
\ll \varepsilon_{K1},\varepsilon_{X1}$.
Of course, one should keep in mind that the above estimates are
heavily dependent on the dynamics above the cut--off scale.

We may speculate on a possible scenario above the cut--off scale
which may lead to (part of) the
desirable hierarchical relations.
Suppose that the symmetry breaking
$SU(9)\times U(1)\to U(3)\times SU(2)$
is induced by a condensate of a scalar field
$T^{\alpha\beta}_{\rho\sigma}$, which is
a 4th-rank tensor 
under $SU(9)$.
Indeed if 
$\langle T^{\alpha\beta}_{\rho\sigma} \rangle \sim
{\rm tr}(T^\alpha{T^\beta}^*{T^\rho}^*T^\sigma)$,
this symmetry breaking takes place.
\begin{figure}[t]\centering
\psfrag{Phi}{$\Phi$}
\psfrag{T}{$\langle T \rangle$}
\psfrag{X}{$X$}
\psfrag{F1}{$\displaystyle
\frac{\langle T^{\alpha\beta}_{\rho\sigma}\rangle}{M}
\, \Phi^\alpha\Phi^\beta{\Phi^\rho}^*{\Phi^\sigma}^*
\sim
\varepsilon_{\Phi 3}\,
{\rm tr}(\Phi\,\Phi^T\,\Phi^*\,\Phi^\dagger)$}
\psfrag{F2}{$\displaystyle
\varepsilon_{X1} \sim \frac{\langle T \rangle^2}{\Lambda^2}
\simgt {\cal O}(1)
$}
\includegraphics[width=7cm]{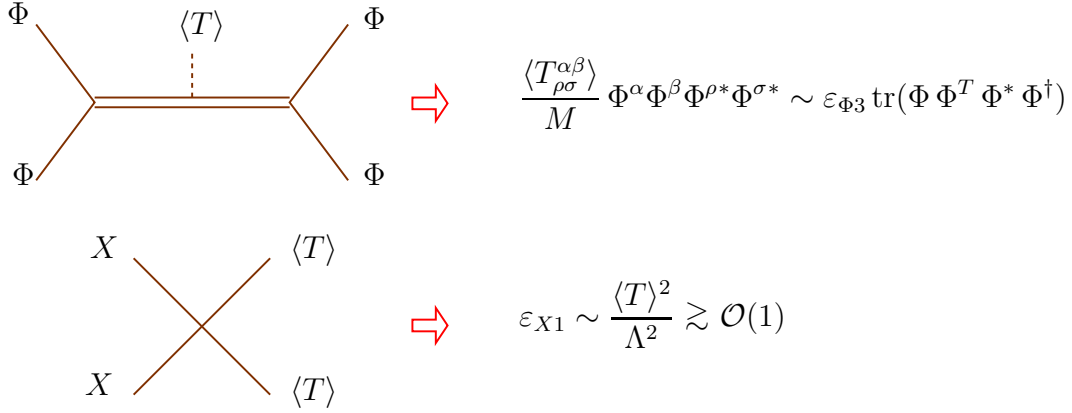}
\hspace*{70mm}
\caption{\small 
Speculation on underlying physics
that may generate $SU(9)\times U(1)$ non-invariant
operators.
\label{SpeculationGenPot2}}
\end{figure}
Through the first diagram shown in Fig.~\ref{SpeculationGenPot2},
the operator $\varepsilon_{\Phi 3}\,
{\rm tr}(\Phi\,\Phi^T\,\Phi^*\,\Phi^\dagger)$
may be induced; the double line denotes a heavy degree of freedom
with an $SU(9)\times U(1)$-invariant mass scale $M$.
Since $\langle T^{\alpha\beta}_{\rho\sigma} 
\rangle \sim {\cal O}(\Lambda)$, the coefficient
$\varepsilon_{\Phi 3}\sim \Lambda/M$ would be a small parameter provided $M\gg \Lambda$.
$\varepsilon_{\Phi 2}$ is even more suppressed, since the operator 
$\varepsilon_{\Phi 2}\,{\rm tr}(\Phi^\dagger\,\Phi\,\Phi^\dagger\,\Phi)$
cannot be generated by a single insertion of $\langle T^{\alpha\beta}_{\rho\sigma} \rangle$
at tree level.
Either two insertions of $\langle T^{\alpha\beta}_{\rho\sigma} \rangle$ or a loop correction
is necessary,
which leads to additional suppression factors.
The second diagram in Fig.~\ref{SpeculationGenPot2} would induce
the operator 
$\varepsilon_{X1}\,{\rm
tr}(T^\alpha\,T^\rho\,T^\beta\,T^\sigma)\,
X^{\alpha\beta}\,{X^{\rho\sigma}}^*$
(together with other operators).
Since there is no intermediate heavy degree of freedom,
the induced coupling $\varepsilon_{X1}$, when normalized by $\Lambda$,
would be order 1.
In order to generate $\varepsilon_{\Phi X1}$ 
with a desired order of magnitude, we need to suppose a more complicated scenario,
but we do not pursue this further here, since anyway the argument is
quite hand--waving, without any explicit model above the cut--off scale.

To end this section, let us comment on the fine tuning problem in 
maintaining a large hierarchy between the scales, 
which we mentioned in Sec.~\ref{s1}.
In the derivation of the potential of $\varphi$, it appears
unnatural that $v_{\rm ew}(\ll v)$ determines the scale,
since the natural scales involved in the effective
potential are $\Lambda$ and $v$
before substituting
$\Phi\approx\Phi_0$ and $X\approx X_0$.
Currently we do not have any reasonable 
idea on how this hierarchy problem may be resolved.
%Similarly, it would be unnatural 
%if there is a large hierarchy between
%$v_0$ and $\Lambda$, or between $f_X$ and $v_0$.
%We will discuss this issue further in a moment.
%\medbreak

\section{Inclusion of Another Scalar Field}
\label{s5}
\clfn

Our goal is to generate the spectrum of the charged leptons
$(m_e,m_\mu,m_\tau)$
such that it satisfies Koide's formula with a high accuracy
and is 
proportional to $(v_1^2,v_2^2,v_3^2)$ approximately,
where $v_i$'s are given in eq.~(\ref{eigenvaluesvi}).
For this purpose,
we need to introduce yet another scalar field.
This is because, if we construct the higher--dimensional operator
${\cal O}^{(\ell)}$
only from the fields $\psi_L$, $e_R$, $\Phi$, $X$ and $\varphi$,
the corresponding charged lepton
mass matrix cannot be brought to a
diagonal form with any choice of basis allowed by $U(3)\times SU(2)$
gauge symmetry.
(Note that $\Phi_0$ is not diagonal.)
The radiative corrections
discussed in Sec.~\ref{s2} will be altered if the mass matrix
cannot be brought to a diagonal form,
and the QED correction will not be canceled.

Thus, we introduce a 
(dimensionless) scalar field
$\Sigma_Y$ which is in the $({\bf 6},1,Q_Y)$ under 
$SU(3)\times SU(2)\times U(1)$.
It is given as a 3--by--3 symmetric matrix and
transforms as $\Sigma_Y \to U\Sigma_Y U^T$.
Consider the potentials
\bea
&&
V_{\Sigma_Y}=-\varepsilon_{Y1}\,v^4\,{\rm tr}
\left( \Sigma_Y^\dagger\Sigma_Y \right) +
\varepsilon_{Y2}\,v^4\, {\rm tr}
\left( \Sigma_Y^\dagger\Sigma_Y\Sigma_Y^\dagger\Sigma_Y \right) +
\varepsilon_{Y3}\,v^4\,
\left[ {\rm tr}
\left( \Sigma_Y^\dagger\Sigma_Y \right)
\right]^2 \, ,
\label{VSigmaY}
\\&&
V_{\Phi \Sigma_Y}=-\varepsilon_{\Phi Y1}\,
{\rm tr}\left(
\Sigma_Y^\dagger\Phi\Phi^\dagger\,\Sigma_Y \Phi^*\Phi^T
\right) \, .
\label{VPhiSigmaY}
\eea
We take all the parameters $\varepsilon_{Y1}$, $\varepsilon_{Y2}$, 
$\varepsilon_{Y3}$, $\varepsilon_{\Phi Y1}$ to be positive.
One can show that, for a given $\Phi$
and in the limit
$\varepsilon_{\Phi Y1}\ll \varepsilon_{Y1},\varepsilon_{Y2},\varepsilon_{Y3}$, $V_{\Sigma_Y}+V_{\Phi\Sigma_Y}$
is minimized at
\bea
\Sigma_Y = \sigma \, U_\Phi\, U_\Phi^T
~~~~;~~~~~
\sigma = \sqrt{\frac{\varepsilon_{Y1}}{2\,(\varepsilon_{Y2}+3\,\varepsilon_{Y3})}}
\, .
\label{vacSigmaY}
\eea
Here, 
$U_\Phi$ is a unitary matrix which diagonalizes $\Phi\Phi^\dagger$,
i.e.,\ $U_\Phi^\dagger \Phi\Phi^\dagger U_\Phi$ is a diagonal matrix;
see Appendix~\ref{appD1}.
In the case that $\Phi=\Phi_0$, the corresponding unitary matrix is given by
\bea
U_0 = \frac{1}{\sqrt{2}}\,
\left(\begin{array}{rrr}
1&0&-i\\
-i&0&1\\
0&\sqrt{2}&0
%\frac{1}{\sqrt{2}}&0&-\frac{i}{\sqrt{2}}\\
%-\frac{i}{\sqrt{2}}&0&\frac{1}{\sqrt{2}}\\
%0&1&0
\end{array}\right)
~~~~~~;~~~~~~~~
U_0^\dagger \, \Phi_0\, U_0 = \Phi_d = 
\left(\begin{array}{rrr}
v_1&0&0\\
0&v_2&0\\
0&0&v_3
%\frac{1}{\sqrt{2}}&0&-\frac{i}{\sqrt{2}}\\
%-\frac{i}{\sqrt{2}}&0&\frac{1}{\sqrt{2}}\\
%0&1&0
\end{array}\right)
\, .
\eea
Therefore, we may incorporate $\Sigma_Y$ in 
the operator ${\cal O}^{(\ell)}$ to diagonalize the lepton mass matrix.

%But the potential considered above is not sufficiently general.
As in the previous section, we  embed $\Sigma_Y$ in a representation of
a larger symmetry group, which is valid above the cut--off scale $\Lambda$.
We could find a reasonable potential only when we embed $\Sigma_Y$ to
a second--rank antisymmetric representation, and this is not possible
with $SU(9)\times U(1)$.
We find a way out by enlarging the gauge group.
Instead of $SU(9)\times U(1)$ we assume that 
$SU(nm)\times U(1)$ ($n\geq 4$, $m\geq 5$) gauge symmetry is exact
above the cut--off scale.\footnote{
$SU(nm)$ includes $SU(n)\times SU(m)$ as a maximal subgroup. 
Below the cut--off $\Lambda$, the symmetry is broken down to $SU(3)\times
SU(2)\times U(1)$, where
$SU(3)$ and $SU(2)$, respectively, are  subgroups of 
$SU(n)$ and $SU(m)$:
$SU(3)$ is embedded trivially in $SU(n)$, i.e.,\ 
the {\boldmath $n$} decomposes
into a ${\bf 3}$ and $n-3$ singlets;
$SU(2)$ is a maximal subgroup of $SU(3)'$, which is embedded in $SU(m)$
trivially.
}
Under this symmetry group, $\Phi$ is embedded in the 
({\boldmath $nm$},\,1).
We denote the field in the latter representation
by $\overline{\Phi}^\xi$ $(0\leq
\xi\leq nm-1)$
and identify $\overline{\Phi}^\xi=\Phi^\xi$ for $0\leq \xi \leq 8$.
$\overline{\Phi}$
decomposes into a $({\bf 3},{\bf 3},1)$ ($=\Phi$), $n-3$ $({\bf 1},{\bf 3},1)$'s,
$m-3$ $({\bf 3},{\bf 1},1)$'s, and $(n-3)(m-3)$ singlets 
after the symmetry is broken down to
$SU(3)\times SU(2)\times U(1)$.
Similarly $X$ is embedded in the second--rank symmetric representation of
$SU(nm)$, denoted by $\overline{X}$, and 
$\overline{X}^{\xi\eta}=X^{\xi\eta}$
 for $0\leq \xi,\eta \leq 8$.
$\Sigma_Y$ is embedded in the second--rank antisymmetric representation of
$SU(nm)$, denoted by $\overline{Y}$;
see Appendix~\ref{appD2} 
for the explicit relation between $\overline{Y}$ and $\Sigma_Y$.
Both $\overline{X}$ and $\overline{Y}$ are
unitary fields.
The kinetic terms of
$\overline{X}$ and $\overline{Y}$
are normalized as $f_{\overline{X}}^2\,|(D_\mu \overline{X})^{\xi\eta}|^2$
and $f_{\overline{Y}}^2\,|(D_\mu \overline{Y})^{\xi\eta}|^2$, respectively,
where $f_{\overline{X}}$ and $f_{\overline{Y}}$ 
are assumed to be much smaller than $v$.
%, as in the previous section.

We examine the general potential of
$\overline{\Phi}$, $\overline{X}$ and  $\overline{Y}$ which is
invariant under $SU(nm)\times U(1)$.
In particular, we would like to see if the potential can be minimized at
\bea
&&
\overline{\Phi}^\xi=
\left\{
\begin{array}{lcc}
\Phi_0^\xi &~& (0\leq\xi\leq 8)\\
0 && (\xi >8)
\end{array}
\right. \, ,
\label{vacconfigSUnm1}
\\ &&
\overline{X}^{\xi\eta} = 
-2\, \delta^{\xi 0}\, \delta^{\eta 0}+\delta^{\xi\eta} \, ,
\\ &&
\Sigma_Y = \sigma\, U_0 U_0^T\, ,
\label{vacconfigSUnm3}
\eea
without fine tuning of parameters,
where $\Sigma_Y$ is embedded in $\overline{Y}$ appropriately.
The general potential can be written in the following form:\footnote{
For instance, the right-hand side of eq.~(\ref{SU9invPot}),
after replacing $\Phi$ by $\overline{\Phi}$ and
$X$ by $\overline{X}$, is included in this expression; 
it corresponds to
the terms for which $q_1,q_1',p_2,p_2',p_3,p_3',q_4,q_4'=0$.
}
\bea
&&
V^{SU(nm)\times U(1)} _{\,\overline{\Phi}\, \overline{X}\, \overline{Y}}=
\sum_{\substack{p_i,p'_i,q_i,q'_i\geq 0\\Q_{\rm tot}=0}} 
C(p_i,p'_i,q_i,q'_i)\,
\prod_{i=1}^4 z_i(p_i)^{q_i}\,
\{ z_i(p'_i)^{q'_i} \}^* \, .
\label{GenSUnmU1Pot}
\eea
$z_i(p_i)$ denote $SU(nm)$--invariant operators\footnote{
The dot ($\cdot$) denotes contraction of $SU(nm)$ indices $\xi,\eta,\dots$;
$\overline{X}^{\,p}=
\underbrace{\overline{X}\! \cdot \!
\overline{X}\cdots \overline{X}}_p$, 
${\rm Tr}(\overline{X})=\overline{X}^{\xi\xi}$, etc.
}
\bea
&&
z_1(p_1)= {\rm Tr}\biggl[ 
\bigl(\overline{X}^\dagger \! \cdot \! \overline{Y}\bigr)^{2p_1}
\biggr],
~~~~~~~~~~~~
z_2(p_2)=
\overline{\Phi} \! \cdot \! 
\bigl(\overline{X}^\dagger\! \cdot \! \overline{Y}\bigr)^{p_2} \! \cdot \!
\overline{\Phi}^{\,*} 
,
\\&&
z_3(p_3)=
\overline{\Phi} \! \cdot \! 
\bigl(\overline{X}^\dagger \! \cdot \! \overline{Y}\bigr)^{2p_3}
\! \cdot \!\,
\overline{X}^\dagger \! \cdot \!
\overline{\Phi}
\,,
~~~~~~
z_4(p_4)=
\overline{\Phi} \! \cdot \! 
\bigl(\overline{Y}^\dagger \! \cdot \! \overline{X}\bigr)^{2p_4+1}
\! \cdot \!\,
\overline{Y}^\dagger \! \cdot \!
\overline{\Phi}
\,.
\eea
The summation is constrained to the sector with vanishing $U(1)$ charge
by the condition
\bea
Q_{\rm tot}
\equiv q_i\sum_i Q(z_i(p_i)) - q'_i\sum_i Q(z_i(p'_i)) 
=0 \, ,
\eea
where $Q(z)$ represents the $U(1)$ charge of the operator $z$.
Due to complexity of the potential, we were unable to clarify
if the configuration 
eqs.~(\ref{vacconfigSUnm1})--(\ref{vacconfigSUnm3})
can be a classical vacuum 
in a sufficiently general region of the parameter space
spanned by $\{ C(p_i,p'_i,q_i,q'_i) \}$.
We only confirmed this in a restricted region of the parameter space:
For definiteness, we set $(n,m)=(4,5)$; we consider the parameter
space spanned by
$C(p_i,p'_i,q_i,q'_i)$ for $p_i,p'_i\leq 1$ and arbitrary $q_i,q_i'$,
while all other $C(p_i,p'_i,q_i,q'_i)$ are set equal to zero.
In this restricted parameter space,
there exists a domain with a finite volume (non--zero measure),
in which
$V^{SU(nm)\times U(1)} _{\,\overline{\Phi}\, \overline{X}\, \overline{Y}}$
is minimized at the configuration 
eqs.~(\ref{vacconfigSUnm1})--(\ref{vacconfigSUnm3})
by appropriately choosing $\overline{Y}$.
Namely, the desired configuration is a vacuum
(in fact, one of many degenerate vacua) in this domain.
See Appendix~\ref{appD3} for details.
This feature may indicate that the configuration 
eqs.~(\ref{vacconfigSUnm1})--(\ref{vacconfigSUnm3})
can be a vacuum 
of $V^{SU(nm)\times U(1)} _{\,\overline{\Phi}\, \overline{X}\, \overline{Y}}$
without fine tuning of the parameters in the potential.

Operators non--invariant under $SU(nm)\times U(1)$
are induced at $\mu\leq\Lambda$.
Suppose the following operators are induced:
\bea
&&
%V_{\overline{X}1} = \varepsilon_{\overline{X}1} \, v^4 \, {\rm
%tr}({\overline{T}^\xi}^\dagger
%\,\overline{T}^\rho\,{\overline{T}^\eta}^\dagger\,\overline{T}^\sigma)\,
%\overline{X}^{\xi\eta}\,{\overline{X}^{\rho\sigma}}^* \, ,
%\\ && 
%\rule[0mm]{0mm}{5mm}
V_{\overline{\Phi},{\rm resid}}=\varepsilon_{\overline{\Phi}}\,
v^2\,\sum_{\xi\geq 9}
\bigl|\overline{\Phi}^\xi\bigr|^2\,,
\label{BreakingOp1}
\\ && 
V_{X1},\,
V_{\Phi 3},\,
V_{\Phi X 1}~~
\mbox{as defined in eqs.~(\ref{VX1}),(\ref{VPhi3}),(\ref{VPhiX1})} \, ,
\\&&
\rule[0mm]{0mm}{5mm}
V_{\Sigma_Y},\,V_{\Phi\Sigma_Y}~~
\mbox{as defined in eqs.~(\ref{VSigmaY}),(\ref{VPhiSigmaY})} \, .
\label{BreakingOp4}
\eea
%where $\{\overline{T}^\xi\}$ denotes an orthonormal basis
%of $n$--by--$m$ matrices; see Appendix~.
Then, with appropriate hierarchy of the parameters, which
we already discussed in this and previous sections, we have
the configuration 
eqs.~(\ref{vacconfigSUnm1})--(\ref{vacconfigSUnm3}) 
as a global minimum of the potential.\footnote{
There are a number of
unwanted massless modes 
% in the potential
% $V(\overline{\Phi},\overline{X},\overline{Y})$
at this minimum.
They are included in $\overline{X}$, $\overline{Y}$ and 
do not couple directly to $\psi_L$,
$e_R$, $\varphi$, $\Phi$ and $\Sigma_Y$.
Although it is straightforward to write down $U(3)\times SU(2)$--invariant
operators which give masses to these massless modes,
we do not include those operators, for the sake of simplicity.
In particular, they do not affect the formulas given in the following discussion.
} 
% , up to
% degrees of freedom corresponding to the
% $U(3)\times SU(2)$ transformation.
Generally, it depends on the dynamics above the cut--off scale which 
$SU(nm)\times U(1)$--breaking operators
are induced, and a set of operators more general than 
eqs.~(\ref{BreakingOp1})--(\ref{BreakingOp4}) can also lead
to the same vacuum configuration;
see the discussion in the previous section.

For later convenience, we may take the potential
\bea
V(\overline{\Phi},\overline{X},\overline{Y})=V_{\overline{\Phi} 1} + V_{\overline{K}1} 
+V_{\overline{\Phi},{\rm resid}}+ V_{X1}+ V_{\Phi 3} 
 + V_{\Phi X 1}
+V_{\Sigma_Y}+V_{\Phi\Sigma_Y}
\, 
\label{PotEssence2}
\eea
with
\bea
&&
V_{\overline{\Phi} 1}= {\lambda }
\left( \,\frac{1}{2}\,%\Bigl(
\overline{\Phi}^{\,\xi}{\overline{\Phi}^\xi}^* 
%\Bigr)
- v^2 \right)^2 \, ,
\label{VPhi1bar}
\\&&
V_{\overline{K}1} = \varepsilon_{K1} \, \bigl|
\overline{\Phi}^\xi \, {\overline{X}^{\xi\eta}}^* \, \overline{\Phi}^\eta 
\bigr|^2 \, ,
\label{VK1bar}
\eea
as a reference potential,
instead of 
$V(\Phi,X)$ defined in
eq.~(\ref{PotEssence}).
For definiteness, we set $(n,m)=(4,5)$.
In this case,
we obtain 
the desired vacuum configuration,
eqs.~(\ref{vacconfigSUnm1})--(\ref{vacconfigSUnm3}),
in the limit 
$\varepsilon_{\Phi 3},\varepsilon_{\Phi X1}\ll
\varepsilon_{K1}, \varepsilon_{X1} $
and 
$\varepsilon_{\Phi Y1}\ll \varepsilon_{Y1},\varepsilon_{Y2},\varepsilon_{Y3}$,
and with the additional conditions
eqs.~(\ref{RealityCond}) and (\ref{Cond-sigma}).

\section{\boldmath Higher--dimensional Operator ${\cal O}^{(\ell)}$}
\label{s6}
\clfn

We present a candidate of the higher--dimensional operator
${\cal O}^{(\ell)}$ which generates the charged lepton spectrum.
The VEV $\Phi_0$, given by eq.~(\ref{Phi0}), 
cannot be brought to a diagonal form (in 3--by--3 matrix
representation) using the $U(3)\times SU(2)$ transformation.
Hence, the operator such as the one in eq.~(\ref{exampleO1})
is inappropriate.
In terms of the fields which we introduced, a simplest possibility may be
given by
\bea
{\cal O}^{(\ell)}_3=
\frac{\kappa^{(\ell)}(\mu)}{\Lambda^2}\,
\bar{\psi}_{L}\, \Phi\, X_A^T\, \Phi \, X_A^T\,\Sigma_Y\, \varphi \, e_{R} \, .
\label{exampleO3}
\eea
In this case $2Q_X+Q_Y=0$ is required, such that this operator
becomes $U(1)$--invariant.
Since $\langle X_A \rangle \approx -\frac{5}{3}\,{\bf 1}$
in the basis
where $\langle \Phi \rangle \approx \Phi_0$ [see eq.~(\ref{VEVofXA}) in
Appendix~\ref{appB}], 
the above operator can be 
approximately rendered to the form of
eq.~(\ref{DiagonalMassMat}) by the change of basis,
$\psi_L \to U_0\,\psi_L$ and $e_R \to U_0^* \, e_R$.
The corresponding charged lepton mass matrix reads
\bea
{\cal M}_\ell =
\frac{25\,\kappa^{(\ell)}(\mu)\,\sigma\,v_{\rm ew}}{9\sqrt{2}\,\Lambda^2} \, 
\left(\begin{array}{ccc}
v_1(\mu)^2&0&0\\
0&v_2(\mu)^2&0\\
0&0&v_3(\mu)^2
\end{array}\right)\, ,
\label{MassMatrixOfModel}
\eea
up to corrections of ${\cal O}(\varepsilon_\Phi/\varepsilon_K)$, ${\cal O}(\varepsilon_\Phi/\varepsilon_{X})$,
${\cal O}(\varepsilon_{\Phi 2}/\varepsilon_{\Phi 3})$ or
${\cal O}(\varepsilon_{\Phi Y1}/ \varepsilon_{Yi})$,
where $\varepsilon_\Phi$ represents $\varepsilon_{\Phi X1} $, 
$\varepsilon_{\Phi 2} $, $\varepsilon_{\Phi 3} $, etc.
%$\varepsilon_K$ and $\varepsilon_X$ 
%parametrize typical magnitudes of
%the second derivatives of $V^{SU(9)\times U(1)} _{\,\Phi X}$
%and $V_X$, respectively, at their minima.

%\subsection{Radiative corrections by scalar bosons}

\begin{figure}[t]\centering
\psfrag{Phi}{\hspace{0mm}$\Phi$}
\psfrag{psiL}{\hspace{0mm}$\psi_L$}
\psfrag{eR}{\hspace{0mm}$e_R$}
\includegraphics[width=4cm]{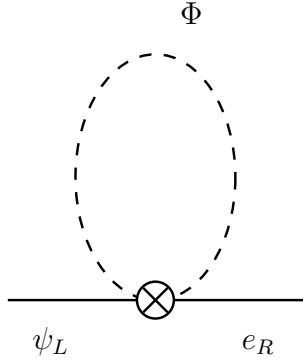}
\caption{\small 
Scalar loop diagram which contributes to charged lepton masses.
Dashed line represents all the mass eigenstates in $\Phi$ which 
couple to the operator ${\cal O}^{(\ell)}_3$
(represented by $\otimes$).
\label{FigScalarLoop}}
\end{figure}

We should check whether the radiative correction induced by
exchange of $\Phi$ (Fig.~\ref{FigScalarLoop}) 
violates Koide's mass formula or not.
With the reference potential eq.~(\ref{PotEssence2}),
we consider the limit $\varepsilon_{X1}\to\infty$
and $\varepsilon_{\Phi 3},\varepsilon_{\Phi X1},\varepsilon_{\Phi Y1}\to 0$
consistently with the assumed hierarchy of the parameters.
Physical modes of ${X_A}$ decouple in the former limit.
Thus, we consider only $V_{\overline{\Phi} 1}$ and
$V_{\overline{K} 1}$.
One may determine the scalar mass eigenstates explicitly
and find
\bea
\delta_\Phi m_i^{\rm pole} =- \frac{2\,{\lambda}}{(4\pi)^2}
\left[
\log\left(\frac{\mu^2}{4{\lambda}v_0^2}\right)+1
\right]\, m_i(\mu)\, .
\label{CorrByScalarLoop}
\eea
Since it has a form\, ${\rm const.}\times m_i$,
Koide's formula will not be affected.
This result may be
more non--trivial than one might think at a first glance, since
the diagram in
Fig.~\ref{FigScalarLoop} 
corresponds to incorporating the class of (infinite number of)
1--loop diagrams shown in
Fig.~\ref{1LoopAnalyInEFT}.
As a cross check, we also computed the coefficient of $\log\mu^2$ 
through
renormalization of the
operator ${\cal O}^{(\ell)}_3$ in the symmetric phase
($v_0=0$).

There are only three physical modes of $\Phi$ which gain
masses of order $v_0$;
these are $\delta \Phi^\alpha \equiv \Phi^\alpha - \Phi_0^\alpha$
which are proportional to $\Phi_0^\alpha$, $X_0^{\alpha\beta} \, \Phi_0^\beta$
and $i \,X_0^{\alpha\beta} \, \Phi_0^\beta$.
The first mode gives the correction eq.~(\ref{CorrByScalarLoop}), while 
the contributions of the second and third modes cancel.
Other modes have masses suppressed by 
$\varepsilon_{\Phi 3},\varepsilon_{\Phi X1},\varepsilon_{\Phi Y1}$,
so that
their contributions to the loop diagram in Fig.~\ref{FigScalarLoop}
are suppressed.

In fact the same features apply to the general
potential of $\Phi$ and $X$, if the parameters of the potential
satisfies the hierarchical relations required to realize the
vacuum configuration $\Phi = \Phi_0$ and $X=X_0$
(as discussed in Secs.~\ref{s4} and ~\ref{s5}).
Namely, absence of radiative
corrections to Koide's formula induced by scalar exchanges
can be shown in the limit
$\varepsilon_{\Phi 3},\varepsilon_{\Phi X1},\varepsilon_{\Phi Y1},
\mbox{etc.} \to 0$
(assuming that contributions from physical modes of $X$
decouple also in this case).
% with a
% more general potential, but we relegate a detailed 
% analysis to future studies.

As we discussed in Sec.~2 with an example of underlying
mechanism,
it is assumed that operators other than
${\cal O}^{(\ell)}_3$, which 
contribute to the charged lepton masses
at higher orders of
$1/\Lambda$, are absent
(or strongly suppressed) at $\mu=\Lambda$.
Since these operators are non--invariant under
$SU(nm)\times U(1)$,
sizes of these operators are determined by the physics above the scale $\Lambda$.
Within the $U(3)\times SU(2)$ effective theory starting from
this boundary condition,
other operators
are not induced radiatively 
at lower energy scales and the relation (\ref{RelPoleMassVEVs}) is 
preserved
(see also the discussion in Sec.~\ref{s2}).\footnote{
A simpler operator such as
$\bar{\psi}_{L}\, \Phi\, \Phi^\dagger  \Sigma_Y\varphi \, e_{R} $
%and
%$\bar{\psi}_{L}\, \Phi\, X_A^T\, \Phi \, X_A^\dagger\, \varphi \, e_{R} $
would be inappropriate for a candidate of ${\cal O}^{(\ell)}$, 
even though it gives the desired spectrum at tree level:
This operator induces a mass matrix
$\delta{\cal M}_\ell \propto {\bf 1}$
% and
%$\delta{\cal M}_\ell \propto \Phi_0\Phi_0^T$ 
radiatively, upon contraction of 
$\Phi$ and $\Phi^\dagger$.
%, and
%$X_A^T$ 
%and $X_A^\dagger$, respectively.
}

%\subsection{\boldmath ${\cal O}(\varepsilon_\Phi)$ corrections to Koide's formula}

In the limit 
$\varepsilon_{\Phi 3},\varepsilon_{\Phi X1}\ll
\varepsilon_{K1}, \varepsilon_{X1} $
and 
$\varepsilon_{\Phi Y1}\ll \varepsilon_{Y1},\varepsilon_{Y2},\varepsilon_{Y3}$,
the root--mass--ratios of the charged leptons are given by
$\sqrt{m_i/m_0}=v_i/v_0$,
where $m_0=m_1+m_2+m_3$.
They are in reasonable agreement with the corresponding
experimental values as we have seen in
eqs.~(\ref{eigenvaluesvi}) and (\ref{exprootmi}).
It would be instructive to see how much corrections are
induced to these values by the small parameters in the potential.
For simplicity, let us compute 
${\cal O}(\varepsilon_\Phi)$ corrections to
the charged lepton spectrum $(m_1,m_2,m_3)$,
corresponding to the potential eq.~(\ref{PotEssence2}) and
the higher--dimensional operator eq.~(\ref{exampleO3}).
The ${\cal O}(\varepsilon_\Phi)$ corrections read
\bea
&&
\delta \left(\!\!\sqrt{\frac{m_1}{m_0}}\right) \approx  \left(
-\frac{0.00209}{\varepsilon_{K1}}-\frac{0.00756}{\varepsilon_{X1}}
\right) \varepsilon_{\Phi 3} + \left(
\frac{0.0312}{\varepsilon_{K1}}+\frac{0.152}{\varepsilon_{X1}}
\right) \varepsilon_{\Phi X1} \, ,
\\ &&
\delta \left(\!\!\sqrt{\frac{m_2}{m_0}}\right) \approx  \left(
-\frac{0.00152}{\varepsilon_{K1}}-\frac{0.00406}{\varepsilon_{X1}}
\right) \varepsilon_{\Phi 3} + \left(
\frac{0.0227}{\varepsilon_{K1}}+\frac{0.0833}{\varepsilon_{X1}}
\right) \varepsilon_{\Phi X1} \, ,
\\ &&
\delta \left(\!\!\sqrt{\frac{m_3}{m_0}}\right) \approx  \left(
\frac{0.000406}{\varepsilon_{K1}}+\frac{0.00112}{\varepsilon_{X1}}
\right) \varepsilon_{\Phi 3} + \left(
-\frac{0.00607}{\varepsilon_{K1}}-\frac{0.0229}{\varepsilon_{X1}}
\right) \varepsilon_{\Phi X1} \, .
\eea
As can be seen, 
the magnitude of the 
correction is larger for smaller
mass eigenvalues,
reflecting the nature of a hierarchical spectrum, as we 
discussed in Sec.~\ref{s3}.
%We assume the relation eq.~(\ref{relalphas}), so that the
%QED corrections and the corrections by the $U(3)$
%family gauge interaction to 
%$\sqrt{m_i/m_0}$
%cancel at ${\cal O}(\alpha)={\cal O}(\alpha_F)$.
%Other corrections are expected to be negligible.
Comparing to eqs.~(\ref{eigenvaluesvi}) and (\ref{exprootmi}),
one finds constraints on typical orders of magnitude of
the parameters as
$\varepsilon_{\Phi 3}/\varepsilon_{K1}\simlt 10^{0}$,
$\varepsilon_{\Phi 3}/\varepsilon_{X1}\simlt 10^{-1}$,
$\varepsilon_{\Phi X1}/\varepsilon_{K1}\simlt 10^{-1}$,
$\varepsilon_{\Phi X1}/\varepsilon_{X1}\simlt 10^{-2}$,
provided there is
no correlation or fine tuning among these parameters, or
with ${\cal O}(
\varepsilon_{\Phi Y1}/ \varepsilon_{Yi})$ corrections.
On the other hand, the overall normalization, 
\bea
m_1 + m_2 + m_3 = 
\frac{25\,\kappa^{(\ell)}(\mu)\,\sigma\,v_{\rm ew}}{9\sqrt{2}\,\Lambda^2} \, v_0^2
\, \left[ 1 + \mbox{corr.} \right] \, ,
\eea
is subject to radiative corrections
induced by electroweak gauge interaction (including QED),
family gauge interaction and scalar exchanges, in addition to the
${\cal O}(\varepsilon_\Phi)$ and
${\cal O}(
\varepsilon_{\Phi Y1}/ \varepsilon_{Yi})$ corrections.

We define the following quantity as a measure of the degree
of violation of Koide's mass relation:
\bea
\Delta \equiv \frac{2(\sqrt{m_1}+\sqrt{m_2}+\sqrt{m_3})^2}
{3\, (m_1+m_2+m_3)}-1 \, .
\eea
This quantity vanishes if Koide's relation is satisfied.
With the reference potential and ${\cal O}^{(\ell)}_3$, the ${\cal O}(\varepsilon_\Phi)$
correction reads
\bea
&&
\Delta =
\left(\frac{1}{8\,\varepsilon_{K1}} + 
\frac{33-15\sqrt{2}\,x_0}{65\,\varepsilon_{X1}}\right)
\bar{\varepsilon}_{\Phi}
+\frac{67-75\sqrt{2}\,x_0}{390\,\varepsilon_{X1}}\,\varepsilon_{\Phi X1}
\\ &&
~~~
\approx
\left(
-\frac{0.00523}{\varepsilon_{K1}}-\frac{0.0171}{\varepsilon_{X1}}
\right) \varepsilon_{\Phi 3} + \left(
\frac{0.0781}{\varepsilon_{K1}}+\frac{0.346}{\varepsilon_{X1}}
\right) \varepsilon_{\Phi X1} \, .
\label{DeltaNum}
\eea
Comparing to the present experimental value
$\Delta^{\rm exp}=(1.1\pm 1.4)\times 10^{-5}$,
we obtain constraints on typical sizes of the parameters
more stringent than the previous ones:
$\varepsilon_{\Phi 3}/\varepsilon_{K1}\simlt 10^{-3}$,
$\varepsilon_{\Phi 3}/\varepsilon_{X1}\simlt 10^{-3}$,
$\varepsilon_{\Phi X1}/\varepsilon_{K1}\simlt 10^{-4}$,
$\varepsilon_{\Phi X1}/\varepsilon_{X1}\simlt 10^{-4}$.

It is easy to adjust the root--mass--ratios
$\sqrt{m_i/m_0}$ to be consistent with the 
current experimental values 
without violating Koide's relation,
as we discussed in Sec.~\ref{s3}.
For instance, it is achieved by
incorporating $V_{\Phi 2}$ with
$\varepsilon_{\Phi 2}/\varepsilon_{\Phi 3} \approx
-6\times 10^{-3}$ into the potential.

\section{Relevant Scales and Further Assumptions}
\label{s7}

Let us discuss the energy scales, $\Lambda$, $v_3(\sim v_0)$, $f_X$, 
involved in the present model.
It would be unnatural 
if there is a large hierarchy between
$v_3$ and $\Lambda$, or between $f_X$ and $v_3$.
As we speculated in Sec.~\ref{s2}, the 
scale of $U(3)$ symmetry breaking,
typically given by $v_3$, may be at $10^2$--$10^3$~TeV,
such that the QED correction is cancelled within a 
scenario of unification of the electroweak $SU(2)_L$ and 
family $U(3)$ symmetries.
There are two indications that 
the cut--off scale $\Lambda$
and the $U(3)$ symmetry breaking scale $v_3$ are not too far apart.
One indication
is the importance of the universality of the $U(1)$ and
$SU(3)$ gauge coupling constants eq.~(\ref{UnivCouplings}).
This universality may be protected above the cut--off scale by embedding 
$SU(3)$ and $U(1)$ in a simple group.
If $v_3$ is very different from $\Lambda$, however, 
the values of the two coupling
constants at $\mu \sim v_3$ would become too different.
Another indication consists in the relation (\ref{MassMatrixOfModel}), 
from which one derives 
\bea
\frac{v_3}{\Lambda}=
\left(
\frac{9\sqrt{2}\,m_\tau}{25\,\kappa^{(\ell)}\,\sigma\,v_{\rm ew}}
\right)^{1/2}
\approx 
\frac{1}{17\sqrt{\kappa^{(\ell)}\,\sigma}} \, 
\eea
up to electroweak corrections, etc.
$\sigma$ is smaller than $1/2$ and is expected to be not very much smaller.
Although it depends on the mechanism how the higher--dimensional operator
${\cal O}^{(\ell)}$ is generated,
if $\kappa^{(\ell)}$ is not large (as one naively expects), 
hierarchy between $v_3$ and $\Lambda$ is mild.
As for the scale $f_X$, it is required to be smaller than $\langle \Phi \rangle$
in order not to alter the spectrum of the family gauge bosons.
Numerically $f_X \simlt 0.3\, v_1\sim 0.005\, v_3$ is required from the present
constraint on
Koide's formula.
%Thus, stabilization of $f_X$ or $v_3$
%against $\Lambda$ does not seem to require severe fine tuning.

There are a few more assumptions implicit in the present model, which we have
not discussed so far.
We assume that the $SU(2)$ family gauge symmetry is broken spontaneously
at a scale higher than $\langle \Phi \rangle$.
This is required to 
protect the symmetry breaking pattern 
eq.~(\ref{SymBreakPat}), which constrains the form of the radiative
correction by the $U(3)$ gauge bosons.
To achieve this, we need additional fields or dynamics, such as
an $SU(2)$ doublet scalar field whose VEV breaks $SU(2)$.
As yet,
we have not succeeded to incorporate such a mechanism consistently
into our model.
Here, we simply assume that the breakdown of $SU(2)$ has occurred
without affecting the properties of our model described above.

We also assume cancellation of gauge anomalies and
decoupling of unwanted fermions.
Namely, we assume cancellation of anomalies introduced by the couplings
of fermions to family gauge bosons, at the scale where 
$U(3)\times SU(2)$ symmetry is unbroken.
This means that we need
fermions other than the SM fermions.
Fermions other than the SM fermions
are requisite in our model also because
$\psi_L$ and $e_R$ are embedded into larger multiplets
of $SU(nm)$.
At lower energy scales, $\mu \ll v$, all the additional fermions are
assumed to aquire
masses of order $v$ or larger, so that
they decouple from the SM sector at and below the electroweak scale.
Only the SM fermions remain at these scales.
Presently we do not have a model which fully explains 
these features.

%(Since the vacuum has no degeneracy up to the symmetry transformation,
%physical modes of $\Phi$, $X$ and $\Sigma_Y$ acquire
%masses of order $\langle \Phi \rangle$ or larger.)

%We also assume decoupling of unwanted particles.
%Since
%$\psi_L$, $e_R$, $\Phi$, $X$ and $\Sigma_Y$ 
%are embedded into larger multiplets
%of $SU(nm)$ above
%the cut--off scale $\Lambda$,
%we assume that other particles in the multiplets have acquired
%masses of order $\langle \Phi \rangle$ or larger, so that
%they decouple from the SM sector at and below the electroweak scale.
%Some of them do indeed acquire such masses with the
%potential under consideration, but not all of them do.
%For instance, no mechanism
%is incorporated for the
%mass generation of unwanted fermions. 
%Presently we do not have a model which explains 
%this mass generation mechanism.

\section{Lepton Flavor Violating Processes and Other Predictions}
\label{s8}

A most characteristic 
prediction of the present model is the
existence of lepton--flavor violating processes
induced by the family gauge interaction.
In the scenario, in which the
$U(3)$ family gauge symmetry and $SU(2)_L$ weak
gauge symmetry are unified
 at 
$10^2$--$10^3$~TeV scale, the family gauge bosons
have masses of the order of the unification scale.

As it is clear from eq.~(\ref{fmuTmu}),
flavor violating decays of a charged lepton
with only charged
leptons and/or photons in the final state, such
as 
$\mu\to 3\,e$ or $\mu \to e\,\gamma$,
are forbidden.
Flavor violating leptonic decays which involve neutrinos, such
as $\mu^-\to e^- \nu_e\bar{\nu}_\mu$,
are allowed,
but the present experimental sensitivities
are very low.
Presumably, the most sensitive process is 
$K_L\to e\,\mu$, although we need 
to make assumptions on the quark sector.
For instance, assuming that the down--type
quarks are in the same representation
of $U(3)$ as the charged leptons, 
and that the mass matrices of the charged leptons and
down--type quarks
are simultaneously diagonalized in an appropriate basis,
this process is induced by an effective 4--Fermi
interaction connecting the first and second generations:
\bea
&&
{\cal L}_{4f}^{(1,2)}=\frac{1}{2\,(v_1^2+v_2^2)}\,
\biggl[\left(
\bar{d}\,\gamma^\nu\gamma_5\,s+\bar{s}\,\gamma^\nu\gamma_5\,d
\right)
\left(
\bar{e}\,\gamma_\nu\gamma_5\,\mu+\bar{\mu}\,\gamma_\nu\gamma_5\,e
\right) 
\nonumber\\&&
~~~~~~~~~~~~~
~~~~~~~~~~~~~~~~
-
\left(
\bar{d}\,\gamma^\nu\,s-\bar{s}\,\gamma^\nu\,d
\right)
\left(
\bar{e}\,\gamma_\nu\,\mu-\bar{\mu}\,\gamma_\nu\,e
\right) \biggr]
+ \cdots
\, .
\label{4Fermi}
\eea
We  find
\bea
\Gamma(K_L\to e\mu)\approx
\frac{m_\mu^2m_{K_L}f_K^2}
{16\pi v_2^4}\,.
\eea
Comparing to the present experimental bound 
${\rm Br}(K_L\to e\mu)<4.7\times 10^{-12}$ \cite{Amsler:2008zz},
we obtain a limit $v_2\simgt 5\times 10^2$~TeV.
Naively this limit may already be marginally in
conflict with the estimated
unification scale in the above scenario.
We should note, however, that this depends 
rather heavily on our assumptions on the quark
sector.
In the case that there exist additional 
factors in the quark sector which suppress the decay width
by a few orders of magnitude,
we may expect
 a signal
for $K_L\to e\mu$ not far beyond the present experimental
reach.
Similarly
the process $K^+ \to \pi^+ e^- \mu^+$
may also be observable in the future.
%Another interesting possibility may be to search for excessive
%opposite--sign
%$e\,\mu$ pair productions at the LHC, through
%4--Fermi interactions such as eq.~(\ref{4Fermi}).

Another interesting
observation,
although it is much more model dependent,  is the following.
In order to stabilize Koide's formula, in our model, 
it is necessary to suppress
$SU(nm)\times U(1)$ non--invariant operators
in the potential of $\Phi$.
This indicates that $\Phi$ includes physical modes which are
much lighter than $v \sim 10^2$--$10^3$~TeV.
In particular, the lightest one, being singlet under the SM
gauge group, may decay into leptons through
the family gauge interaction or the operator ${\cal O}^{(\ell)}$
with a significant branching ratio.
Hence, if this lightest scalar boson happens to be produced
at the LHC, an excess in multi-lepton final states may be observed.

\section{Summary and Discussion}
\label{s9}

In this paper, we propose a model of charged lepton sector,
in the context of an EFT valid below the cut--off scale
$\Lambda$,
which predicts a charged lepton spectrum consistently
with the experimental values. 
In particular, we implement specific
mechanisms into the model, such that
the spectrum satisfies Koide's
mass formula within the present experimental accuracy.
In this model 
radiative corrections as well as other corrections to Koide's formula
are kept under control,
and this feature primarily differentiates the
present model from the other
models in the literature which predict Koide's formula.
By studying within EFT, we circumvent many problems,
at the price of introducing the cut--off scale at
$10^2$--$10^3$~TeV scale, while 
non--trivial relations between family symmetries
and observed charged lepton spectrum can
still be investigated.

In our model, we adopt a mechanism, through which
the charged lepton mass matrix
becomes proportional to the square of the VEV of a scalar field
$\Phi$ 
\cite{Koide:1989jq}.
On the basis of this mechanism,
we incorporate two new mechanisms in the model which are worth emphasizing:
\begin{enumerate}
\renewcommand{\labelenumi}{(\roman{enumi})}
\item
The radiative correction to Koide's formula
induced by family gauge interaction
has the same form as the QED correction with opposite sign.
This form is determined by the symmetry breaking pattern
eq.~(\ref{SymBreakPat}) and the representations of $\psi_L$
and $e_R$.
Within a unification scenario, cancellation of the QED correction
can take place.
\item
A charged lepton
spectrum, 
which has a hierarchical structure and
approximates the experimental values, 
follows from a simple potential $V_{\Phi 3}$, under the
condition that Koide's formula is protected.
\end{enumerate}
Existence of such simple mechanisms may indicate
relevance of $U(3)\times SU(2)$ family gauge symmetry
in relation to the charged lepton spectrum.

Our model is constructed as an effective theory
valid below the cut--off scale $\Lambda$
respecting this symmetry.
We introduce scalar fields $\overline{\Phi}$,
$\overline{X}$ and $\overline{Y}$ as multiplets of
$SU(nm)\times U(1)$, in which $U(3)\times SU(2)$ is embedded.
It is assumed that $SU(nm)\times U(1)$ is spontaneously
broken to $U(3)\times SU(2)$ below the scale $\Lambda$.
We minimize the potential of the scalar fields and
determine its classical vacuum.
The charged lepton masses are related to the VEVs of
the scalar fields 
at scale $\mu=\Lambda$,
$m_i^{\rm pole}\propto v_i(\Lambda)^2$;
at this scale  radiative
corrections to the VEVs essentially vanish within the
effective theory.
Then, the mass matrix
of the charged leptons are given in terms of the VEVs, such
that Koide's mass formula is stabilized, and that the spectrum
agrees with the experimental values.
This is achieved 
formally without fine tuning of parameters in the model,
except for (a) the tuning required for stabilization of the
electroweak scale $v_{\rm ew}$, and
(b) the tuning required for the cancellation of the QED correction,
that is, realizing $\alpha_F=\frac{1}{4}\alpha$ at relevant scales.
We argue that the latter tuning can be replaced by 
a tuning of the unification scale, within a scenario in which
$U(3)$ family gauge symmetry and $SU(2)_L$ weak
gauge symmetry  are unified
at $10^2$--$10^3$~TeV scale.

In addition our model may contain following fine tuning.
We were unable to explore the parameter space
of the $SU(nm)\times U(1)$--invariant potential
$V^{SU(nm)\times U(1)} _{\,\overline{\Phi}\, \overline{X}\,
\overline{Y}}$
sufficiently, due to
technical complexity.
It may be the case that certain fine tuning is necessary
to realize the configuration
eqs.~(\ref{vacconfigSUnm1})--(\ref{vacconfigSUnm3}) as a classical vacuum.

Evidently the present model is incomplete, since it is restricted to the
charged lepton sector.
The model should be implemented in a larger framework which
incorporates at least the following aspects missing in the present 
model:
(i) Including the quarks and explaining the
masses and mixings of the quarks and neutrinos;
(ii) Cancellation of anomalies introduced by the couplings
of fermions to family gauge bosons; 
(iii) Unification of 
$U(3)$ and $SU(2)_L$ gauge symmetries
at $10^2$--$10^3$~TeV scale.
Possibly these problems are solved simultaneously
in some model, and one anticipates that
such a model would necessarily contain a large number of new particles,
for the following reasons:
(a) all the particles are embedded into multiplets of large groups,
especially if one also requires to unify hypercharge $U(1)$ and
color $SU(3)$ gauge groups together with $U(3)$ and $SU(2)_L$;
(b) additional fermions are necessary to cancel anomalies; and
(c) scalar fields would be necessary to give masses of order $\langle \Phi \rangle$
to fermions (apart from
the SM fermions) through their VEVs \cite{FuturePub}.

%Our model contains unnatural features,
%even if they do not lead to logical inconsistencies.
%In particular, the following features are
%specific to our model:
%(1) It is unnatural that the unification scale 
%of $U(3)$ and $SU(2)_L$
%should be tuned to $10^2$--$10^3$~TeV 
%in order to realize $\alpha_F=\frac{1}{4}\alpha$, and
%(2) it is unclear why only ${\cal O}_3^{(\ell)}$ is
%induced as an ${\cal O}(1/\Lambda^2)$ operator
%that generates the lepton masses;
%for instance, ${\cal O}_1^{(\ell)}$ appears to be simpler, but
%it should be absent.
%Overall, the model is rather complicated, and the 
%source of complexity is
%mostly conspiracy to realize Koide's formula.

Although our model predicts a realistic lepton
spectrum, 
%formally without fine tuning of parameters
%(besides two exceptions),
in fact many of the questions are simply reassigned to 
physics above the cut--off scale
and remain unanswered:
While we replaced the conditions on
the lepton spectrum
by the boundary conditions of the effective potential,
we do not address which dynamics leads to these boundary
conditions.
(Only a speculation is given.)
We may nevertheless state that not only did we circumvent
fine tuning but also 
the problems actually simplified.
The required boundary conditions are
certain hierarchical structure among the couplings 
of the effective potential.
These conditions would be simpler to realize 
than, for instance,
to realize Koide's relation among the lepton Yukawa 
couplings with
$10^{-5}$ accuracy {\it a priori}.

Phenomenologically our model predicts existence of lepton
violating processes at $10^2$--$10^3$~TeV scale, assuming
the unification scenario at this scale.
The processes $K_L\to\mu e$ and $K^+\to\pi^+e^-\mu^+$
are expected to be sensitive to the predictions of our model,
although we need additional assumptions on the quark sector.
Stability of Koide's formula indicates existence of light modes
in $\Phi$, and
the lightest mode may decay into leptons with
a significant branching ratio;
they may generate an interesting signal at the LHC.

It is unlikely that the present model describes 
Nature correctly to the details,
since we can easily construct variants of the present
model with similar complexity.
Overall, the present model is rather complicated, and the 
source of complexity is
conspiracy to realize Koide's formula with a
high accuracy.
Hence, we place more emphasis on
the major mechanisms incorporated in the model, which 
look appealing and may reflect physics that governs
the spectrum of  the charged leptons.

%The characteristic of the present scenario is that no
%fine tuning of parameters is needed
%to produce the lepton spectrum close to a realistic one.
%If certain parameters are sufficiently small, 
%$m_e/m_0$, $m_\mu/m_0$, $m_\tau/m_0$ are automatically
%close to the experimental values, as well as Koide's mass formula is
%satisfied.

\section*{Acknowledgements}

The author is grateful to K.~Tobe for discussion.
This work is supported in part by Grant-in-Aid for
scientific research No.\ 17540228 from
MEXT, Japan.

\newpage

\appendix
\clfn
\section*{Appendices}

\section{\boldmath Generators of $U(3)$}
\label{appA}

The generators for the representation $({\bf 3},1)$
of $U(3)\simeq SU(3)\times U(1)$ are given by
\bea
&&
T^0= \frac{1}{\sqrt{6}} \left( \begin{array}{ccc}
1&
0&
0\\
0&
1&
0\\
0&
0&
1\\
\end{array} \right) ,~~~~~
T^1= \frac{1}{2} \left( \begin{array}{ccc}
0&
1&
0\\
1&
0&
0\\
0&
0&
0\\
\end{array} \right) ,~~~~~~
T^2= \frac{1}{2} \, \left( \begin{array}{ccc}
0&
-i&
0\\
i&
0&
0\\
0&
0&
0\\
\end{array} \right) , \nonumber \\ &&
T^3= \frac{1}{2} \left( \begin{array}{ccc}
1&
0&
0\\
0&
-1&
0\\
0&
0&
0\\
\end{array} \right) ,~~~~~
T^4= \frac{1}{2} \left( \begin{array}{ccc}
0&
0&
1\\
0&
0&
0\\
1&
0&
0\\
\end{array} \right) ,~~~~~~
T^5= \frac{1}{2} \, \left( \begin{array}{ccc}
0&
0&
-i\\
0&
0&
0\\
i&
0&
0\\
\end{array} \right) , 
\label{Texplicit}
\\ &&
T^6= \frac{1}{2} \left( \begin{array}{ccc}
0&
0&
0\\
0&
0&
1\\
0&
1&
0\\
\end{array} \right) ,~~~~~~~
T^7= \frac{1}{2} \left( \begin{array}{ccc}
0&
0&
0\\
0&
0&
-i\\
0&
i&
0\\
\end{array} \right) ,~~~~
T^8= \frac{1}{2\sqrt{3}}
\left( \begin{array}{ccc}
1&
0&
0\\
0&
1&
0\\
0&
0&
-2\\
\end{array} \right) ,
\nonumber
\eea
which satisfy eq.~(\ref{U3generators}).
\\

\section{\boldmath Decomposition of $X$ under $U(3)\times SU(2)$}
\label{appB}

$X$, which is in the $({\bf 45},Q_X)$ of $SU(9)\times U(1)$,
decomposes into
$X_S^1({\bf 6},{\bf 1},Q_X)\oplus X_S^5({\bf 6},{\bf 5},Q_X)\oplus
X_A(\bar{\bf 3},{\bf 3},Q_X)$
under $SU(3) \times SU(2) \times U(1)$.
Explicitly they can be constructed as follows:
\bea
\tilde{X}_{ik;jl}=X^{\alpha\beta}\,T^\alpha_{ij}\,T^\beta_{kl}\, 
~~~~~
\stackrel{\rm equiv.}{\Longleftrightarrow}
~~~~~
X^{\alpha\beta}=4\,\tilde{X}_{ik;jl}\,T^\alpha_{ji}\,T^\beta_{lk}\, ,
\eea
\bea
&&
\left( \! \! \begin{array}{c}\tilde{X}_S\\ \tilde{X}_A \end{array} 
\! \! \right)_{ik;jl}
=\frac{1}{4}\left(
\tilde{X}_{ik;jl} \pm \tilde{X}_{ki;jl} \pm \tilde{X}_{ik;lj} + \tilde{X}_{ki;lj}
\right)
\\ &&
~~~~~~~~~~~~~~
=\frac{1}{2}\left(
\tilde{X}_{ik;jl} \pm \tilde{X}_{ki;jl}
\right) \, ,
\eea
\bea
&&
(X_A)_{mn}=\epsilon_{mik}\,\epsilon_{njl}\,(\tilde{X}_A)_{ik;jl} \, ,
\\ &&
\rule[0mm]{0mm}{5mm}
(X_S^1)_{ik}=(\tilde{X}_S)_{ik;mm}\, ,
\\ &&
( X_S^5)_{ik;jl}=
(\tilde{X}_S)_{ik;jl}-\frac{1}{3}\,(\tilde{X}_S)_{ik;mm}\,\delta_{jl} \, .
\eea
When $\langle X\rangle =X_0$, 
the corresponding VEVs 
of $X_A$, $X_S^1$ and $X_S^5$ are given, respectively, by
\bea
&&
\langle X_A \rangle_{mn}=-\frac{5}{3}\,\delta_{mn} \, ,
\label{VEVofXA}
\\ &&
\langle X_S^1\rangle_{ik}=\frac{1}{6}\,\delta_{ik}\, ,
\\ &&
\langle X_S^5\rangle_{ik;jl}=
\frac{1}{12}(\delta_{il}\delta_{kj}+\delta_{ij}\delta_{kl})-\frac{1}{18}
\delta_{ik}\delta_{jl} \, .
\eea
\\

\section{\boldmath Properties of $V(\Phi,X)$}
\label{appC}

\subsection{\boldmath Maximizing $|\Phi^0|^2$}
\label{appC1}

\begin{figure}[t]\centering
\psfrag{z0}{\hspace{0mm}$z^0$}
\psfrag{z1}{\hspace{0mm}$z^1$}
\psfrag{z8}{\hspace{0mm}$z^8$}
\psfrag{z02}{\hspace{0mm}$z^0$}
\psfrag{zatot}{\hspace{0mm}$\sum_a z^a$}
\psfrag{rightarrow}{\hspace{5mm}$\Longrightarrow$}
\includegraphics[width=12cm]{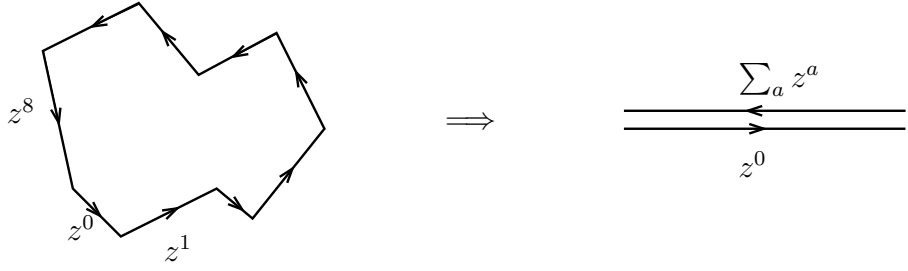}
\caption{\small 
\protect{Eq.~(\ref{condnonagon})} corresponds to a nonagon
 with a fixed length of circumference in the complex plane.
If we maximize $|z^0|$, the nonagon collapses to a line;
after overall phase rotation,
all $z^\alpha$'s can be made real, where $z^0=v_0^2>0$ and $z^a\leq 0$.
\label{nonagon}}
\end{figure}
We impose the conditions
\bea
{\Phi^\alpha}^*\Phi^\alpha=2\,v_0^2>0\, ,
~~~~~~~~~~~~~~~
(\Phi^0)^2=\Phi^a\,\Phi^a \, .
\eea
If we maximize $|\Phi^0|^2$ under these conditions, 
all $\Phi^\alpha$'s  can be made real simultaneously
by a common phase rotation.
Namely, there exists a phase $\theta$ such that
$e^{-i\theta}\Phi^\alpha \in {\bf R}$ for all $\alpha$. 
\\
{\it Proof}:~
Let 
\bea
z^0=(\Phi^0)^2\, ,
~~~~~~~
z^a=-(\Phi^a)^2 \, .
\eea
Then $z^\alpha$'s satisfy
\bea
\sum_{\alpha=0}^8 z^\alpha = 0\, ,
~~~~~~~~
\sum_{\alpha=0}^8 |z^\alpha|=2\,v_0^2 >0 \, .
\label{condnonagon}
\eea
These equations represent a nonagon with a fixed
length of circumference in the complex plane.
$|z^0|=|\Phi^0|^2$ is maximized when the nonagon collapses to
a line, where all $z^a$'s are parallel 
to one another and antiparallel to $z^0$ in the complex plane
 with
\bea
%z^0=-\sum_{a=1}^8 z^a\, ,
%~~~~~~~~~~
|z^0|=\sum_{a=1}^8 |z^a|={v_0^2}\, .
\eea
See Fig.~\ref{nonagon}.
\\

\subsection{\boldmath Variation of $V_{\Phi 3}$ at $\Phi=\Phi_0$}
\label{appC2}

The variation of $V_{\Phi 3}$, defined by eq.~(\ref{VPhi3}), 
is positive semi--definite at $\Phi=\Phi_0$ under
$SU(3)\times SU(3)\times U(1)$ transformation.
Namely,
if $\Phi_0'=U_1\Phi_0 U_2^\dagger$ with
$U_1U_1^\dagger=U_2U_2^\dagger={\bf 1}$, 
$V_{\Phi 3}(\Phi_0')\geq V_{\Phi 3}(\Phi_0)$.
\\
{\it Proof}:~
Let 
\bea
\Phi_1\equiv U_2U_1^\dagger\Phi_0'=U_2\Phi_0 U_2^\dagger \, .
\eea
Then, noting
\bea
(\Phi_0^0)^2=\Phi_0^a\,\Phi_0^a=v_0^2\, ,
%~~~~~~
%\Phi_0=\Phi_0^\dagger\, ,
\eea
$\Phi_1$ satisfies the same relation:
\bea
(\Phi_1^0)^2=\Phi_1^a\,\Phi_1^a=v_0^2\,  .
%~~~~~~
%\Phi_1=\Phi_1^\dagger\, .
\eea
Since $\Phi_0^\alpha$'s are real, $\Phi_1^\alpha$'s are also real.
According to Sec.~\ref{s3}, $\Phi_0$ is a configuration which
minimizes $V_{\Phi 3}(\Phi)$ under the condition
$(\Phi^0)^2=\Phi^a\Phi^a=v_0^2$ and $\Phi^\alpha \in {\bf R}$.
Therefore, $V_{\Phi 3}(\Phi_1)\geq V_{\Phi 3}(\Phi_0)$.
Since $\Phi_1$ and $\Phi_0'$ are connected by a
$U(3)\times SU(2)$ transformation,
$V_{\Phi 3}(\Phi_1)= V_{\Phi 3}(\Phi_0')$.
It follows $V_{\Phi 3}(\Phi_0')\geq V_{\Phi 3}(\Phi_0)$.
\medbreak

Due to this property, 
$V_{X1}+V_K+V_{\Phi 3}$ is minimized at
$X=X_0$ and $\Phi=\Phi_0$
under the
constraint $\Phi^\alpha \in {\bf R}$ and $\Phi^\alpha\Phi^\alpha=2\,v_0^2$.

\subsection{\boldmath $CP$ transformations of $\Phi$ and $X$}
\label{appC3}

$CP$ transformation of $\Phi$ is defined by
\bea
(CP)\,\Phi(x)\,(CP)^\dagger =\Phi({\cal P}x)^*\, ,
\eea
where ${\cal P}x=(x^0,-\vec{x})$.
Equivalently,
\bea
&&
(CP)\,\Phi^\alpha(x)\,(CP)^\dagger =[{\cal C}^{\alpha\beta}\,\Phi^\beta({\cal P}x)]^*\, 
\eea
with
%\\&&
%\rule[0mm]{0mm}{6mm}
\bea
{\cal C}^{\alpha\beta} = [{\rm diag.}(+1,+1,-1,+1,+1,-1,+1,-1,+1)]_{\alpha\beta} \, .
\eea
Similarly $CP$ transformation of $X$ is defined by
\bea
(CP)\,X^{\alpha\beta}(x)\,(CP)^\dagger =
[{\cal C}^{\alpha\alpha'}\,X^{\alpha'\beta'}({\cal P}x)\,
{\cal C}^{\beta'\beta}]^*\, ,
\label{CPtransfX}
\eea
or
\bea
&&
(CP)\,(X_A)_{ij}\,(CP)^\dagger = (X_A)_{ij}^* \, ,
\\&&
(CP)\,(X_S^1)_{ij}\,(CP)^\dagger = (X_S^1)_{ij}^* \, ,
\\&&
(CP)\,(X_S^5)_{ik;jl}\,(CP)^\dagger = (X_S^5)_{ik;jl}^* \, .
\eea
$CP$ transformations of other fields are the same as those of the SM.

For example, $V_{X1}$ defined by eq.~(\ref{VX1}) is $CP$--invariant.
An example of $CP$ non--invariant operator is 
\bea
V_{}=\frac{1}{2i}(f-f^*)
\eea
with
\bea
f=[\epsilon_{ijk}\epsilon_{lmn}\,(X_A)_{il}\,(X_A)_{jm}\,(X_A)_{kn}]
[\epsilon_{i'j'k'}\epsilon_{l'm'n'}\,(X_S^1)_{i'l'}\,(X_S^1)_{j'm'}\,(X_S^1)_{k'n'}]^*
\, .
\eea
\\

\subsection{\boldmath Stability of $V_X$ at $X=X_0$}
\label{appC4}

When $V_X(X)$ is invariant under $U(3)\times SU(2)$ and $CP$,
its first derivative vanishes 
$\partial V_X/\partial X^{\alpha\beta}=
\partial V_X/\partial {X^{\alpha\beta}}^*=0$
at $X=X_0$.
\\
{\it Proof}:~
\bea
&&
\delta V_X = 
\frac{\partial V_X}{\partial (X_A)_{ij}}\, (\delta X_A)_{ij}
+\frac{\partial V_X}{\partial (X_A)^*_{ij}}\, (\delta X_A)^*_{ij}
+\frac{\partial V_X}{\partial (X_S^1)_{ij}}\, (\delta X_S^1)_{ij}
\nonumber
\\&&
~~~~~~~~
+\frac{\partial V_X}{\partial (X_S^1)^*_{ij}}\, (\delta X_S^1)^*_{ij}
+\frac{\partial V_X}{\partial (X_S^5)_{ik;jl}}\, (\delta X_S^5)_{ik;jl}
+\frac{\partial V_X}{\partial (X_S^5)^*_{ik;jl}}\, (\delta X_S^5)^*_{ik;jl}\,.
\label{deltaVX1}
\eea
Due to the residual $SU(2)_V$ symmetry of the VEV $\langle X\rangle =X_0$,
the differential coefficients evaluated at $X=X_0$ take following forms:
\bea
&&
\frac{\partial V_X}{\partial (X_A)_{ij}}\, ,
\frac{\partial V_X}{\partial (X_A)^*_{ij}} \, ,
\frac{\partial V_X}{\partial (X_S^1)_{ij}}\, ,
\frac{\partial V_X}{\partial (X_S^1)^*_{ij}}
\Biggr|_{X=X_0}
\, \propto
\delta_{ij} \, ,
\\&&
\frac{\partial V_X}{\partial (X_S^5)_{ik;jl}}\, ,
\frac{\partial V_X}{\partial (X_S^5)^*_{ik;jl}}
\Biggr|_{X=X_0}
\, = C_1\, \delta_{ik}\delta_{jl} +
C_2\, \delta_{ij}\delta_{kl} +
C_3\, \delta_{il}\delta_{jk} \, ,
\eea
where $C_i$'s are constants.
Substituting to eq.~(\ref{deltaVX1}),
we have
\bea
&&
\delta V_X \biggr|_{X=X_0}= 
C_1'\,(\delta X_A)_{ii} +
C_2'\,(\delta X_A)^*_{ii} +
C_3'\, (\delta X_S^1)_{ii} +
 C_4'\, (\delta X_S^1)^*_{ii}
\nonumber
\\&&
~~~~~~~~~~~~~~~~
 +
C_5'\, (\delta X_S^5)_{ik;ik} +
 C_6'\, (\delta X_S^5)^*_{ik;ik}+
  C_7'\, (\delta X_S^5)_{ik;ki}+
  C_8'\,  (\delta X_S^5)^*_{ik;ki} \, ,
\eea
where we used $(\delta X_S^5)_{ii,jj}=0$.

An arbitrary infinitesimal variation of $X$, which
is symmetric and unitary, 
can be parametrized by
\bea
X^{\alpha\beta}+\delta X^{\alpha\beta} = 
W^{\alpha\alpha'}\, X^{\alpha'\beta'}\, {W^{\beta\beta'}}\, 
\eea
with
\bea
W\simeq \left(
\begin{array}{cccc}
1+i\epsilon_{00}^R & i(\epsilon_{01}^R+i\epsilon_{01}^I) & \cdots &
i(\epsilon_{08}^R+i\epsilon_{08}^I)\\
 i(\epsilon_{01}^R-i\epsilon_{01}^I) & 1+i\epsilon_{11}^R &\cdots &
 i(\epsilon_{18}^R+i\epsilon_{18}^I)\\
 \vdots & \vdots & \ddots & \vdots\\
 i(\epsilon_{08}^R-i\epsilon_{08}^I)& i(\epsilon_{18}^R-i\epsilon_{18}^I) 
&\cdots & 1+i\epsilon_{88}^R 
\end{array}
\right) \, ,
\eea
neglecting ${\cal O}(\epsilon^2)$ terms.
An explicit calculation shows that, for a variation $X=X_0+\delta X$, 
$(\delta X_A)_{ii} $,
$(\delta X_S^1)_{ii}$, $(\delta X_S^5)_{ik;ik}$ and $ (\delta X_S^5)_{ik;ki}$
depend only on $\epsilon^R_{\alpha\alpha}$ for
$0\leq \alpha\leq 8$.
(In this proof, no sum is taken over $\alpha$ in
$\epsilon^R_{\alpha\alpha}$
without explicit summation symbol $\sum_\alpha$.)
Hence, 
\bea
\delta V_X \biggr|_{X=X_0} \simeq
\sum_{\alpha=0}^8 \, \epsilon^R_{\alpha\alpha}\,
\frac{\partial}{\partial \epsilon^R_{\alpha\alpha}}
V_X(X_0+\delta X)\Biggr|_{\epsilon_{\alpha\beta}^R,\epsilon_{\alpha\beta}^I=0}
\, .
\eea
On the other hand, applying $CP$ transformation eq.~(\ref{CPtransfX}) 
to $X=X_0+\delta X$, one finds that $X_0$ is $CP$--even, whereas all
the coefficients of 
$\epsilon_{\alpha\alpha}^R$ 
in $\delta X$ are $CP$--odd.  
This means, if $V_X$ is $CP$--invariant,
\bea
\frac{\partial}{\partial \epsilon^R_{\alpha\alpha}}
V_X(X_0+\delta X)\Biggr|_{\epsilon_{\alpha\beta}^R,\epsilon_{\alpha\beta}^I=0}
=0 
~~~~~~\mbox{for}~~0\leq\alpha\leq 8
\, ,
\eea
so that the first derivative vanishes,
$\delta V_X |_{X=X_0}=0$.
\\

\section{\boldmath $\overline{Y}$, $\Sigma_Y$ and Their Potential}
\label{appD}

\subsection{\boldmath Minimum of $V_{\Sigma_Y}+V_{\Phi \Sigma_Y}$}
\label{appD1}

We show that $V_{\Sigma_Y}+V_{\Phi \Sigma_Y}$, given by
eqs.~(\ref{VSigmaY}) and (\ref{VPhiSigmaY}),
is minimized at the configuration eq.~(\ref{vacSigmaY})
in the limit
$\varepsilon_{\Phi Y1}\ll \varepsilon_{Y1},\varepsilon_{Y2},\varepsilon_{Y3}$.

It is known \cite{Li:1973mq} that, for
$\varepsilon_{Y1},\varepsilon_{Y2},\varepsilon_{Y3}>0$,
 $V_{\Sigma_Y}$ is minimized at 
\bea
\Sigma_Y = \sigma \, U\, U^T
~~~~~~;~~~~~~~~
\sigma = \sqrt{\frac{\varepsilon_{Y1}}{2\,(\varepsilon_{Y2}+3\,\varepsilon_{Y3})}}
\, ,
\label{SigmaYapp}
\eea
where $U$ is an arbitrary 3--by--3 unitary matrix.
Let
\bea
U_\Phi^\dagger\, \Phi\,\Phi^\dagger \,U_\Phi
=
\left(\begin{array}{rrr}
u_1^2&0&0\\
0&u_2^2&0\\
0&0&u_3^2
\end{array}
\right)
\equiv M_d^2
~~~~~~~;~~~~~~~~
u_i>0\, .
\label{Mdsq}
\eea
We assume that all $u_i$'s are different.
Substituting eqs.~(\ref{SigmaYapp}),(\ref{Mdsq}) to $V_{\Phi \Sigma_Y}$,
it is expressed as
\bea
V_{\Phi \Sigma_Y}=-\varepsilon_{\Phi Y1}\,\sigma^2\,
{\rm tr}
\left(
W^\dagger\,M_d^2\,W\,M_d^2
\right)
~~~~~~~;~~~~~~~~
W=U_\Phi^\dagger\,U\,U^T\,U_\Phi^*\, .
\eea
$W$ is unitary.
Define
\bea
M_d^2={\cal A}^\alpha\,T^\alpha
\,,
~~~~~~~
W^\dagger\,M_d^2\,W ={\cal B}^\alpha\,T^\alpha
~~~~~~~;~~~~~~~~
{\cal A}^\alpha,\,{\cal B}^\alpha \in {\bf R}\,.
\eea
Then ${\cal A}^\alpha{\cal A}^\alpha={\cal B}^\alpha{\cal B}^\alpha$,
since 
$\displaystyle
{\rm tr}\left[\left(W^\dagger\,M_d^2\,W\right)^2\right]
={\rm tr}\left(M_d^4\right)$.
Hence, $V_{\Phi \Sigma_Y}=
-\frac{1}{2}\,
\varepsilon_{\Phi Y1}\,\sigma^2\,{\cal A}^\alpha{\cal B}^\alpha$
is minimized when ${\cal A}^\alpha={\cal B}^\alpha$.
This means $W=U_d$ and
$UU^T= U_\Phi U_d U_\Phi^T$, where
$U_d$ is an arbitrary diagonal unitary matrix defined in
eq.~(\ref{U1VcubeTr});
it can be absorbed into a redefinition of $U_\Phi$ as
$U_\Phi'=U_\Phi U_d^{1/2}$.

\subsection{\boldmath Relation between $\overline{Y}$ and ${\Sigma_Y}$}
\label{appD2}

$\overline{Y}$ is in the 
({\boldmath $_{nm}C_2$},\,$Q_Y$) under
$SU(nm)\times U(1)$, where
{\boldmath $_{nm}C_2$} stands for the
second-rank antisymmetric representation of $SU(nm)$.
$\overline{Y}$ is defined to be unitary.
Thus,
\bea
\overline{Y}^{\xi\eta}=-\overline{Y}^{\eta\xi}
~~~~~~;~~~~~~
\overline{Y}^{\xi\eta}\,
{\overline{Y}^{\zeta\eta}}^*=\delta^{\xi\zeta}\,.
\eea
The indices take values
$0\leq\xi,\eta,\zeta,\dots\leq nm-1$. 

An orthonormal basis of $n$--by--$m$ matrices
is denoted by
$\{\overline{T}^\xi\}$ with  the normalization condition
\bea
{\rm tr}\Bigl( {\overline{T}^\xi}^\dagger
\overline{T}^\eta \Bigr)
=
{\rm tr}\Bigl( {\overline{T}^\xi}\,
{\overline{T}^\eta}^\dagger \Bigr)
=\frac{1}{2}\,\delta^{\xi\eta} \, .
\eea
In particular, the first 9 bases are taken as
\bea
\overline{T}^\xi_{ij}
=\left\{
\begin{array}{ll}
T^\xi_{ij}~~& \mbox{$1\leq i,j\leq 3$}\\
\rule[0mm]{0mm}{5mm}
~0&\mbox{otherwise}
\end{array}
\right.
~~~~~~~~
(0\leq\xi\leq 8)
\, .
\eea
We may identify $\Sigma_Y({\bf 6},1,Q_Y)$ embedded in $\overline{Y}$
as follows.
\bea
&&
\tilde{Y}_{ik;jl}=\overline{Y}^{\xi\eta}\,\overline{T}^\xi_{ij}\,
\overline{T}^\eta_{kl}\, 
~~~~~
\stackrel{\rm equiv.}{\Longleftrightarrow}
~~~~~
\overline{Y}^{\xi\eta}=4\,\tilde{Y}_{ik;jl}\,
\overline{T}^{\xi\,*}_{ij}\,\overline{T}^{\eta\,*}_{kl}\, ,
\\&&
\rule[0mm]{0mm}{9mm}
\left( \! \! \begin{array}{c}\tilde{Y}_{SA}\\ \tilde{Y}_{AS} \end{array} 
\! \! \right)_{ik;jl}
=\frac{1}{4}\left(
\tilde{Y}_{ik;jl} \pm \tilde{Y}_{ki;jl} \mp \tilde{Y}_{ik;lj} - \tilde{Y}_{ki;lj}
\right)
\\ &&
~~~~~~~~~~~~~~~
=\frac{1}{2}\left(
\tilde{Y}_{ik;jl} \pm \tilde{Y}_{ki;jl}
\right) \, ,
\\&&
\rule[0mm]{0mm}{6mm}
(\Sigma_Y)_{ik}=(\tilde{Y}_{SA})_{ik;45}
~~~~~~
\mbox{for $1\leq i,k\leq 3$}
\,.
\eea

\subsection{\boldmath A vacuum of 
the $SU(nm)\times U(1)$--invariant potential}
\label{appD3}

We analyze a vacuum configuration of the
$SU(nm)\times U(1)$--invariant potential
$V^{SU(nm)\times U(1)} _{\,\overline{\Phi}\, \overline{X}\, \overline{Y}}$
given by
eq.~(\ref{GenSUnmU1Pot}).
We restrict our analysis to the case
$(n,m)=(4,5)$ and consider only
$C(p_i,p'_i,q_i,q'_i)$ for $p_i,p'_i\leq 1$ and arbitrary $q_i,q_i'$,
while all other $C(p_i,p'_i,q_i,q'_i)$ are set equal to zero.
In this restricted parameter space
spanned by $\{C(p_i,p'_i,q_i,q'_i)\}$, we examine if
the configuration given by
eqs.~(\ref{vacconfigSUnm1})--(\ref{vacconfigSUnm3})
can minimize
$V^{SU(nm)\times U(1)} _{\,\overline{\Phi}\, \overline{X}\, \overline{Y}}$.
We assume that the $U(1)$ charge vanishes,
\bea
Q_{\rm tot}
\equiv q_i\sum_i Q(z_i(p_i)) - q'_i\sum_i Q(z_i(p'_i)) 
=0 \, ,
\eea
only in the sector for which $\sum_i (q_i+q_i')>1$.
This is not a strong condition: Except when
$Q_X$ and $Q_Y$ satisfy specific relations, this condition is met.

We have checked the following two properties.
(I) At each point of the parameter space,
the first derivative of 
$V^{SU(nm)\times U(1)} _{\,\overline{\Phi}\, \overline{X}\, \overline{Y}}$
vanishes at the configuration 
eqs.~(\ref{vacconfigSUnm1})--(\ref{vacconfigSUnm3}),
if 
\bea
\sigma \leq \frac{1}{2} \, .
\label{Cond-sigma}
\eea
(II) The configuration 
eqs.~(\ref{vacconfigSUnm1})--(\ref{vacconfigSUnm3})
minimizes 
$V^{SU(nm)\times U(1)} _{\,\overline{\Phi}\, \overline{X}\, \overline{Y}}$,
if the condition (\ref{Cond-sigma}) is satisfied
and at each point in a
hypersurface $S$ in the parameter space;
the hypersurface $S$ is defined by the condition
$C(p_i,p_i',q_i,q_i')\geq 0$ if
$p_i= p_i'$ and $q_i= q_i'$ for all $i$,
while $C(p_i,p'_i,q_i,q'_i)=0$ if 
$p_i\neq p_i'$ or $q_i\neq q_i'$ for any $i$.
These two properties (I)(II) ensure that, of each point
in $S$, there exists
a neighborhood, which has a non--zero volume, and in which 
$V^{SU(nm)\times U(1)} _{\,\overline{\Phi}\, \overline{X}\, \overline{Y}}$
is minimized by the configuration in question.
Namely, there exists a finite volume (non--zero measure)
in the parameter space
(at least) in a neighborhood of
$S$, in which the desired configuration
becomes a vacuum.

The above properties (I)(II) are verified in the following manner.
It suffices to show that all $z_i(p_i)$ for $p_i\leq 1$
can be brought to
zero simultaneously at the configuration 
eqs.~(\ref{vacconfigSUnm1})--(\ref{vacconfigSUnm3})
by appropriately adjusting components of $\overline{Y}$
except for $\Sigma_Y$.
In fact, in this case,
$V^{SU(nm)\times U(1)} _{\,\overline{\Phi}\, \overline{X}\, \overline{Y}}$
as well as its first derivative vanish at any point of the parameter space.
Thus, property (I) follows.
Since
$V^{SU(nm)\times U(1)} _{\,\overline{\Phi}\, \overline{X}\, \overline{Y}}
\geq 0$ in $S$,
the property (II) follows as well.
We have checked numerically that all $z_i(p_i)$ can be brought to
zero at the configuration 
eqs.~(\ref{vacconfigSUnm1})--(\ref{vacconfigSUnm3})
by explicitly constructing the corresponding $\overline{Y}$
for a given value of $\sigma$.
This turned out to be possible (at least) if
the condition (\ref{Cond-sigma}) is met, since there are quite large
degrees of freedom in the choice of $\overline{Y}$.
(If $\sigma$ is too large, it conflicts the unitarity condition of
$\overline{Y}$.)


\newpage
\begin{thebibliography}{0}
%\cite{Koide:1982wm}
\bibitem{Koide:1982wm}
  Y.~Koide,
  %``Exactly Solvable Model Of Relativistic Wave Equation And Meson Spectra,''
  Nuovo Cim.\  A {\bf 70} (1982) 411
  [Erratum-ibid.\  A {\bf 73} (1983) 327].
  %%CITATION = NUCIA,A70,411;%%

%\cite{Amsler:2008zz}
\bibitem{Amsler:2008zz}
  C.~Amsler {\it et al.}  [Particle Data Group],
  %``Review of particle physics,''
  Phys.\ Lett.\  B {\bf 667} (2008) 1.
  %%CITATION = PHLTA,B667,1;%%

%\cite{Foot:1994yn}
\bibitem{Foot:1994yn}
  R.~Foot,
  %``A Note on Koide's lepton mass relation,''
  arXiv:hep-ph/9402242.
  %%CITATION = HEP-PH/9402242;%%

%\cite{Koide:1983qe}
\bibitem{Koide:1983qe}
  Y.~Koide,
  %``A New View Of Quark And Lepton Mass Hierarchy,''
  Phys.\ Rev.\  D {\bf 28} (1983) 252;
  %%CITATION = PHRVA,D28,252;%%
%\cite{Esposito:1995bw}
%\bibitem{Esposito:1995bw}
  S.~Esposito and P.~Santorelli,
  %``A Geometric Picture For Fermion Masses,''
  Mod.\ Phys.\ Lett.\  A {\bf 10} (1995) 3077.
  %[arXiv:hep-ph/9603369].
  %%CITATION = MPLAE,A10,3077;%%

\bibitem{Koide:2005nv}
  For a review, see Y.~Koide,
  %``Challenge to the mystery of the charged lepton mass formula,''
  arXiv:hep-ph/0506247. 
  %%CITATION = HEP-PH/0506247;%%

%\cite{Li:2006et}
\bibitem{Li:2006et}
  N.~Li and B.~Q.~Ma,
  %``Energy scale independence of Koide's relation for quark and lepton
  %masses,''
  Phys.\ Rev.\  D {\bf 73} (2006) 013009.
  %[arXiv:hep-ph/0601031].
  %%CITATION = PHRVA,D73,013009;%%

%\cite{Xing:2006vk}
\bibitem{Xing:2006vk}
  Z.~z.~Xing and H.~Zhang,
  %``On the Koide-like relations for the running masses of charged leptons,
  %neutrinos and quarks,''
  Phys.\ Lett.\  B {\bf 635} (2006) 107.
  %[arXiv:hep-ph/0602134].
  %%CITATION = PHLTA,B635,107;%%

%\cite{Ma:2006ht}
\bibitem{Ma:2006ht}
  E.~Ma,
  %``Lepton family symmetry and possible application to the Koide mass
  %formula,''
  Phys.\ Lett.\  B {\bf 649} (2007) 287.
  %[arXiv:hep-ph/0612022].
  %%CITATION = PHLTA,B649,287;%%

%\cite{Rosen:2007rt}
\bibitem{Rosen:2007rt}
  For recent works, see
  G.~Rosen,
  %``Heuristic development of a Dirac-Goldhaber model for lepton and quark
  %structure,''
  Mod.\ Phys.\ Lett.\  A {\bf 22} (2007) 283;
  %%CITATION = MPLAE,A22,283;%%
%\cite{Koide:2008zz}
%\bibitem{Koide:2008zz}
  Y.~Koide,
  %``O(3) Flavor Symmetry And An Empirical Neutrino Mass Matrix,''
  Phys.\ Lett.\  B {\bf 665} (2008) 227;
  %%CITATION = PHLTA,B665,227;%%
%\cite{Koide:2008sj}
%\bibitem{Koide:2008sj}
  %Y.~Koide,
  %``Empirical Neutrino Mass Matrix Related to Up-Quark Masses,''
  J.\ Phys.\ G {\bf 35} (2008) 125004;
  %[arXiv:0803.3101 [hep-ph]].
  %%CITATION = JPHGB,G35,125004;%%
%\cite{Koide:2008qm}
%\bibitem{Koide:2008qm}
  %Y.~Koide,
  %``Phenomenological Meaning of a Neutrino Mass Matrix Related to Up-Quark
  %Masses,''
  Phys.\ Rev.\  D {\bf 78} (2008) 093006;
  %[arXiv:0809.2449 [hep-ph]].
  %%CITATION = PHRVA,D78,093006;%%
%\cite{Koide:2008tr}
%\bibitem{Koide:2008tr}
  %Y.~Koide,
  %``Charged Lepton Mass Relations in a Supersymmetric Yukawaon Model,''
  arXiv:0811.3470 [hep-ph];
  %%CITATION = ARXIV:0811.3470;%%
%\cite{Haba:2007vy}
%\bibitem{Haba:2007vy}
  N.~Haba and Y.~Koide,
  %``New Origin of a Bilinear Mass Matrix Form,''
  Phys.\ Lett.\  B {\bf 659} (2008) 260;
  %[arXiv:0708.3913 [hep-ph]].
  %%CITATION = PHLTA,B659,260;%%
%\cite{Haba:2008wr}
%\bibitem{Haba:2008wr}
  %N.~Haba and Y.~Koide,
  %``F-term Induced Flavor Mass Spectrum,''
  JHEP {\bf 0806} (2008) 023,
  %[arXiv:0801.3301 [hep-ph]].
  %%CITATION = JHEPA,0806,023;%%
and references therein.

%\cite{Koide:1989jq}
\bibitem{Koide:1989jq}
  Y.~Koide,
  %``Charged lepton mass sum rule from U(3) family Higgs potential
  %model,''
  Mod.\ Phys.\ Lett.\  A {\bf 5} (1990) 2319.
  %%CITATION = MPLAE,A5,2319;%%

%\cite{Sumino:2008hu}
\bibitem{Sumino:2008hu}
  Y.~Sumino,
  %``Family Gauge Symmetry and Koide's Mass Formula,''
  Phys.\ Lett.\  B {\bf 671}, 477 (2009).
  %[arXiv:0812.2090 [hep-ph]].
  %%CITATION = PHLTA,B671,477;%%

%\cite{Koide:1995xk}
\bibitem{Koide:1995xk}
  Y.~Koide and M.~Tanimoto,
  %``U(3)-family nonet Higgs boson and its phenomenology,''
  Z.\ Phys.\  C {\bf 72}, 333 (1996).
  %[arXiv:hep-ph/9505333].
  %%CITATION = ZEPYA,C72,333;%%

%\cite{Koide:1995pb}
\bibitem{Koide:1995pb}
  Y.~Koide and H.~Fusaoka,
  %``Top quark mass enhancement in a seesaw type quark mass matrix,''
  Z.\ Phys.\  C {\bf 71} (1996) 459;
  %[arXiv:hep-ph/9505201].
  %%CITATION = ZEPYA,C71,459;%%
%\cite{Koide:1995xk}
%\bibitem{Koide:1995xk}
  Y.~Koide and M.~Tanimoto,
  %``U(3)-family nonet Higgs boson and its phenomenology,''
  Z.\ Phys.\  C {\bf 72} (1996) 333.
  %[arXiv:hep-ph/9505333].
  %%CITATION = ZEPYA,C72,333;%%
  \\
See also \cite{Koide:2005nv}
and references therein.

%\cite{Koide:2005ep}
\bibitem{Koide:2005ep}
  Y.~Koide,
  %``Permutation symmetry S(3) and VEV structure of flavor-triplet Higgs
  %scalars,''
  Phys.\ Rev.\  D {\bf 73}, 057901 (2006).
  %[arXiv:hep-ph/0509214].
  %%CITATION = PHRVA,D73,057901;%%
  
%\cite{Antusch:2007re}
\bibitem{Antusch:2007re}
There are a large number of papers on the fermion flavor
structure based on $SU(3)$ or $SO(3)$ family symmetry.
See, for instance,
%\cite{Berezhiani:1990wn}
%\bibitem{Berezhiani:1990wn}
  Z.~G.~Berezhiani and M.~Y.~Khlopov,
  %``The theory of broken gauge symmetry of generations,''
  Sov.\ J.\ Nucl.\ Phys.\  {\bf 51} (1990) 739;
  %[Yad.\ Fiz.\  {\bf 51} (1990) 1157].
  %%CITATION = YAFIA,51,1157;%%
%\cite{King:2005bj}
%\bibitem{King:2005bj}
  S.~F.~King,
  %``Predicting neutrino parameters from SO(3) family symmetry and  quark-lepton
  %unification,''
  JHEP {\bf 0508} (2005) 105;
  %[arXiv:hep-ph/0506297].
  %%CITATION = JHEPA,0508,105;%%
%\cite{deMedeirosVarzielas:2005ax}
%\bibitem{deMedeirosVarzielas:2005ax}
  I.~de Medeiros Varzielas and G.~G.~Ross,
  %``SU(3) family symmetry and neutrino bi-tri-maximal mixing,''
  Nucl.\ Phys.\  B {\bf 733} (2006) 31;
  %[arXiv:hep-ph/0507176].
  %%CITATION = NUPHA,B733,31;%%
%\cite{Appelquist:2006qq}
%\bibitem{Appelquist:2006qq}
  T.~Appelquist, Y.~Bai and M.~Piai,
%  %``Neutrinos and SU(3) Family Gauge Symmetry,''
  Phys.\ Rev.\  D {\bf 74} (2006) 076001;
%  [arXiv:hep-ph/0607174].
  %%CITATION = PHRVA,D74,076001;%%
  S.~Antusch, S.~F.~King and M.~Malinsky,
  %``Solving the SUSY Flavour and CP Problems with SU(3) Family Symmetry,''
  JHEP {\bf 0806} (2008) 068,
  %[arXiv:0708.1282 [hep-ph]].
  %%CITATION = JHEPA,0806,068;%%
and references therein.

%\cite{Appelquist:2006ag}
%\bibitem{Appelquist:2006ag}
%  T.~Appelquist, Y.~Bai and M.~Piai,
%  %``Quark mass ratios and mixing angles from SU(3) family gauge symmetry,''
%  Phys.\ Lett.\  B {\bf 637} (2006) 245
%  [arXiv:hep-ph/0603104].
  %%CITATION = PHLTA,B637,245;%%
%\cite{Appelquist:2006qq}
%\bibitem{Appelquist:2006qq}
%  T.~Appelquist, Y.~Bai and M.~Piai,
%  %``Neutrinos and SU(3) Family Gauge Symmetry,''
%  Phys.\ Rev.\  D {\bf 74} (2006) 076001
%  [arXiv:hep-ph/0607174].
  %%CITATION = PHRVA,D74,076001;%%



%\cite{DelCima:1999gg}
\bibitem{DelCima:1999gg}
  O.~M.~Del Cima, D.~H.~T.~Franco and O.~Piguet,
  %``Gauge independence of the effective potential revisited,''
  Nucl.\ Phys.\  B {\bf 551} (1999) 813.
  %[arXiv:hep-th/9902084].
  %%CITATION = NUPHA,B551,813;%%

%\cite{Dolan:1974gu}
\bibitem{Dolan:1974gu}
Earlier works are 
  L.~Dolan and R.~Jackiw,
  %``Gauge invariant signal for gauge symmetry breaking,''
  Phys.\ Rev.\  D {\bf 9}, 2904 (1974);
  %%CITATION = PHRVA,D9,2904;%%
%\cite{Nielsen:1975fs}
%\bibitem{Nielsen:1975fs}
  N.~K.~Nielsen,
  %``On The Gauge Dependence Of Spontaneous Symmetry Breaking In Gauge
  %Theories,''
  Nucl.\ Phys.\  B {\bf 101}, 173 (1975);
  %%CITATION = NUPHA,B101,173;%%
%\cite{Fukuda:1975di}
%\bibitem{Fukuda:1975di}
  R.~Fukuda and T.~Kugo,
  %``Gauge Invariance In The Effective Action And Potential,''
  Phys.\ Rev.\  D {\bf 13}, 3469 (1976).
  %%CITATION = PHRVA,D13,3469;%%
See also \cite{DelCima:1999gg} and references therein.

%\cite{Li:1973mq}
\bibitem{Li:1973mq}
  L.~F.~Li,
  %``Group Theory Of The Spontaneously Broken Gauge Symmetries,''
  Phys.\ Rev.\  D {\bf 9} (1974) 1723.
  %%CITATION = PHRVA,D9,1723;%%

\bibitem{FuturePub}
Y.~Sumino, in preparation.

\end{thebibliography}
\end{document}